\documentclass[12pt]{article}
\usepackage[margin=1in]{geometry}
\usepackage[pdftex,colorlinks=false]{hyperref}
\usepackage{graphicx}
\usepackage{amsmath}
\usepackage{amsthm}
\usepackage{amssymb}
\usepackage{bm}
\usepackage{color}
\usepackage[nosort]{cite}

\DeclareMathOperator{\Tr}{Tr}
\newcommand{\lieg}{\mathfrak{g}}
\newcommand{\lieh}{\mathfrak{h}}
\newcommand{\bbZ}{\mathbb{Z}}
\newcommand{\bbR}{\mathbb{R}}

\def\ket#1{\left\lvert #1\right\rangle}

\DeclareMathOperator{\hol}{hol}
\DeclareMathOperator{\ev}{ev}
\newcommand{\ptop}{\mathcal{P}}

\newcommand{\myvec}[1]{\bm{#1}}
\newcommand{\loops}{\mathcal{L}}
\newcommand{\loopsfix}{\Omega(M,x_{0})}
\newcommand{\spatial}{\myvec{S}}
\newcommand{\tangent}{\myvec{T}}
\newcommand{\deform}{\myvec{N}}
\newcommand{\xxxcA}{\mathcal{A}}
\newcommand{\wil}{W(\sigma)}
\newcommand{\half}{\frac{1}{2}}

\theoremstyle{plain}
\newtheorem{theorem}{Theorem}[section]
\newtheorem{lemma}[theorem]{Lemma}
\newtheorem{corollary}[theorem]{Corollary}

\theoremstyle{remark}
\newtheorem{remark}{Remark}[section]
\def\rf#1{(\ref{eq:#1})}
\def\lab#1{\label{eq:#1}}
\def\br{\begin{eqnarray}}
\def\er{\end{eqnarray}}
\def\be{\begin{equation}}
\def\ee{\end{equation}}
\def\({\left(}
\def\){\right)}
\def\pa{\partial}

\def\IR{\mathbb{R}}
\def\IZ{\mathbb{Z}}
\def\IC{\mathbb{C}}
\newcommand{\sbr}[2]{\left\lbrack\,{#1}\, ,\,{#2}\,\right\rbrack}

\def\vp{\varphi}
\def\u2{\mid u\mid^2}
\def\ck{{\cal K}}
\DeclareMathOperator{\tr}{tr}
\def\one{\hbox{{1}\kern-.25em\hbox{l}}}
\def\nonu{\nonumber}

\numberwithin{equation}{section}

\begin{document}


\begin{titlepage}

\vspace{.2in}
\begin{center}
{\large\bf Integrable theories and loop spaces: fundamentals, applications and
new developments }
\end{center}

\vspace{.5cm}

\begin{center}
Orlando Alvarez$^{a,}$\footnote{email: \tt oalvarez@miami.edu},
L. A. Ferreira$^{b,}$\footnote{email: {\tt laf@ifsc.usp.br} 
[email correspondent]} 
and J. S\'anchez-Guill\'en$^{c,}$\footnote{email: {\tt joaquin@fpaxp1.usc.es}}

\vspace{.5 in}
\small

$^a$ Department of Physics\\
    University of Miami\\
    P.O. Box 248046\\
    Coral Gables, FL 33124, USA

\par \vskip .2in \noindent
$^b$ Instituto de F\'\i sica de S\~ao Carlos; IFSC/USP;\\
Universidade de S\~ao Paulo  \\ 
Caixa Postal 369, CEP 13560-970, S\~ao Carlos-SP, BRAZIL

\par \vskip .2in \noindent
$^c$ Departamento de Fisica de Particulas, Universidad
     de Santiago
     \\
      and Instituto Galego de Fisica de Altas Enerxias (IGFAE)
     \\ E-15782 Santiago de Compostela, SPAIN

\normalsize
\end{center}

\vspace{.5in}

\begin{abstract}

We review our proposal to generalize the standard two-dimensional
flatness construction of Lax-Zakharov-Shabat to relativistic field
theories in $d+1$ dimensions.  The fundamentals from the theory of
connections on loop spaces are presented and clarified.  These ideas
are exposed using mathematical tools familiar to physicists.  We
exhibit recent and new results that relate the locality of the loop
space curvature to the diffeomorphism invariance of the loop space
holonomy.  These result are used to show that the holonomy is abelian
if the holonomy is diffeomorphism invariant.  

These results justify in part and set the limitations of the local
implementations of the approach which has been worked out in the last
decade.  We highlight very interesting applications like the
construction and the solution of an integrable four dimensional field
theory with Hopf solitons, and new integrability conditions which
generalize BPS equations to systems such as Skyrme theories.
Applications of these ideas leading to new constructions are
implemented in theories that admit volume preserving diffeomorphisms
of the target space as symmetries.  Applications to physically
relevant systems like Yang Mills theories are summarized.  We also
discuss other possibilities that have not yet been explored.

\end{abstract} 
\end{titlepage}

\newpage
\setcounter{page}{2}
\small
\tableofcontents
\normalsize

\newpage

\section{Introduction}
\label{sec:introduction}

Symmetry principles play a central role in Physics and other sciences.
The laws governing the four fundamental interactions of Nature are
based on two beautiful implementations of such ideas.  The
electromagnetic, weak and strong nuclear interactions have their basic
structures encoded into the gauge principle that leads to the
introduction of a non-integrable phase in the wave functions of the
particles.  The gravitational interaction originates from the
principle of equivalence that leads to the general covariance of the
dynamics under coordinate transformations.

An important and crucial issue in the success of these principles is
the identification of the fundamental objects that are acted upon by
the symmetry transformation.  Those objects belong to a representation
space of the transformation groups where dynamical consequences can be
studied.  Our understanding of the atomic, nuclear and particle
phenomena has led us to a description of the world in terms of objects
that have the dual character of point particles and waves.  The
symmetries we understand are formulated in terms of unitary
representations on the Hilbert space of the quantum  theory.  The
great lacuna in that formulation is Einstein's Theory of General
Relativity.  We have not yet identified the quantum objects that could
fully illuminate the symmetry principles of the gravitational interaction.

The other lacuna in our understanding of physical phenomena is the
strong coupling regime or the non-perturbative regime of many theories.
Even though this has apparently a more technical flavor it may hide
some very important pieces of information in our quest for the
description of Nature.  The lack of non-perturbative methods has
prevented or at most delayed developments in many fronts of our
knowledge, from condensed matter physics to the confinement of quarks
and gluons, weather prediction and a variety of mechanisms in
biological systems.  We may have the microscopic theories for
practically all these phenomena but we just do not know how to solve
them.  There have been many successes in the strong coupling regime
though the methods and the models are quite diverse.  The discussions
in this review are motivated by successes in certain low dimensional
models that play an important role in many areas of Physics from
condensed matter to high energy physics.  We want to be concrete and
to make clear the points that will be discussed in this review.
Consider the simple and very well known sine-Gordon model.  This is a
$1+1$ dimensional field theory of a real scalar field $\phi$ with
equation of motion 
\be 
\partial_t^2\phi - \partial_x^2\phi +
\frac{m^2}{\beta} \sin \(\beta\,\phi\)=0 \,, \lab{sgeq} 
\ee 
where
$\beta$ is a dimensionless coupling constant and $m$ a mass parameter
(in natural units where $c=\hbar =1$).  The symmetries of this theory
are the Poincar\'e group in two dimensions, and the discrete
transformations $\phi\rightarrow \phi+\frac{2\,\pi\, n}{\beta}$, with
$n$ an integer.  Certainly these symmetries are important in the study
of the model, but they are far from being responsible for the amazing
properties this theory presents.  In order to understand the full
symmetry group of the model we have to describe the dynamics in terms
of additional objects besides the scalar field $\phi$.  You can easily
verify that the sine-Gordon equation \rf{sgeq} is equivalent to a zero
curvature condition \br \partial_{+}\, A_{-}- \partial_{-}\, A_{+} +
\sbr{A_{+}}{A_{-}}=0 \lab{zcsg} \er for a connection that is a
functional of the scalar field $\phi$: \br
A_{+}=-\frac{m^2}{4}\(\begin{array}{cc} 0& e^{i\beta\,\phi}\\
\lambda\, e^{-i\beta\,\phi}&0
\end{array}\)
\qquad
\qquad
A_{-}= \(\begin{array}{cc}
-\frac{i\beta}{2}\,\partial_{-}\phi & 1/\lambda\\
1 & \frac{i\beta}{2}\,\partial_{-}\phi
\end{array}\).
\er
Here we are using light cone coordinates $x_{\pm}=\(x_0\pm x_1\)/2$, 
and $A_{\pm}=A_{0}\pm A_{1}$, with $x_0$ and $x_1$ being the time and
space coordinates respectively. The parameter $\lambda$ is arbitrary
and may be chosen to be complex. It is called the spectral parameter and it
makes the algebra, where the connection $A_{\mu}$ lives to be the infinite
dimensional $sl(2)$ loop algebra. In terms of these new
objects you can see that the sine-Gordon theory has an infinite dimensional
group of symmetries given by the gauge transformations
\be
A_{\mu}\rightarrow g\, A_{\mu}\, g^{-1}-\partial_{\mu}g\, g^{-1}
\lab{hiddensym}
\ee
with $g$ being elements of the $sl(2)$ loop group ($\lambda$
dependent).  These are called hidden symmetries because they are not
symmetries of the equations of motion or of the Lagrangian. They play
an important role in the development of exact methods in integrable
field theories and practically all that is known in soliton theory is
derived from the zero curvature condition \rf{zcsg}. Exact solutions can be
constructed using techniques like the inverse scattering or dressing
methods. In addition, in $1+1$ dimensions, the flatness condition
\rf{zcsg} is in fact a conservation law. One can show that the
conserved quantities after the imposition of appropriate boundary conditions 
are given by the eigenvalues of the operator 
\be
W_{\Gamma} = P\exp\left[ \int_{\Gamma} d\sigma \, A_{\mu}\,
  \frac{dx^{\mu}}{d\sigma}\right] ,
\lab{wgamma}
\ee
where $\Gamma$ is the spatial sub-manifold of space-time, and where $P$
means path ordering. By expanding in positive and negative powers of
the spectral parameter $\lambda$ you get an infinite number
of conserved charges. The evaluation of those charges can be better
done by using the central extension of the loop algebra, the affine
Kac-Moody algebra, using for instance the methods of \cite{simplecharges}. 

For soliton theories belonging to the class of the sine-Gordon model,
the so-called affine Toda theories, the hidden symmetries
\rf{hiddensym} can be understood in simpler terms. Such theories can
be obtained by Hamiltonian reduction of the so-called two-loop WZNW
model \cite{twoloop} that is invariant under the local symmetry group
$G_L\times G_R$ where $G_{L/R}$ are two copies of the loop group
mentioned above. By Hamiltonian reduction the corresponding Noether
charges can be related to the conserved charges coming from the
eigenvalues of the operator \rf{wgamma}. The algebra of the reduced
currents of the two-loop WZNW model is of the so-called
$W$-algebra type and the associated symmetries seem to mix in a
non-trivial way internal and space-time transformations. 

The hidden symmetries associated to \rf{hiddensym} are known to exist
in theories defined on a two dimensional space-time.  Of course the
existence of similar structures in dimensions higher than two would be
very important to understand the non-perturbative aspects of many
physical phenomena.  It is natural to ask if a change of the basic
objects used to represent the dynamics of a theory could aid in
investigating such structures.  Some years ago we proposed an
approach to construct what could perhaps be integrable theories in
higher dimensions \cite{Alvarez:1997ma}.  Since we are basically
interested in finding symmetries beyond those already known in
ordinary field theories, the main idea is to ponder how the conserved
charges of the type \rf{wgamma} would look like in higher dimensions.
We expect them to involve integrations on the spatial sub-manifold,
and so in a $d+1$ dimensional space-time it would be an integration in
$d$ dimensions.  The conservation laws associated to \rf{wgamma}
follow from the fact that the path ordered integrals of the connection
$A_{\mu}$ are path independent, and that in  turn follows, via
the non-abelian Stokes theorem, from the flatness of the connection.
So we need the generalization of the concept of a flat connection
that can be integrated on a $d$ dimensional surface in space-time.  As
discussed in \cite{Alvarez:1997ma}, the key concept is that of connections
on loop spaces.  Take the case of a $2+1$ dimensional space-time,
where the relevant surface is two dimensional.  We can fix a point
$x_0$ on such surface and scan it with closed loops, starting and
ending at $x_0$.  The surface can therefore be seen as a collection of
loops.  By ordering the loops, the surface becomes therefore a path in
the space of all loops.  What we need therefore is a one-form
connection on loop space.  The path ordered integral of such
connection on loop space will replace the operator \rf{wgamma} and
its flatness condition leads to the conservation laws.

In order to implement such ideas we need to connect the objects in
loop space with those in space-time. The proposal put forward in
\cite{Alvarez:1997ma}  was to construct a connection in loop space.
For example, in the case of a $2+1$ dimensional space-time we 
suggested that 
a rank two
antisymmetric tensor $B_{\mu\nu}$ and a one form connection
$A_{\mu}$ were the necessary ingredients. 
The connection in loop space introduced in
\cite{Alvarez:1997ma} was then
\be
{\cal A}\(x^{\mu}\(\sigma\)\) = \int_{\gamma} d\sigma \, W^{-1} \, B_{\mu\nu}\,
W\, \frac{d\,x^{\mu}\(\sigma\)}{d\sigma}\, \delta x^{\nu}\(\sigma\)
\lab{conectionloopspace}
\ee
where the integral is made on a loop $\gamma$ in space-time,
parametrized by $\sigma$. The quantity $W$ is obtained from the
connection $A_{\mu}$ through the differential equation
\be
\frac{d\, W}{d\sigma}+A_{\mu}\,
\frac{d\,x^{\mu}\(\sigma\)}{d\sigma}\,W=0
\ee
Notice that the quantity $W$ in \rf{conectionloopspace} implements a
parallel transport of $B_{\mu\nu}$, and that leads to better behaviour under 
gauge transformations. 
In order to obtain conservation laws we imposed the flatness
condition on the connection on loop space
\be
\delta {\cal A}+ {\cal A}\wedge {\cal A}=0
\lab{flatloop}
\ee

For a  space-time of dimension $d+1$ we had to consider generalized
loop spaces, \emph{i.e.}, the space of maps from the sphere $S^{d-1}$ to the
space-time. The connection will then be defined in terms of an 
antisymmetric tensor of rank $d$, and possibly additional lower rank tensors. 

The purpose of the present paper is twofold. First we make a
review of the proposal of \cite{Alvarez:1997ma} for the implementation
of zero curvature conditions on loop spaces that lead to conservation
laws and hidden symmetries for theories defined in a space-time of any
dimension. We also review and discuss the developments that have followed
from such approach giving many examples. Second we present new
results about the method.     

The most important new results are given in Section~\ref{sec:conn}, 
and they are concerned with the concept of
$r$-flatness. We have stated that the conserved charges are given by path
ordered integrals of the connection in loop space. Such paths
correspond to a surface in space-time. The charges should depend only
the physical surface and not on the way we scan it with loops. In
other words the charges should not depend upon the parametrization of
the surface. Therefore, the path ordered integral of the connection on
loop space should be re-parametrization invariant. A connection
satisfying this is called $r$-flat. The most important result of section
\ref{sec:conn} is to show that a $r$-flat connection ${\cal A}$ in
loop space must satisfy  
\be
{\cal A}\wedge {\cal A}=0
\ee
Therefore, in order to have conservation laws we need the two summands in
\rf{flatloop} to vanish separately. 
The second important result of Section~\ref{sec:conn} is that the
holonomy group of $r$-flat connections in loop spaces is always
abelian. These conditions drastically reduce
the possible non-trivial structures we can have for the 
implementation of hidden
symmetries for physical theories in a space-time of dimension higher
than two. Our results have the character of a \emph{No-Go Theorem}. 

In Section~\ref{sec:localcurv} we discuss the local conditions in
space-time which are sufficient for the vanishing of the curvature of
the connection in loop space.  Such local conditions are the ones that
have been used in the literature to construct physical theories with
an infinite number of conservation laws in any dimension.  In
Section~\ref{sec:examples} we provide some examples of such theories.
The possibilities of using the approach for developing methods for the
construction of exact solutions is discussed in
Section~\ref{sec:solutions} and many examples are given.  Further
applications of our approach to integrable theories in any dimension
are given in Section~\ref{sec:app} including some examples possibly
relevant for the low energy limit (strong coupling) of gauge theories.


\section{Connections in loop space}
\label{sec:conn}

\subsection{Philosophy}
\label{sec:philosophy}

In this Section we discuss the theory of connections on loop spaces
from a physicists viewpoint using geometrical and topological concepts
at the level of the text by Nakahara~\cite{Nakahara:book}.
Connections on loop spaces is an old subject in the physics
literature, see for example \cite{Kalb:1974yc,Nepomechie:1982rb}.  We
restrict our discussion in this Section to what is needed in
applications to integrable models.  We present the ideas developed in
\cite{Alvarez:1997ma} and new results in a slightly different way that
is faithful to the original presentation.  The motivation for that
work was the generalization to higher dimensions of the ideas and of
the technology that was developed around the Lax-Zakharov-Shabat
framework\cite{lax} for integrable systems in $(1+1)$-dimensions.  The
basic idea is that the equations of motion may be formulated as a
flatness condition with an appropriate connection.  The holonomy of
the connection is independent of the loop used and the holonomy can be
massaged to construct an infinite number of conservation laws.  The
$(1+1)$ conservation laws involve traces of the holonomy, $\Tr
\exp\left(\int A\right)$, using an ordinary connection.  The
generalization to a $(2+1)$-dimensional spacetime should be an object
of the type $\Tr \exp\left(\int B \right)$ where $B$ is now a $2$-form
since space is two dimensional.  Continuing in this fashion you would
require an $n$-form $B^{(n)}$ in the $(n+1)$-dimensional case where
the answer would be $\Tr \exp\left(\int B^{(n)} \right)$.

What is the meaning of these integrals when the $B^{(n)}$ take values
in a non-abelian Lie algebra?  Our approach was to indirectly address
this question by writing down a differential equation whose solution
would be the desired integral.  This is analogous to using the
parallel transport equation, a differential equation for the Wilson
line, as way of defining $\exp\left(\int A\right)$.  In fact we well
know that if $A$ is non-abelian $\exp\left(\int A\right)$ is not the
correct expression.  The solution to the differential equation is a
path ordered exponential.  We wanted to use the same philosophy in
higher dimension and let the differential equation\footnote{We also
felt that the differential equation would also control how wild things
could get.} tell us what is supposed to replace $\exp\left(\int
B^{(n)} \right)$.  To get conservation laws in higher dimensions we
needed an analog of the flatness condition on the connection $A$ and
an analog of holonomy.  The idea is to use the differential equation
to define the holonomy.  Requiring that the holonomy be independent of
the submanifold led to ``zero curvature'' conditions that were local and
non-local.  

The original framework we developed for an
$(n+1)$-dimensional spacetime led to an inductive solution.  First we
solved the problem for a $1$-form $A$.  Second, introduce a $2$-form
$B$ and use $A$ and the already solved problem to solve the problem on
a $2$-manifold.  Third, introduce a $3$-form $B^{(3)}$ and use the
already solved $2$-dimensional problem to solve the new $3$-manifold
problem.  This procedure continues all the way to an $n$-form
$B^{(n)}$. In this way we defined a holonomy associated with the 
$n$-manifold. Requiring that the holonomy be invariant with respect to 
deformations of this manifold led to local and non-local 
``zero curvature'' conditions.

The procedure just described had a very unsatisfactory aspect.  First
what we are doing is constructing a submanifold.  We start with a
point and move it to construct a curve.  The curve is developed in
$1$-dimension to get a surface.  This surface is then developed into a
$3$-manifold, \emph{etc.}, until we get a $n$-manifold.  Our
differential equations are integrated precisely in the order of this
construction.  This means that the holonomy will in general depend on
the parametrization used to develop the manifold.  If we perform a
diffeomorphism on this manifold, \emph{i.e.}, a reparametrization and
effectively a redevelopment, then we do not expect to get the same
holonomy.  This was not a problem for us because the ``zero
curvature'' conditions we needed to study integrable systems solved
the problem.  Still, the procedure is very unsatisfactory and we
searched for a better formulation.  Additionally, there was a great
simplification.  In many models we considered and from the general form
of the field equations, we noted that the $1$-form $A$ and the
$n$-form $B^{(n)}$ sufficed.

The framework of the inductive procedure strongly suggested that we
had connections on some appropriate path space.  As soon as we
restricted to a $1$-form $A$ and $n$-form $B^{(n)}$ it became clear
how to write down a connection like object on an appropriate loop
space and to show that the change in holonomy when deforming a path in
the loop space gave the curvature of the connection.  Flatness of the
curvature led to a holonomy that did not change if the manifold was
deformed.  Also, the holonomy was automatically invariant under the
action of the diffeomorphism group of the manifold for ``zero
curvature'' connection thus restoring ``relativistic invariance''.

There is a large mathematical body of literature devoted to developing
a theory of non-abelian connections on loop spaces.  It is not clear
whether this is the correct approach for a theory of connections on
loop spaces.  The mathematical concepts used in these approaches are
much more sophisticated than what we require to discuss applications
of connections on higher loop spaces to integrable models.  The
subject is very Category Theory oriented and beyond the charge of this
review.  Explaining concepts such as abelian gerbes, non-abelian
gerbes, abelian gerbes with connection\footnote{An early application
of these to physics is \cite{Alvarez:1984es,Alvarez:1985vk}.},
non-abelian gerbes with connection, $2$-groups, $2$-bundles,
$2$-connections, \emph{etc.}, would be a long review article in itself.
Here we provide a selection of papers that are relevant to our
applications and provide contemporary viewpoints on the subject
\cite{attal-2001,attal-2004-29,Hofman:2002ey,Baez:2004in,Schreiber:2004ma,
Baez:2005qu,Baez:2005sn,Breen:2005,schreiber-2008}.

In summary, connections on loop spaces provides a suitable but not
totally satisfying generalization of the Lax-Zakharov-Shabat scenario.
The current framework suffers from non-locality and does not appear to
have enough structure to give a satisfactory construction of
conservation laws.  It is our belief that the current theory of
connections on loop spaces is not quite correct because there are too
many non-localities even at step one.  We do not know what the final set up
will be but we do know that the current set up is good enough for some
applications.

\subsection{Curvature and holonomy}
\label{sec:curv-hol}

Locality plays a very important role in physics and for this reason we
require all our constructions to use local data.  We have a spacetime
manifold $M$ where all the action takes place.  For simplicity we
always assume that $M$ is connected and simply connected.  For example
in the Lax-Zakharov-Shabat case $M$ may be taken to be a $(1+1)$
dimensional lorentzian cylinder.  In a higher dimensional example $M$
may be $\bbR \times S^{d}$.

For future reference we note that if $\alpha$ and $\beta$ are
$1$-forms and if $u$ and $v$ are vector fields then
$(\alpha\wedge\beta)(u,v) = \alpha(u) \beta(v) - \alpha(v)\beta(u)$.
We use the standard physics notation for the Heaviside
step function:
\begin{equation*}
    \theta(x) =
    \begin{cases}
	1, & x>0\,,  \\
	0, & x<0\,.
    \end{cases}
\end{equation*}

The geometric data we manipulate comes from a principal
bundle~\cite{Nakahara:book,KN:I}.  Assume you have a manifold $X$ and
a principal fiber bundle $\pi: P \to X$ with connection $A$ and
curvature $F$.  The structure group $G$ of the principal bundle is a
connected finite dimensional Lie group with Lie algebra $\lieg$.  The
fiber over $x\in X$ will be denoted by $P_{x}$.  Locally, a
connection\footnote{Technically, a connection $\omega$ on $P$ is a
$\lieg$-valued $1$-form on $P$ that transforms via the adjoint
representation under the action of $G$ and restricts on the fiber to
the Maurer-Cartan form for $G$.  If $s:X \to P$ is a local section
then $A=s^{*}\omega$ is the pullback of the connection.} $A$ is a
$\lieg$-valued $1$-form on the manifold $X$.

For $x_{0} \in X$, let $\Omega(X,x_{0})$ be the space of parametrized
loops with basepoint $x_{0}$.  We view a loop in $\gamma \in
\Omega(X,x_{0})$ as a map $x: S^{1} \to X$ with a basepoint condition.  To
make this more explicit, the circle $S^{1}$ is parametrized by the
interval $[0,2\pi]$.  The basepoint condition on the map $x$ is
$x(2\pi) = x(0)=x_{0}$. The space of all loops will be denoted by
$\displaystyle \loops(X) = \cup_{x\in X} \Omega(X,x)$.

Let $\gamma \in \Omega(X,x_{0})$ and let $W$ be parallel transport
(Wilson line) along $\gamma$ associated with the connection $A$.  We
solve the parallel transport equation\footnote{To avoid confusion in
the future, this equation will be used in a variety of settings.  The
meanings of $W$ and $A$ will change but the equation is the same.  For
example we will discuss parallel transport in an infinite dimensional
setting using
this equation.}%
\begin{equation}
    \frac{d}{d\tau} W(\tau) + A_{\mu}(x(\tau)) \,
    \frac{dx^{\mu}}{d\tau}\; W(\tau) = 0\,,
    \label{eq:def-W}
\end{equation}
along the loop $\gamma$ with initial condition $W(0)=I$.  Let
$f:[0,2\pi] \to [0,2\pi]$ be a diffeomorphism of the interval that
leaves the endpoints fixed, \emph{i.e.}, $f(0)=0$ and $f(2\pi) =
2\pi$.  If $\tilde{\tau} = f(\tau)$ is a new parametrization of the
interval then the holonomy is independent of the parametrization.
This is easily seen by inspecting \eqref{eq:def-W}.  You can
generalize this to allow for back tracking.  Any two loops that differ
by backtracking give the same holonomy.  The holonomy depends only on
the point set of the loop and not on the parametrization of the loop.

Consider a deformation of
$\gamma$ prescribed by a vector field $\deform$.  The change in holonomy, see
\cite[eq.  (2.7)]{Alvarez:1997ma}, is given by
\begin{equation}
    W^{-1}(2\pi) \bigl(\delta W(2\pi) \bigr) = \int_{0}^{2\pi} d\tau\;
    W^{-1}(\tau) F\bigl(\tangent(\tau),\deform(\tau)\bigr)
    W(\tau)\,,
    \label{eq:holonomy}
\end{equation}
where $\tangent$ is the tangent vector to the curve $\gamma$.  This
variational formula requires the basepoint to be kept fixed otherwise
there is an additional term.  We obtain the standard result that the
holonomy does not vary under a homotopy (continuous deformation) if
and only if the curvature vanishes.

It is also worthwhile to connect \eqref{eq:holonomy} to the holonomy
theorem of Ambrose and Singer \cite{KN:I}.  The intuition developed
here is useful in understanding some of the ideas we will pursue in
loop spaces.  Pick a basepoint $x_{0}$ on a finite dimensional
manifold $X$.  Let $\gamma \in \Omega(M,x_{0})$ be a contractible
loop.  Parallel transport around $\gamma$ gives a group element
$W(2\pi)= W(\gamma)$ called the holonomy of $\gamma$.  The set $\{
W(\gamma)\}$ for all contractible $\gamma\in\Omega(M,x_{0})$ is a
subgroup of $G$ called the restricted holonomy group and denoted by
$H_{x_{0}}$.  A result of the Ambrose-Singer theorem is that the Lie
algebra of $H_{x_{0}}$ is related to the curvature $F$ of the
connection in a way we now make more precise.
\begin{figure}[tbp]
    \centering
    \includegraphics[width=0.7\textwidth]{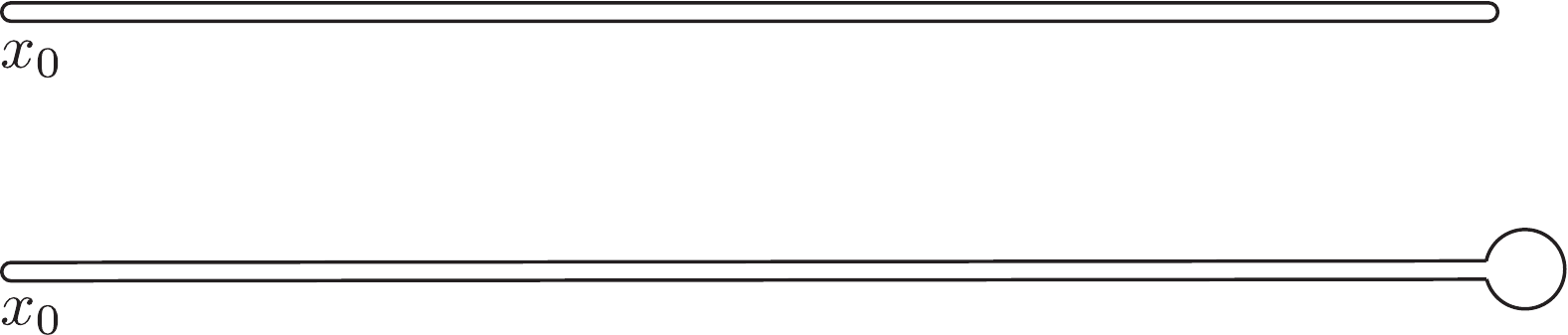}
    \begin{quote}
	\caption[AS]{\small The top figure schematically represents a
	loop based at $x_{0}$ where you go out to a point $x$ and
	return along the same curve.  In the bottom figure you go to
	$x$ and then go around a small ``infinitesimal circle'' and
	return along the initial curve.  A mathematician's
	\emph{lasso} is a loop that looks like the bottom figure.  We
	are interested in comparing the holonomies between the top and
	the bottom loops.}
	\label{fig:AS}
    \end{quote}
\end{figure}
Consider a point $x\in X$ and a path $p$ from $x_{0}$ to $x$.  On
every tangent $2$-plane at $x$ evaluate the curvature and parallel
transport that to $x_{0}$ along $p$ giving you Lie algebra elements in
$\lieg$.  Consider the collection of all such elements as you consider
all all paths connecting $x_{0}$ to $x$ and all $x\in X$.  The set of all
these elements spans a vector subspace $\lieh \subset \lieg$.  A
consequence of the Ambrose-Singer theorem is that $\lieh$ is the Lie
algebra of the restricted holonomy group $H_{x_{0}}$.  In
Figure~\ref{fig:AS} we have two loops.  The bottom loop is a small
deformation of the top one.  The holonomy for the top loop is
$W(2\pi)=I$.  The difference between the holonomies of the top and
bottom loops may be computed using \eqref{eq:holonomy}.  We see that
$\delta W(2\pi)$ only gets a contribution from the small infinitesimal
circle.  This contribution is the curvature of the $2$-plane
determined by the deformation of the loop.  That curvature is
subsequently parallel transported back to $x_{0}$ and gives us the Lie
algebra element $\delta W(2\pi)$.

\begin{remark}
    \label{rem:loop}
    Observe that $\varpi:\loops(X) \to X$ is a fiber bundle
    with projection given by the starting point map $\varpi(\gamma) =
    \gamma(0)$ where $\gamma : [0,2\pi] \to X$ is a loop.  The fiber
    over $x\in X$ is $\Omega(X,x)$.  We can use the map $\varpi:
    \loops(X) \to X$ to pullback the principal bundle $\pi:P \to X$ to
    a principal bundle $\tilde{\pi} : \widetilde{P} \to \loops(X)$.
    The fiber over $\gamma \in \loops(X)$ is isomorphic to
    $P_{\gamma(0)}$.  The structure group is still the finite
    dimensional group $G$.
\end{remark}

\begin{remark}
    \label{rem:finite}
    Next we notice that equation~\eqref{eq:holonomy} is valid even if
    the manifold $X$ is itself a loop space.  To see how this works
    assume we have a finite dimensional target space $M$ and we
    consider a loop in $M$ that we are going to deform.  Usually we
    look at a finite $r$ parameter deformation.  So we have a family
    of loops that we parametrize as $x(\sigma; q^{1},q^{2}, \ldots,
    q^{r})$ where $\sigma$ is the loop parameter.  We can view this as
    a map $\phi: Q \to \loops(M)$ where $Q$ is an $r$-dimensional
    manifold.  We use $\phi$ to pullback the bundle $\widetilde{P}$
    back to $Q$ and we are in a finite dimensional situation where we
    know that \eqref{eq:holonomy} is valid.
\end{remark}

\subsection{Connections on loop space}%
\label{sec:conn-loop}

Now we specialize to the first case of interest where $X = \loops(M)$
for some finite dimensional target space $M$.  For technical reasons
related to the behavior of the holonomy when you vary the endpoints of
a path it is convenient to fix the basepoint of the loops and this is
the explanation of why we restrict ourselves to $\loopsfix
\hookrightarrow \loops(M)$.  We have a principal fiber bundle
$\tilde{\pi} : \widetilde{P} \to \loops(M)$ with fiber isomorphic to
$P_{\gamma(0)}$ that we can restrict to $\loopsfix$.  This is the
bundle we will be using and we will also call it $\widetilde{P}$.  Let
$B= \tfrac{1}{2} B_{\mu\nu}(x) dx^{\mu}\wedge dx^{\nu}$ be a
$\lieg$-valued $2$-form on $M$ that transforms via the adjoint
representation under gauge transformation of $P$.  Let $\delta$ be the
exterior derivative on the space $\loopsfix$, we know that
$\delta^2=0$ and
\begin{equation*}
	\delta x^\mu(\sigma)\wedge \delta x^\nu(\sigma') =- \delta
	x^\nu(\sigma')\wedge\delta x^\mu(\sigma)\;.
\end{equation*}
Here $\sigma \in [0,2\pi]$ is the parameter along the loop.  Because
the basepoint $x_{0}$ is fixed we must have $\delta x^{\mu}(0) =
\delta x^{\mu}(2\pi)=0$.  We use $A$ and $B$ to
construct~\cite{Alvarez:1997ma} a connection $\mathcal{A}$ on
$\tilde{\pi} : \widetilde{P} \to \loopsfix$ by
\begin{equation}
	\mathcal{A}[\gamma] = \int_0^{2\pi} d\sigma 
	\; W(\sigma)^{-1}B_{\mu\nu}(x(\sigma))W(\sigma)
	\frac{dx^\mu}{d\sigma}
	\delta x^\nu(\sigma)\,.
	\label{eq:calA}
\end{equation}
In the above, the loop $\gamma \in \loopsfix$ is described by the map
$x: [0,2\pi] \to M$, the parallel transport (Wilson line) $W(\sigma)$
is along $\gamma$ from $x(0)=x_{0}$ to $x(\sigma)$ using the connection $A$
on $P$.  Note that at each $\sigma$,
\begin{equation*}
    B_{\mu\nu}(x(\sigma)) \frac{dx^\mu}{d\sigma} \delta x^\nu(\sigma)
\end{equation*}
is a $\lieg$-valued $1$-form at $x(\sigma)$ that is parallel
transported back using $W(\sigma)$ to $x(0)$ where all the parallel
transported objects are added together at the common endpoint.
Parallel transport gives an identification of the different fibers of
$P \to M$ and this is used to add together the various Lie algebra
elements in \eqref{eq:calA}.  Note the the connection $\mathcal{A}$ is
a $\lieg$-valued $1$-form on $\loopsfix$.  The Lie algebra element is
associated with the fiber $P_{\gamma(0)} \approx G$.

Morally, our definition of $\mathcal{A}$ is motivated by the following
canonical construction.  Let $\ev: S^{1} \times \loopsfix \to M$ be
the evaluation map defined by $\ev: (\sigma,\gamma) \mapsto
\gamma(\sigma)$.  Given a $2$-form $B$ on $M$ you can construct a
$1$-form $\mathcal{A}$ on $\loopsfix$ via pullback and integration.
The pullback $\ev^{*}B$ is a $2$-form on $S^{1} \times \loopsfix$ and
therefore integrating over the circle
\begin{equation*}
    \int_{S^{1}} \ev^{*}B
\end{equation*}
reduces the degree by one and gives a $1$-form on $\loopsfix$.

The connection defined by \eqref{eq:calA} is reparametrization
(diffeomorphism) invariant.  To show this choose a diffeomorphism
$g:[0,2\pi] \to [0,2\pi]$ that leaves the endpoints fixed and let
$\tilde{\gamma}$ be the parametrized loop given by the map $\tilde{x}
= x \circ g$.  The it is easy to see that $\mathcal{A}[\tilde{\gamma}]
= \mathcal{A}[\gamma]$.  For future reference, the new parametrization
of the loop will be denoted by $\tilde{\sigma} = g(\sigma)$.

The following notational convention is very useful.  For any object
$X$ which transforms under the adjoint representation, parallel
transport it from $x(\sigma)$ to $x(0)$ along $\gamma$ and denote this
parallel transported object by
\begin{equation}
        X^W(\sigma) = \wil^{-1} X(x(\sigma)) \wil \;.
    \label{eq:def-Xw}
\end{equation}
An elementary exercise shows that
\begin{equation}
    \frac{d}{d\sigma} X^W = \wil^{-1}(D_\mu X) \bigl(x(\sigma)\bigr)
    \wil \; \frac{dx^\mu}{d\sigma}\;.
    \label{eq:der-XW}
\end{equation}
The curvature $\mathcal{F} = \delta\xxxcA + \xxxcA\wedge\xxxcA$ is
given by, see \cite[Sec.~5.3]{Alvarez:1997ma},
\begin{align}
	\mathcal{F}
	 & =  - \half\int_0^{2\pi} d\sigma \;
	 \wil^{-1} 
	 \left[ D_\lambda B_{\mu\nu}
	 + D_\mu B_{\nu\lambda}
	 + D_\nu B_{\lambda\mu}\right](x(\sigma)) 
	 \wil  \nonumber \\
	 &   \mathstrut\hspace{.5in} \times
	 \frac{dx^\lambda}{d\sigma}
	 \delta x^\mu(\sigma)\wedge \delta x^\nu(\sigma) 
	 \nonumber \\
	& \quad -  \int_0^{2\pi} d\sigma \int_0^\sigma d\sigma' \;
	 \left[ F^W_{\kappa\mu}(x(\sigma')) ,
	 B^W_{\lambda\nu}(x(\sigma)) \right] 
	\frac{dx^\kappa}{d\sigma'} \frac{dx^\lambda}{d\sigma}
	\delta x^\mu(\sigma') \wedge\delta x^\nu(\sigma) 
	\nonumber \\
	& \quad + \half \int_0^{2\pi} d\sigma \int_0^{2\pi} d\sigma'
	\left[ B^W_{\kappa\mu}(x(\sigma')),
	B^W_{\lambda\nu}(x(\sigma)) \right]
	\frac{dx^\kappa}{d\sigma'}\frac{dx^\lambda}{d\sigma} \delta
	x^\mu(\sigma') \wedge \delta x^\nu(\sigma)\;.
	\label{eq:F-loop-1}
\end{align}
The last summand above may be rewritten as
\begin{equation*}
    \int_0^{2\pi} d\sigma \int_0^{\sigma} d\sigma'
    \left[ B^W_{\kappa\mu}(x(\sigma')),
    B^W_{\lambda\nu}(x(\sigma)) \right]
    \frac{dx^\kappa}{d\sigma'}\frac{dx^\lambda}{d\sigma} \delta
    x^\mu(\sigma') \wedge \delta x^\nu(\sigma) 
\end{equation*}
by observing that the integrand is symmetric under the interchange of
$\sigma$ and $\sigma'$.  You have to take into account both the
anti-symmetry of the Lie bracket and the anti-symmetry of the wedge
product. This means that \eqref{eq:F-loop-1} takes the form
\begin{align*}
	\mathcal{F}
	 & =  - \half\int_0^{2\pi} d\sigma \;
	 \wil^{-1} 
	 \left[ D_\lambda B_{\mu\nu}
	 + D_\mu B_{\nu\lambda}
	 + D_\nu B_{\lambda\mu}\right](x(\sigma)) 
	 \wil   \\
	 &   \qquad\qquad \times
	 \frac{dx^\lambda}{d\sigma} \;
	 \delta x^\mu(\sigma)\wedge \delta x^\nu(\sigma) 
	  \\
	& \quad +  \int_0^{2\pi} d\sigma \int_0^\sigma d\sigma' 
	 \\
	& \qquad \times
	 \left[ B^W_{\kappa\mu}(x(\sigma')) - F^W_{\kappa\mu}(x(\sigma')) ,
	 B^W_{\lambda\nu}(x(\sigma)) \right] 
	\frac{dx^\kappa}{d\sigma'} \frac{dx^\lambda}{d\sigma} \;
	\delta x^\mu(\sigma') \wedge\delta x^\nu(\sigma) \,.
\end{align*}
We write the components of the curvature $2$-form $\mathcal{F}$ in a 
more skew symmetric way
\begin{subequations}\label{eq:F-loop}
\begin{align}
	 \mathcal{F} & = - \half\int_0^{2\pi} d\sigma \; \wil^{-1}
	 \left[ D_\lambda B_{\mu\nu} + D_\mu B_{\nu\lambda} + D_\nu
	 B_{\lambda\mu}\right](x(\sigma)) \wil \nonumber \\
	 &   \qquad\qquad \times
	 \frac{dx^\lambda}{d\sigma} \;
	 \delta x^\mu(\sigma)\wedge \delta x^\nu(\sigma) 
	 \label{eq:F1} \\
	& \quad + \half \int_0^{2\pi} d\sigma \int_0^{2\pi} d\sigma' \biggl(
	\theta(\sigma-\sigma') \left[ B^W_{\kappa\mu}(x(\sigma')) -
	F^W_{\kappa\mu}(x(\sigma')) , B^W_{\lambda\nu}(x(\sigma))
	\right]
	\nonumber \\
	& \qquad - \theta(\sigma'-\sigma) \left[ B^W_{\lambda\nu}(x(\sigma)) -
	F^W_{\lambda\nu}(x(\sigma)) , B^W_{\kappa\mu}(x(\sigma'))
	\right] \biggr)
	\nonumber \\
	& \qquad \times
	\frac{dx^\kappa}{d\sigma'} \frac{dx^\lambda}{d\sigma} \;
	\delta x^\mu(\sigma') \wedge\delta x^\nu(\sigma) \,.
	\label{eq:F2}
\end{align}
\end{subequations}

This curvature contains two parts. There is a term that is local in 
$\sigma$ and involves the exterior covariant differential of $B$ which 
is the three form $D_{A}B$. There is a term that is non-local in 
$\sigma$ and involves Lie brackets of $B$ and $F$. If you want the 
physics to be local this term is a bit disturbing and we will discuss it 
presently.

There are two basic examples of a flat connection.
There is a simple lemma that follows from \eqref{eq:F-loop} and the
Bianchi identity $D_{A}F=0$.
\begin{lemma}\label{lem:B-F}
    If $B=F$ then $\mathcal{F}=0$, \emph{i.e.}, $\mathcal{A}$ is flat.
\end{lemma}
\noindent
This case was studied\footnote{The combination $B-F$ is sometimes
called the fake curvature \cite{Breen:2005,Baez:2004in}.} in 
\cite{Breen:2005,Schreiber:2004ma}.  Presently there are no known
applications of $B=F$ to integrability.  The other example is the case
we discussed in our studies of integrability \cite{Alvarez:1997ma}.
\begin{lemma}
    If $A$ is a flat connection on $P$, if the values of $B$ belong to
    an abelian ideal $\mathfrak{p}$ of $\lieg$, and $D_{A}B=0$ then
    $\mathcal{F}=0$, \emph{i.e.}, $\mathcal{A}$ is flat.
\end{lemma}
First we note that saying that $B$ takes values in an ideal (not
necessarily abelian) is a gauge invariant statement.  A subalgebra is
generally not invariant under group conjugation and therefore not
compatible with a gauge invariant characterization.  Note that if $M$
is connected and simply connected then we can use the flat connection
to globally trivialize the principal bundle $P$ and we can be very
explicit about what it means to say the $B$ has values in an abelian
ideal.  Namely pick a reference point in $M$ and parallel transport
the abelian ideal $\mathfrak{p}$ to all other points in $M$.  The
flatness of $A$ and $\pi_{1}(M)=0$ tells us that this identification
is independent of the path chosen for parallel transport.  For us,
these conditions were sufficient to construct an infinite number of
conserved charges using holonomy in analogy to integrable models in
$(1+1)$-dimensions.

\begin{figure}[tbp]
    \centering
    \includegraphics[width=0.7\textwidth]{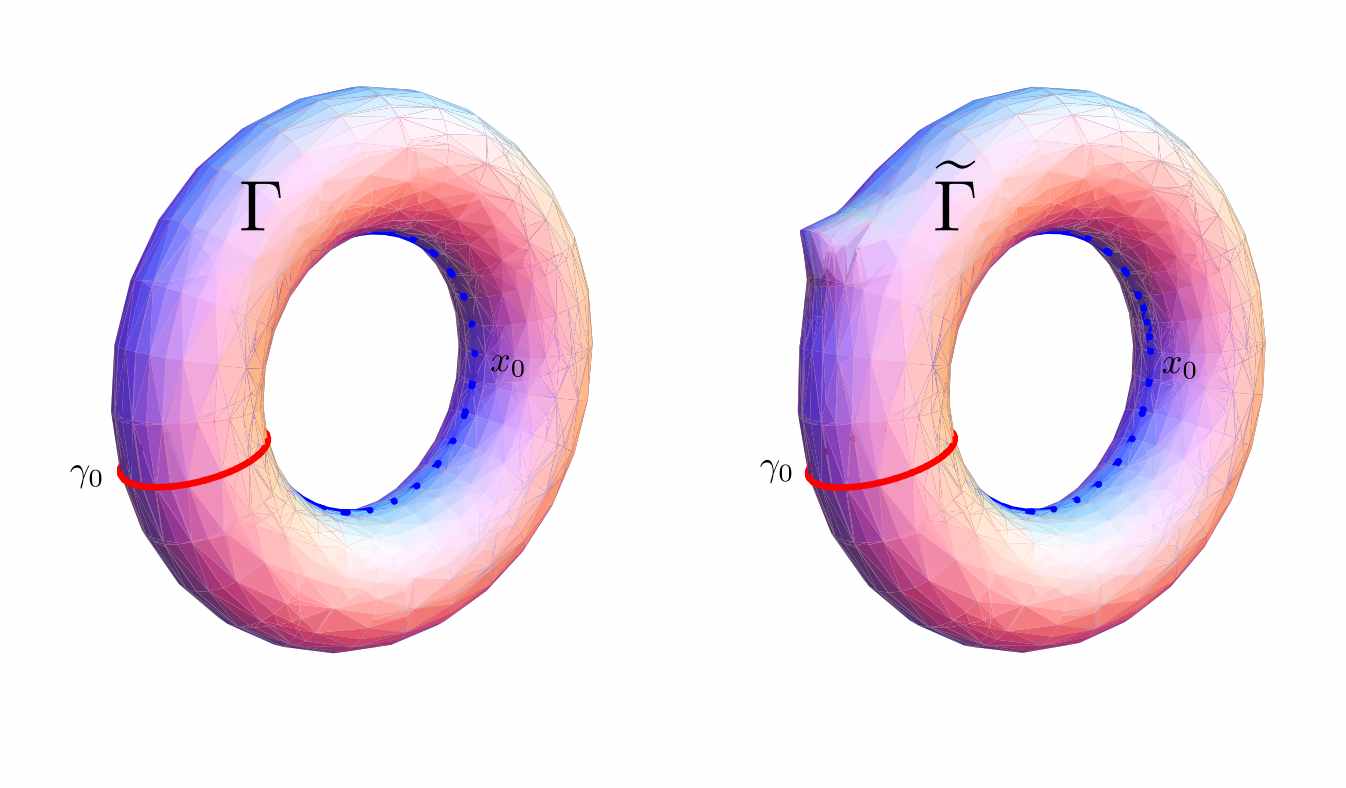}
    \begin{quote}
    \caption[xyz]{\small On the left is the loop of loops $\Gamma$.
    This is a torus with ``basepoint'' $\gamma_{0}$.  The dotted
    vertical circle is mapped to $x_{0}$.  On the right we have
    $\widetilde{\Gamma}$, a localized deformation of $\Gamma$.  We are
    interested in comparing the difference in holonomy between the
    two.  You should shrink the vertical interior circle (representing
    $x_{0}$) to a point and think of the above as doughnuts with
    very small holes.}
    \label{fig:torii}
    \end{quote}
\end{figure}

We can apply \eqref{eq:holonomy} to extend the standard result
relating holonomy and curvature to the loop space case.  To compute
the holonomy we need a loop in the loop space $\loopsfix$ with
``basepoint'' $\gamma_{0} \in \loopsfix$.  A loop in $\loopsfix$ is
given by a map $x: [0,2\pi]^{2} \to M$.  If we use coordinates
$(\sigma,\tau)$ on $[0,2\pi]^{2}$ then our map satisfies the boundary
conditions $x(\sigma,0) = x(\sigma,2\pi) = \gamma_{0}(\sigma)$ and
$x(0,\tau)=x(2\pi,\tau)=x_{0}$.  For fixed $\tau$ the map
$x_{\tau}(\sigma): [0,2\pi] \to M$ given by
$x_{\tau}(\sigma)=x(\sigma,\tau)$ is a loop in $M$, \emph{i.e.},
$x_{\tau} \in \Omega(M,x_{0})$.  The parameter $\tau$ gives the
``time'' development of the loop.  We compute the holonomy using
\eqref{eq:def-W} with connection \eqref{eq:calA}.  Let $\Gamma$ denote
the parametrized loop of loops given by the map $x: [0,2\pi]^{2} \to
M$.  $\Gamma$ is the image of a torus, see Figure~\ref{fig:torii}.
Let $\hol(\Gamma)$ be the holonomy of $\Gamma$ computed by integrating
the differential equation using the loop connection $\mathcal{A}$.  If
the torus $\Gamma$ is infinitesimally deformed while preserving the
boundary conditions then $\delta( \hol(\Gamma)) = 0$ for all deformations
if and only if $\mathcal{F}=0$.  This is a consequence of applying
Remark~\ref{rem:finite} and equation~\eqref{eq:holonomy} to $\Gamma$
and its deformations.

\subsection{Holonomy and reparametrizations}
\label{sec:hol-rep}

\begin{figure}[tbp]
    \centering
    \includegraphics[width=0.7\textwidth]{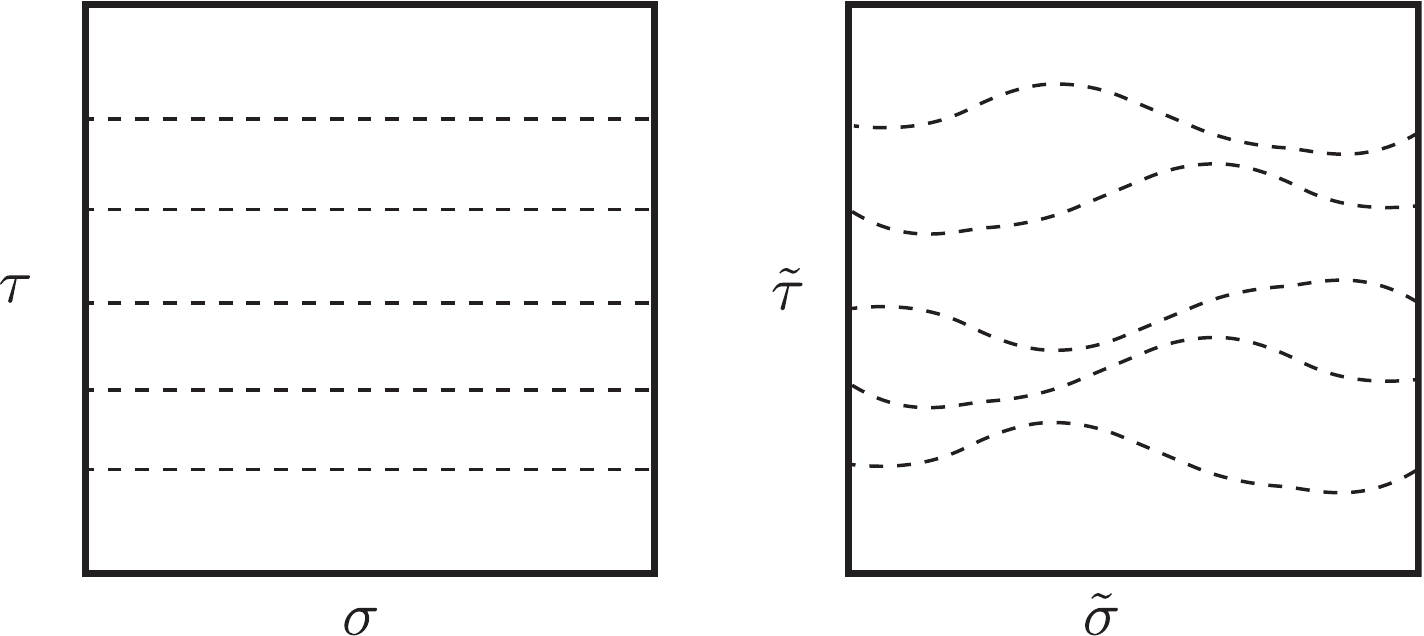}
    \begin{quote}
    \caption[xyz]{\small On the left is parametrization by
    $(\sigma,\tau)$.  The constant $\tau$ segments that determine
    $x_{\tau}$ are the dotted segments on the left and these are used
    to compute the connection $\mathcal{A}[x_{\tau}]$.  A
    diffeomorphism of the square (that leaves the boundary fixed)
    leads to a new parametrization $(\tilde{\sigma},\tilde{\tau})$
    where the horizontal segments from the left become the wavy curves
    on the right.  To compute $\mathcal{A}[\tilde{x}_{\tilde{\tau}}]$
    we need horizontal $\tilde{\tau}$ constant segments so we are
    going to get very different results.}
    \label{fig:squares}
    \end{quote}
\end{figure}

There is a drawback with this framework and that is that the holonomy
depends on the parametrization of $\Gamma$.  To understand this we go
through the mechanics of the computation.  Fix $\tau$ and use
$x_{\tau}(\sigma)$ to compute the connection $\mathcal{A}[x_{\tau}]$.
In doing this computation we have to compute the Wilson line $W$ (with
connection $A$) that clearly depends on the loops $\{x_{\tau}\}$.
Next we integrate differential equation \eqref{eq:holonomy} to get the
$\tau$ development of parallel transport for connection $\mathcal{A}$.
It is clear from this discussion that $\sigma$ and $\tau$ play very
different roles and there is a prescribed order of integrating in the
$\sigma$ and $\tau$ directions.  We expect the holonomy to be
different under two distinct parametrizations $(\sigma,\tau)$ and
$(\tilde{\sigma},\tilde{\tau})$ as discussed in
Figure~\ref{fig:squares}.

A diffeomorphism\footnote{In this Section we assume all
diffeomorphisms of the square are connected to the identity
transformation and act as the identity transformation on the boundary
of the square.} $h: [0,2\pi]^{2} \to [0,2\pi]^{2}$ of the square gives
a new parametrization $(\tilde{\sigma},\tilde{\tau})= h(\sigma,\tau)$
and a new torus $\widetilde{\Gamma}$.  The images $\Gamma$ and
$\widetilde{\Gamma}$ are the same point set in $M$.  We assume for
expositional simplicity that this point set is a smooth
$2$-dimensional submanifold of $M$.  As just discussed we expect that
$\hol(\Gamma) \neq \hol(\widetilde{\Gamma})$ in general.  There are
three basic examples where the holonomy will be the same.  If the
diffeomorphism is of the form $(\tilde{\sigma}, \tilde{\tau}) =
(g(\sigma),\tau)$ then the connection $\mathcal{A}$ does not change
and you get the same result.  If the diffeomorphism is of the form
$(\tilde{\sigma}, \tilde{\tau}) = (\sigma,f(\tau))$ then the
discussion following \eqref{eq:def-W} applies.  A function composition
of these two cases where you have a diffeomorphism of the form
$(\tilde{\sigma}, \tilde{\tau}) = (g(\sigma),f(\tau))$ also leaves the
holonomy fixed.

Many of the models of interest of us are local field theories that can
be made diffeomorphism invariant and for this reason we would like to
try to understand the conditions that lead to a reparametrization
invariant holonomy.  We would like the holonomy to be geometrical and
only depend on the image of the map $x(\sigma,\tau)$ and not on the
details of the parametrization.  If $\hol(\Gamma)= \hol(\widetilde{\Gamma})$
for all diffeomorphisms of $[0,2\pi]^{2}$ then we say that the
connection $\mathcal{A}$ is \emph{r-flat} (reparametrization
flat)\footnote{The concepts and the conditions for $r$-flatness were
developed by OA and privately communicated to U.~Schreiber who very 
gracefully acknowledged OA in \cite{Baez:2004in}.  This
observation was an answer to some email correspondence in trying to
understand the relationship of his flatness conditions on
$\mathcal{F}$, our flatness conditions on $\mathcal{F}$ and how
reparametrization invariance fits in.  In our article
\cite{Alvarez:1997ma} we had mentioned that our flatness condition
leaves the holonomy invariant under reparametrizations.  $R$-flatness
was developed much further and in more generality in
\cite{Baez:2004in}.  With hindsight a better name would have been
something like \emph{h-diff-inv} for ``holonomy is diff invariant''.} A
consequence of the definition of $r$-flatness and
of the holonomy deformation equation \eqref{eq:holonomy} is that a
flat connection $\mathcal{A}$ is automatically $r$-flat.
\begin{lemma}
    \label{lem:flat-imp-r-flat}
    A flat $\mathcal{A}$ connection is $r$-flat.
\end{lemma}

We work out the condition for $r$-flatness for infinitesimal
diffeomorphisms.  We look at the holonomy of our torus $\Gamma$.  What
distinguishes flatness from $r$-flatness is that in the former case
the holonomy is not changed by an arbitrary deformation $\delta
x^{\mu}$ while in the latter case the deformation has to be tangential
to the $2$-manifold $\Gamma$.  The tangent vector $\myvec{T}$ to the
loop $x_{\tau}$ is given by
\begin{equation}
    \tangent = \frac{\partial x^{\mu}}{\partial \tau}\; 
    \frac{\partial}{\partial x^{\mu}}\,.
    \label{eq:def-T}
\end{equation}
The ``spatial tangent vector'' along the loop $x_{\tau}$ is
\begin{equation}
    \spatial = \frac{\partial x^{\mu}}{\partial \sigma}\; 
    \frac{\partial}{\partial x^{\mu}}\,.
    \label{eq:def-S}
\end{equation}
Finally, a general deformation tangential to the $2$-submanifold
$\Gamma$ is given by
\begin{equation}
    \deform = a(\sigma,\tau) \spatial + b(\sigma,\tau) \tangent\,,
    \label{eq:def-N}
\end{equation}
where $a$ and $b$ are arbitrary functions vanishing on the boundary of
the square $[0,2\pi]^{2}$.  The condition for
$r$-flatness will be given by inserting $\tangent$ and $\deform$ into
\eqref{eq:holonomy} and requiring the change in holonomy to vanish.
We remark that $r$-flatness is automatic by construction in the
$1$-dimensional case.  It becomes a new non-trivial phenomenon in the
$2$-dimensional case.  It is really a statement about holonomy that we
try to capture in terms of information contained in the curvature.

\begin{theorem}\label{thm:r-flatness-cond}
    The connection $\mathcal{A}$ is $r$-flat if and only if for all
    maps $x:[0,2\pi]^{2} \to M$ we have
    \begin{align}
	 0 &= \int_0^{2\pi} d\sigma' \; \biggl( \theta(\sigma-\sigma')
	 \left[ \left\{ F^W_{\kappa\mu}(x(\sigma',\tau')) -
	 B^W_{\kappa\mu}(x(\sigma',\tau')) \right\} ,
	 B^W_{\lambda\nu}(x(\sigma,\tau)) \right] \nonumber \\
	& \qquad\qquad - \theta(\sigma'-\sigma) \left[ \left\{
	F^W_{\lambda\nu}(x(\sigma,\tau))-
	B^W_{\lambda\nu}(x(\sigma,\tau))\right\},
	B^W_{\kappa\mu}(x(\sigma',\tau')) \right] \biggr) \nonumber\\
	&\qquad \times \frac{\partial x^{\kappa}}{\partial \sigma'}
	\frac{\partial x^{\mu}}{\partial \tau'} \frac{\partial
	x^{\lambda}}{\partial \sigma} \frac{\partial x^{\nu}}{\partial
	\tau} \quad\text{for $\tau=\tau'$}.
	\label{eq:xr-2}
    \end{align}
\end{theorem}
Note that all the action occurs at ``equal time''.  A key observation
required in proving this theorem is that the tangent space to the
$2$-manifold is two dimensional and spanned by $\spatial$ and
$\tangent$.  This and the fact that in \eqref{eq:F1} everything is at
the same value of $\sigma$ means that
$D_{A}B(\spatial,\tangent,\deform)=0$ automatically because $D_{A}B$
is a $3$-form.  Therefore we will only get a contribution from the
nonlocal \eqref{eq:F2} term.  Even here things simplify because there
is no contribution from the $\spatial$ summand in \eqref{eq:def-N}
because there is always a contraction of two $\spatial$ vectors at the
same $(\sigma,\tau)$ with some $2$-form.  We only have to worry about
the $\tangent$ summand in \eqref{eq:def-N}.  A straightforward
computation gives the equation above.

\begin{corollary}
    If $F=0$ and $B$ takes values in an abelian ideal $\mathfrak{p}$
    of $\lieg$ then the connection $\mathcal{A}$ is $r$-flat.
\end{corollary}
Using \eqref{eq:def-Xw} we see that something that takes values in the
ideal $\mathfrak{p}$ will remain in $\mathfrak{p}$ under parallel
transport.  Notice that we do not have to impose $DB=0$ to obtain
$r$-flatness.  Connections of this type are examples of $r$-flat
connections that are not flat, \emph{i.e.}, $\mathcal{F} \neq 0$.  The
curvature $\mathcal{F}$ which is locally given by $D_{A}B$ will take
values in $\mathfrak{p}$.  There are examples of such connections.  In
reference \cite{Alvarez:1997ma} we constructed some models based on a
non-semisimple Lie algebra $\lieg$ that contains a non-trivial abelian
ideal\footnote{Equivalently we have a non-semisimple Lie group $G$
with an abelian normal subgroup $P$.} $\mathfrak{p}$.
These models are automatically $r$-flat without imposing $DB=0$ which
was necessary for our integrability studies.

The case $F=B$ leads automatically to a $r$-flat connection by
Lemma~\ref{lem:flat-imp-r-flat}. The converse is also true.
\begin{corollary}
    An $r$-flat connection $\mathcal{A}$ satisfying $B=F$ is also flat.
\end{corollary}
The proof is elementary because the Bianchi identity $D_{A}F=0$
automatically implies that $D_{A}B=0$. We remind the reader of 
Lemma~\ref{lem:B-F}

A connection $\mathcal{A}$ is said to be \emph{curvature local} if the
nonlocal commutator term in \eqref{eq:F-loop} for $\mathcal{F}$
vanishes.  In other words, we have
\begin{align} 
    0 &= \biggl( \theta(\sigma-\sigma') \left[
    B^W_{\kappa\mu}(x(\sigma')) - F^W_{\kappa\mu}(x(\sigma')) ,
    B^W_{\lambda\nu}(x(\sigma)) \right] 
    \nonumber \\
    & \qquad - \theta(\sigma'-\sigma) \left[
    B^W_{\lambda\nu}(x(\sigma)) - F^W_{\lambda\nu}(x(\sigma)) ,
    B^W_{\kappa\mu}(x(\sigma')) \right] \biggr)
    \frac{dx^\kappa}{d\sigma'} \frac{dx^\lambda}{d\sigma},
    \label{eq:curv-local-cond}
\end{align}
for all $\sigma,\sigma' \in [0,2\pi]$ and for all loops $x:[0,2\pi]
\to M$.  In this case the curvature is given by the integral of a
local integrand (except for the parallel transport) in $\sigma$:
\begin{align}
    \mathcal{F}  &= - \half\int_0^{2\pi} d\sigma \; \wil^{-1}
    \left[ D_\lambda B_{\mu\nu} + D_\mu B_{\nu\lambda} + D_\nu
    B_{\lambda\mu}\right](x(\sigma)) \wil \nonumber \\
    &   \qquad\qquad \times
    \frac{dx^\lambda}{d\sigma} \;
    \delta x^\mu(\sigma)\wedge \delta x^\nu(\sigma).
    \label{eq:F-local}
\end{align}

To prove a main theorem of this Section we need the lemma below.
\begin{lemma}\label{lem:curv-local}
    A connection $\mathcal{A}$ is curvature local if and only if
    \begin{equation}
	\left[
	    B^W_{\kappa\mu}(x(\sigma')) - F^W_{\kappa\mu}(x(\sigma')) ,
	    B^W_{\lambda\nu}(x(\sigma)) \right] =0, 
	\label{eq:curv-local-x}
    \end{equation}
    for all $\sigma',\sigma \in [0,2\pi]$ and for all loops $x:[0,2\pi] \to M$.
\end{lemma}
\noindent
First we look at \eqref{eq:curv-local-cond} in more detail.
It is convenient to define
\begin{equation}
    C_{\mu\nu}(\sigma',\sigma) = \left[
    B^W_{\kappa\mu}(x(\sigma')) - F^W_{\kappa\mu}(x(\sigma')) ,
    B^W_{\lambda\nu}(x(\sigma)) \right] \frac{dx^\kappa}{d\sigma'}
    \frac{dx^\lambda}{d\sigma}\,.
    \label{eq:def-C}
\end{equation}
In this notation the curvature local condition becomes
\begin{equation} 
    0 =  \theta(\sigma-\sigma') C_{\mu\nu}(\sigma',\sigma)
    - \theta(\sigma'-\sigma) C_{\nu\mu}(\sigma,\sigma') \,,
    \label{eq:curv-local-cond-1}
\end{equation}
\begin{figure}[tbp]
    \centering
    \includegraphics[width=0.7\textwidth]{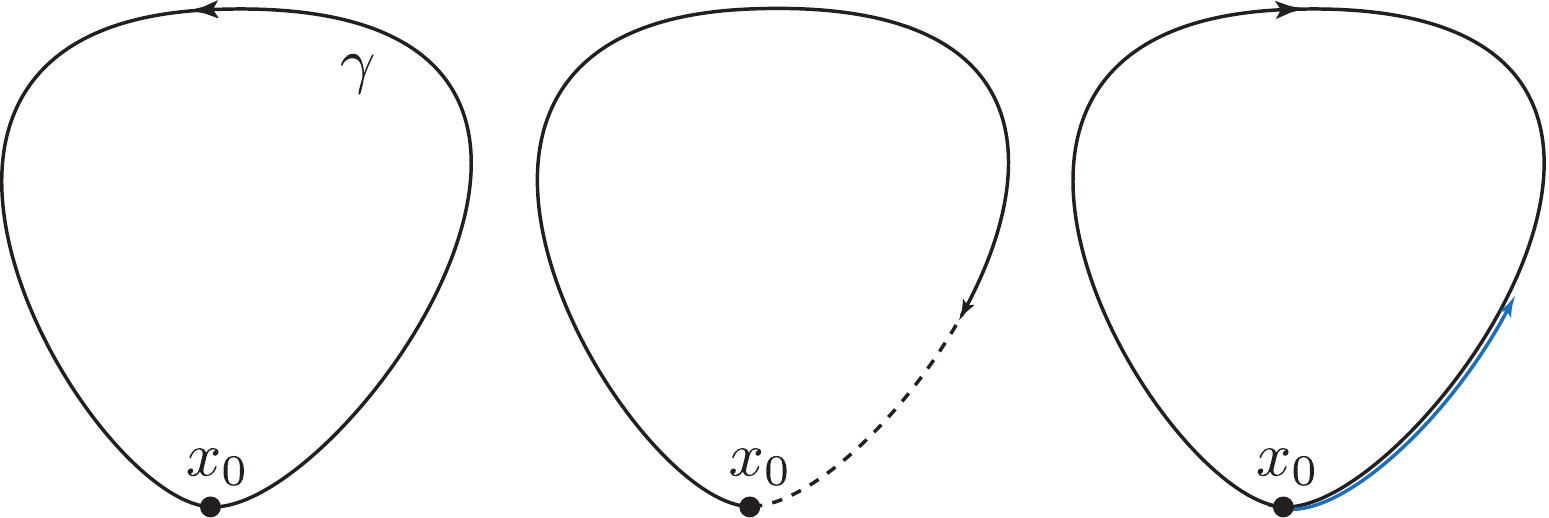}
    \begin{quote}
	\caption[gam]{\small On the left we have the loop $\gamma$
	with basepoint $x_{0}$.  In the center figure we have
	pictorially denoted $W_{\gamma^{-1}}(\sigma)$.  Notice that
	according to the figure on the right this is the same as
	$W_{\gamma}(2\pi -\sigma) W_{\gamma^{-1}}(2\pi)$.}
	\label{fig:hol-conj}
    \end{quote}
\end{figure}
If we choose $\sigma > \sigma'$ then we have
$C_{\mu\nu}(\sigma',\sigma)=0$, if we choose $\sigma < \sigma'$ then
we get $C_{\nu\mu}(\sigma,\sigma')=0$.  We conclude that
$C_{\mu\nu}(\sigma',\sigma)=0$ for all $\sigma' < \sigma$.  This
condition is supposed to be valid for all curves so we conclude that
$\left[ B^W_{\kappa\mu}(x(\sigma')) - F^W_{\kappa\mu}(x(\sigma')) ,
B^W_{\lambda\nu}(x(\sigma)) \right] =0$, for all $\sigma' < \sigma$
and for all loops $x:[0,2\pi] \to M$.  We can strengthen this to
$\sigma' > \sigma$.  Let $\gamma \in \Omega(M,x_{0})$ then the inverse
loop $\gamma^{-1} \in\Omega(M,x_{0})$ is defined by
$\gamma^{-1}(\sigma) = \gamma(2\pi - \sigma)$.  The result we need is
that $W_{\gamma^{-1}}(\sigma) = W_{\gamma}(2\pi -\sigma)
W_{\gamma^{-1}}(2\pi)$ that is most easily obtained by drawing a
picture, see Figure~\ref{fig:hol-conj}.  Assume the connection is
curvature local then schematically we have for $\sigma' < \sigma$ and
loop $\gamma^{-1}$:
\begin{align*}
    0 & = \left[ W_{\gamma^{-1}}^{-1}(\sigma') \, \{B-F\}
    (\gamma^{-1}(\sigma')) \, W_{\gamma^{-1}}(\sigma') ,
    W_{\gamma^{-1}}^{-1}(\sigma) \,B(\gamma^{-1}(\sigma)) \,
    W_{\gamma^{-1}}(\sigma) \right], \\
    & = \biggl[W_{\gamma^{-1}}^{-1}(2\pi) W_{\gamma}^{-1}(2\pi -
    \sigma') \, \{B-F\} (\gamma^{-1}(\sigma')) \, W_{\gamma}(2\pi -
    \sigma') W_{\gamma^{-1}}(2\pi) , \\
    &\qquad W_{\gamma^{-1}}^{-1}(2\pi) W_{\gamma}^{-1}(2\pi -\sigma)
     \,B(\gamma^{-1}(\sigma)) \,
     \,W_{\gamma}(2\pi -\sigma)
    W_{\gamma^{-1}}(2\pi) \biggr], \\
    & = W_{\gamma^{-1}}^{-1}(2\pi) \biggl[W_{\gamma}^{-1}(2\pi -
    \sigma') \, \{B-F\} (\gamma(2\pi - \sigma')) \, W_{\gamma}(2\pi -
    \sigma') , \\
    &\qquad W_{\gamma}^{-1}(2\pi -\sigma) \,B(\gamma(2\pi - \sigma))
    \, \,W_{\gamma}(2\pi -\sigma) \biggr] W_{\gamma^{-1}}(2\pi)\,.
\end{align*}
At the last equality we have effectively interchanged the order of 
$\sigma'$ and $\sigma$ and this concludes our proof of the lemma.

In general, the target space $M$ is curved and  equation
\eqref{eq:curv-local-x} is interpreted as the evaluation of the
respective $2$-forms on pairs of tangent vectors at $T_{x(\sigma)}M$
and $T_{x(\sigma')}M$.

One of the main results we establish in this section is:
\begin{theorem}\label{thm:c-l-iff-r-flat}
    The connection $\mathcal{A}$ is curvature local if and only if
    $\mathcal{A}$ is $r$-flat.
\end{theorem}
\noindent 
The proof ($\Longrightarrow$) follows from the fact that
local condition \eqref{eq:curv-local-cond} implies \eqref{eq:xr-2}.
What is very surprising is that the converse ($\Longleftarrow$) is
true and the proof is more subtle.

The $r$-flat condition using the notation introduced in the proof of 
Lemma~\ref{lem:curv-local} is
\begin{equation}
    0 = \int_0^{2\pi} d\sigma' \; \biggl( \theta(\sigma-\sigma')
    C_{\mu\nu}(\sigma',\sigma) - \theta(\sigma'-\sigma)
    C_{\nu\mu}(\sigma,\sigma') \biggr) T^{\mu}(\sigma') 
    T^{\nu}(\sigma)\,,
    \label{eq:xr-3}
\end{equation}
where $\tangent$ is defined in \eqref{eq:def-T}.  A loop $x(\sigma)$
in may be developed ``in time'' in many possible ways, therefore the
temporal tangent vector $\tangent$ may be taken to be arbitrary at
each point of the curve.  With this in mind we take the functional
derivative
\begin{equation*}
    \frac{\delta}{\delta T^{\rho}(\tilde{\sigma}')}\, 
    \frac{\delta}{\delta T^{\omega}(\tilde{\sigma})}
\end{equation*}
of \eqref{eq:xr-3} and obtain
\begin{align}
    0 & = [ \delta(\tilde{\sigma}-\sigma) - \delta(\tilde{\sigma}'
    -\sigma) ] \theta(\tilde{\sigma}-\tilde{\sigma}')
    C_{\rho\omega}(\tilde{\sigma}',\tilde{\sigma}) \nonumber \\
    &\quad + [ \delta(\tilde{\sigma}' -\sigma) -
    \delta(\tilde{\sigma}-\sigma) ]
    \theta(\tilde{\sigma}'-\tilde{\sigma})
    C_{\omega\rho}(\tilde{\sigma},\tilde{\sigma}')\,.
    \label{eq:xr-4}
\end{align}

There are three variables we can vary independently: $\sigma$,
$\tilde{\sigma}$ and $\tilde{\sigma}'$.  If we choose $\sigma \neq
\tilde{\sigma}$ and $\sigma \neq \tilde{\sigma}'$ then there are no
conclusions we can reach about the $C$s.  First we look at the case
$\tilde{\sigma} > \tilde{\sigma}'$ where the constraint reduces to $0
= [ \delta(\tilde{\sigma}-\sigma) - \delta(\tilde{\sigma}' -\sigma) ]
C_{\rho\omega}(\tilde{\sigma}',\tilde{\sigma})$.  As we vary $\sigma$
such that $\sigma \to \tilde{\sigma}$ we see that we have to require
$C_{\rho\omega}(\tilde{\sigma}',\tilde{\sigma}) =0$ and similarly as
$\sigma \to \tilde{\sigma}'$.  Repeating the argument in the case
$\tilde{\sigma} < \tilde{\sigma}'$ we learn that $
C_{\omega\rho}(\tilde{\sigma},\tilde{\sigma}') =0$.  This result
together the previous argument that we used to reverse the order of
$\tilde{\sigma}'$ and $\tilde{\sigma}$ give us the hypotheses of
Lemma~\ref{lem:curv-local}.  We have proven the converse part of
Theorem~\ref{thm:c-l-iff-r-flat}.

Theorem~\ref{thm:c-l-iff-r-flat} is very satisfying from the physics
viewpoint.  Non-locality in the curvature $\mathcal{F}$ and the lack
of reparametrization invariance of the $\mathcal{A}$ holonomy have a
common origin.  From the viewpoint of physics the central tenet is
probably requiring diffeomorphism invariance.  Requiring that the
physics be diffeomorphism invariant leads to a local curvature
$\mathcal{F}$.  For us the diffeomorphism invariance has additional
important consequences such as the Lorentz invariance of the conserved
charges constructed via holonomy.

\subsection{Connections on higher loop spaces}%
\label{sec:higher-loop}

We now move to the higher dimensional case \cite[Section 
5]{Alvarez:1997ma}, see also \cite{Baez:2004in}.
Instead of using ``toroidal'' loop spaces it is simpler to use
``spherical'' loop spaces.  These are defined inductively by
$\Omega^{n+1}(M,x_0) = \Omega(\Omega^n(M,x_0),x_0)$.  To be more
explicit we have
\begin{align}
	\Omega^n(M,x_0) &= \left\{ f: [0,2\pi]^n \to M \; \big| \;
	f|_{\partial [0,2\pi]^n} = x_0 \right\}, \label{eq:the-square}\\
	&= \left\{ f: S^n \to M \; \big| \;
	f(\text{north pole}) = x_0 \right\}.
	\nonumber
\end{align}
A tangent vector $X$ at $N \in \Omega^{n}(M,x_0)$ is a vector field on $N$
(not necessarily tangential to $N$).  This vector field generates a
one-parameter family of deformations.  Note that the vector field
must vanish at the basepoint because $x_0$ is kept fixed by the
deformation.  The vector field $X$ will replace the role of $\delta
x^\mu(\sigma)$ in our discussion of higher loop spaces.

The construction of a connection $\mathcal{A}$ on $\Omega^{n}(M,x_0)$
is motivated by the evaluation map $\ev: S^{n} \times
\Omega^{n}(M,x_0) \to M$ defined by $\ev: (\sigma, N) \mapsto
N(\sigma)$ where $\sigma \in S^{n}$ and $N \in \Omega^{n}(M,x_0)$.
Let $B$ be a $(n+1)$-form on $M$ then $\ev^{*}B$ is a $(n+1)$-form on
$S^{n} \times \Omega^{n}(M,x_0)$ and therefore integration over 
$S^{n}$
\begin{equation*}
    \int_{S^{n}} \ev^{*}B
\end{equation*}
reduces degree by $n$ and gives a $1$-form $\mathcal{A}$ on
$\Omega^{n}(M,x_0)$.  This is the basic idea but a little
massaging has to take place in order to respect gauge invariance.

Connections on $M$ and $\Omega(M,x_{0})$ constitute the exceptional
cases.  The generic cases are connections on $\Omega^{n}(M,x_{0})$ for
$n \ge 2$ as we now explain.  Assume we take a $\lieg$-valued
$(n+1)$-form and try to mimic \eqref{eq:calA}.  Let $N \in
\Omega^{n}(M,x_{0})$ be represented by a map $x: S^{n} \to M$.  A
typical point in $S^{n}$ will be denoted in local coordinates as
$\sigma = (\sigma^{1},\ldots,\sigma^{n})$.  Let $X$ be a tangent
vector at $N \in \Omega^{n}(M,x_{0})$.  In other words, $X$ is a vector
field on $N$. We write
\begin{equation*}
    \mathcal{A}(X) = \int_{N} \iota_{X}B^{W}\,,
\end{equation*}
where $\iota_{X}$ is interior multiplication with respect to the
tangent vector $X$, \emph{i.e.}, evaluate the $(n+1)$-form on the
first slot and therefore obtaining a $n$-form.  $W$ above represents
parallel transport from $x_{0}$ to $x(\sigma)$.  It is at this stage
that we see that the case $\Omega^{n}(M,x_{0})$ for $n \ge 2$ is
different that previous case because extra data has to be specified.
In the case of a connection on $M$ the parallel transport was not
necessary.  In the case of $\Omega^{1}(M,x_{0})$ the path is
determined by the loop.  Since parallel transport is insensitive to
backtracking and to parametrization everything works automatically in
this case.  In the present case with $n \ge 2$ we see that we have to
specify a path from the north pole to each $\sigma \in S^{n}$.  This
in turn gives us a path from $x_{0}$ to $x(\sigma)$.  Assume we have
made a choice\footnote{There is physically less satisfying alternative
choice of paths that can be used to define connections in
$\Omega^{n}(M,x_{0})$ for $n \ge 1$.  Since $M$ is connected we can
\emph{a priori} choose a fiducial path from $x_{0}$ to $x \in M$ and
use those to define the parallel transport needed in the definition of
the connection $\mathcal{A}$.  This is physically very unsatisfying
because the fiducial paths have nothing to do with the ``$n$-brane''
in $M$ or its temporal evolution.} that we will denote by
$\gamma_{\sigma}$.  We denote the parameter\footnote{We have
deliberately chosen the parameter to be in $[0,1]$ to distinguish a
path from a loop where the parameter is in $[0,2\pi]$.} along
$\gamma_{\sigma}$ by $\lambda \in [0,1]$.  To be more explicit the
equation should be written as
\begin{equation}
    \mathcal{A}(X) = \int_{N}  
    W_{\gamma_{\sigma}}^{-1}(1) \biggl( \iota_{X}B(x(\sigma)) \biggr) 
    W_{\gamma_{\sigma}}(1) \,.
    \label{eq:def-calA-N}
\end{equation}
Here $W_{\gamma_{\sigma}}(\lambda)$ denotes parallel transport from
$x_{0}$ to $\gamma_{\sigma}(\lambda)$.  There are technical issues of
continuity and smoothness that need to be addressed.  For example if
you choose the paths to be the great circles emanating from the north
pole of $S^{n}$ then how do you make sure all is okay, \emph{e.g.},
single valuedness, when you arrive at the south pole.  These are
important issues that have to be analyzed but from a physics point of
view there is a big red flag waving to us at this point.
$\mathcal{A}(X)$ depends on the specification a lot of of extra data,
namely the choice of $\{\gamma_{\sigma}\}$, but in standard local
field theories such data\footnote{The point $x_{0}$ is an extra datum
but of a trivial type.  For example, it could be taken to be the point
at infinity because of finite energy constraints.} does not appear
naturally: it is not in the lagrangian, it is not in the equations of
motion, it is not in the boundary conditions.  Mathematically there is
no canonical choice of paths in $S^{n}$.  If we change the choice of
paths keeping everything else fixed (such as the map $x: S^{n} \to M$)
the connection $\mathcal{A}$ changes.  Under an infinitesimal
deformation of the paths the change may be computed using
\eqref{eq:holonomy} and the result is expressed in terms of the
curvature $F$.  To require that the physics be independent of the
extraneous data suggests that the connection $A$ should be flat.  This
is the choice that was made for the exposition given in Section~5 of
\cite{Alvarez:1997ma}.  The case of a non-flat $A$ was discussed in
detail in \cite{Baez:2004in}.  From now on we assume that $F=0$.

If we define the  curvature as $\mathcal{F} = d\mathcal{A} + 
\mathcal{A} \wedge \mathcal{A}$ then equations (5.8) and (5.10) of 
\cite{Alvarez:1997ma} tell us that if $X,Y$ are two tangent vectors 
to $N \in \Omega^{n}(M,x_{0})$ then
\begin{equation}
    \mathcal{F}(X,Y) = \int_{N} W^{-1}\bigl( \iota_{Y}\iota_{X}D_{A}B 
    \bigr) W + \left[ \int_{N}\iota_{X}B^{W}, \int_{N} \iota_{Y} 
    B^{W} \right].
    \label{eq:curv-n}
\end{equation}
Remember that $W$ depends only on the endpoint because $M$ is simply
connected and $A$ is flat.  The sign difference between the exterior
derivative term in \eqref{eq:curv-n} and the exterior derivative term
in \eqref{eq:F-loop} is due to a sign difference in the respective
definitions \eqref{eq:def-calA-N} and \eqref{eq:calA} in the case
$n=1$.

The notions of curvature local and $r$-flatness\footnote{The allowed
diffeomorphisms of $S^{n}$ are those that are connected to the
identity transformation and also leave $x_{0}$ fixed.} can be extended
to this case and the discussion is simpler because we have chosen
$F=0$.
\begin{lemma}\label{lem:main-prop}
    If $A$ is flat then $\mathcal{A}$ is curvature local if and only
    if
    \begin{equation}
	\left[ \iota_{X(\sigma)}B^{W}(x(\sigma)),  \iota_{Y(\sigma')} 
	    B^{W}(x(\sigma')) \right] =0\,,
        \label{eq:curv-local-n}
    \end{equation}
    for all $N \in \Omega^{n}(M,x_{0})$.  We have that $x(\sigma)$ and
    $x(\sigma')$ are in $N$; and $X(\sigma)$ and $Y(\sigma')$ are
    arbitrary tangent vectors respectively in $T_{x(\sigma)}M$ and
    $T_{x(\sigma')}M$.  Note that $X$ and $Y$ do not have to be
    tangential to $N$.
\end{lemma}
\noindent 
This is just the commutator term of \eqref{eq:curv-n}
written out more explicitly and requiring it to vanish.  
The notation is a bit cryptic and explicitly detailed below:
\begin{equation*}
    \iota_{X(\sigma)}B^{W}(x(\sigma)) = 
    B^{W}_{\mu\nu_{1}\cdots\nu_{n}}(x(\sigma))\; X^{\mu}(\sigma) 
    \frac{\partial x^{\nu_{1}}}{\partial \sigma^{1}} \cdots 
    \frac{\partial x^{\nu_{n}}}{\partial \sigma^{n}}\;
    d\sigma^{1} \wedge \cdots \wedge d\sigma^{n}\,.
\end{equation*}
The conclusions of Lemma~\ref{lem:main-prop} above may be written as
\begin{equation*}
	[B_{\omega\mu_{1}\cdots\mu_{n}}^{W}(x),
	B_{\rho\nu_{1}\cdots\nu_{n}}^{W}(x')]=0 \quad\text{for $x,x'
	\in M$.}
\end{equation*}
The reason is that $W$ only depends on the endpoint because $A$ is 
flat, $x(\sigma)$ and $x(\sigma')$ can be arbitrary points in $M$, 
and the tangent $n$-planes determined by $N \subset M$ can be 
arbitrary. 

\begin{theorem}
    If $A$ is flat then $\mathcal{A}$ is $r$-flat if and only if
    \begin{equation}
	\left[ \iota_{T(\sigma)}B^{W}(x(\sigma)), \int_{N} \iota_{T} 
	    B^{W} \right] =0\,,
	\label{eq:r-flat-n}
    \end{equation}
    for all $N \in \Omega^{n}(M,x_{0})$.  We have that $x(\sigma) \in
    N$, and $T$ is an arbitrary tangent vector giving a deformation of
    $N$, \emph{i.e.}, $T(\sigma) \in T_{x(\sigma)}M$.  Note that $T$
    does not have to be tangential to $N$.
\end{theorem}
\noindent
The proof is along the same lines of Theorem~\ref{thm:r-flatness-cond} 
but the notation is different.
\begin{lemma}\label{lem:main-conv}
    If $A$ is flat then $\mathcal{A}$ is $r$-flat if and only if
    \begin{equation*}
	    [B_{\omega\mu_{1}\cdots\mu_{n}}^{W}(x),
	    B_{\rho\nu_{1}\cdots\nu_{n}}^{W}(x')]=0 \quad\text{for $x,x'
	    \in M$.}
    \end{equation*}
\end{lemma}
\noindent 
Use the method in the proof of the converse part of
Theorem~\ref{thm:c-l-iff-r-flat}.  The following main theorem is a
direct consequence of Lemma~\ref{lem:main-prop} and
Lemma~\ref{lem:main-conv}.
\begin{theorem}
    If $F=0$  then $\mathcal{A}$ is curvature local if and only if 
    $\mathcal{A}$ is $r$-flat.
\end{theorem}

\begin{corollary}
    If $A$ is flat and if $B$ takes values in an abelian ideal
    $\mathfrak{p} \subset \lieg$ then $\mathcal{A}$ is both curvature 
    local and $r$-flat with curvature $\mathcal{F}$ taking values in 
    $\mathfrak{p}$ and given by
    \begin{equation}
	\mathcal{F}(X,Y) = \int_{N} W^{-1}\bigl( \iota_{Y}\iota_{X}D_{A}B 
	    \bigr) W\,.
        \label{eq:cal-F-n}
    \end{equation}
\end{corollary}

\subsection{Is the loop space curvature abelian?}%
\label{sec:curv-abelian}

A classic result of homotopy theory is that $\pi_{n}(M,x_{0})$ is
abelian if $n>1$.  Here we will argue that a $r$-flat connection on a
loop space $\Omega^{n}(M,x_{0})$ with $n \ge 1$ has abelian holonomy.
The main arguments in this section are more topological/geometrical
and are independent of detailed results of the previous sections.

First we sketchily review the abelian nature of the higher homotopy
groups \cite{Hatcher:algebraictopology}.  The $n$-th homotopy group
$\pi_{n}(M,x_{0})$ is defined as follows.  If $\alpha \in
\Omega^{n}(M,x_{0})$ then denote by $[\alpha]$ the set of all elements
of $\Omega^{n}(M,x_{0})$ that are equivalent under homotopy.  Under
the composition of maps, the homotopy equivalence classes of elements
of $\Omega^{n}(M,x_{0})$ becomes a group denoted by
$\pi_{n}(M,x_{0})$.  The construction will be important for us in
applying to our holonomy ideas.  The composition of two elements of
$\Omega^{n}(M,x_{0})$ is defined by
\begin{equation}
    (\alpha_{2} \circ 
    \alpha_{1})(\sigma^{1},\sigma^{2},\ldots,\sigma^{n}) =
    \begin{cases}
        \alpha_{1}(\sigma^{1},\sigma^{2},\ldots,2\sigma^{n}) & 
	\text{for $\sigma^{n} \in [0,\pi]$}, \\
	\alpha_{2}(\sigma^{1},\sigma^{2},\ldots,2\sigma^{n}-\pi) & 
	\text{for $\sigma^{n} \in [\pi,2\pi]$}.	
    \end{cases}
    \label{eq:composition}
\end{equation}
The group product in $\pi_{n}(M,x_{0})$ is defined by $[\alpha_{2}] 
\cdot [\alpha_{1}] = [\alpha_{2}\circ\alpha_{1}]$. The claim is that 
for $n>1$ this product is abelian, \emph{i.e.}, $[\alpha_{2}\circ 
\alpha_{1}] = [\alpha_{1}\circ \alpha_{2}]$.
\begin{figure}[tbp]
    \centering
    \includegraphics[width=0.8\textwidth]{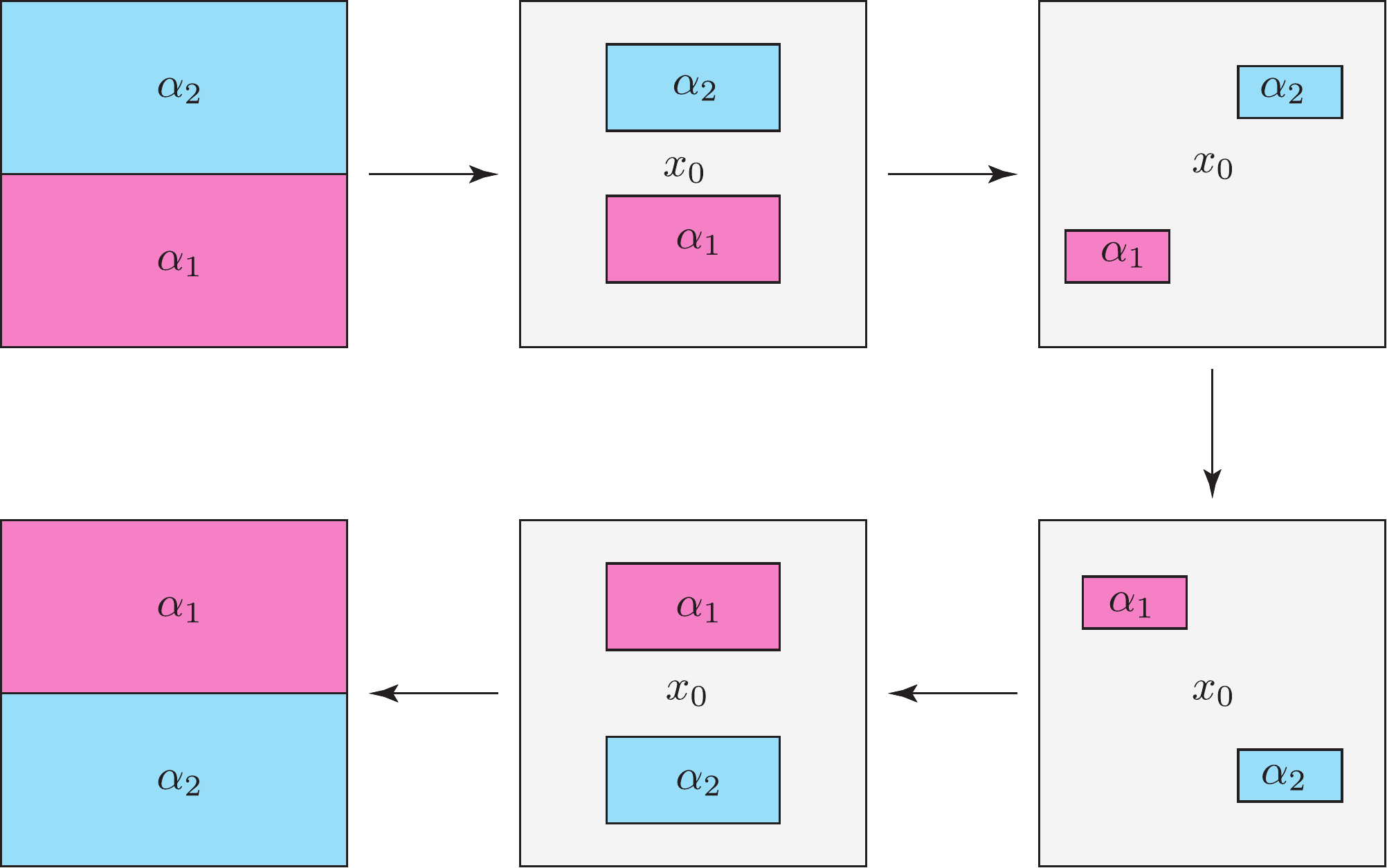}
    \begin{quotation}
	\caption[abel]{\small A flowchart to illustrate that the
	higher homotopy groups are abelian.  With a different
	interpretation the same flowchart may be used to show that the
	holonomy groups associated with the higher loop spaces are
	abelian.}
	\label{fig:abelian}
    \end{quotation}
\end{figure}
To show this we use the flowchart in Figure~\ref{fig:abelian} that is
valid for $n>1$.  We start at the top left hand corner where the
diagram there represents the composition of maps $\alpha_{2}\circ
\alpha_{1}$ as described in \eqref{eq:composition}.  The vertical
direction is the last coordinate and the horizontal direction
represents all the other coordinates.  Note that the entire boundary
and the horizontal segment in the middle get mapped to $x_{0}$.  Next
we move along the arrow to the second box by using a homotopy to
shrink the domains of the maps $\alpha_{1}$ and $\alpha_{2}$.  The
large box is the standard domain for the maps that represent the
loops.  The light gray area is all mapped to $x_{0}$.  We deform again
and move $\alpha_{1}$ and $\alpha_{2}$ around as illustrated in the
various figures and eventually blow up domains to standard size.  In
this way we have constructed a family of homotopies that reverse the
order of $\alpha_{1}$ and $\alpha_{2}$.  Note well that we have
demonstrated $[\alpha_{2}\circ \alpha_{1}] = [\alpha_{1}\circ
\alpha_{2}]$, we have \emph{not} shown that $\alpha_{2}\circ
\alpha_{1} = \alpha_{1}\circ \alpha_{2}$.

Assume we have a flat connection $A$ and we are studying a $r$-flat
connection $\mathcal{A}$ on $\Omega^{n}(M,x_{0})$ for $n \ge 1$. 
We are interested in computing the holonomy associated with the 
connection $\mathcal{A}$ so it is worthwhile being precise about 
exactly what we are going to do. Because we are working in 
$\Omega^{n}(M,x_{0})$ our ``base $n$-loop'' is the constant loop 
$x_{0}$. If we let $\bm{\sigma} = (\sigma^{1},\ldots,\sigma^{n})$ 
then a loop in $\Omega^{n}(M,x_{0})$ is given by a map 
$\gamma: [0,2\pi]^{n} \times [0,2\pi] \to M$ with the following 
properties:
\begin{enumerate}
    \item  $\gamma(\bm{\sigma},0) = \gamma(\bm{\sigma},2\pi) = x_{0}$

    \item  If $\gamma_{\tau}(\bm{\sigma}) = \gamma(\bm{\sigma},\tau)$ 
    then $\gamma_{\tau} \in \Omega^{n}(M,x_{0})$.
\end{enumerate}
From this we see that  $\Omega(\Omega^{n}(M,x_{0}),x_{0}) =
\Omega^{n+1}(M,x_{0})$ which is just the old inductive definition of
the higher loop spaces.  What are the diffeomorphism of the parameter
space that are compatible with the loop structure we have?  We are
looking for diffeomorphisms $f$ of $[0,2\pi]^{n+1}$ that are connected
to the identity transformation, and have the property that $f$
restricted to the boundary $\partial \left( [0,2\pi]^{n+1} \right)$ is a
diffeomorphism of the boundary that is connected to the identity
transformation.  Under such a diffeomorphism the basepoint $x_{0}$ is
kept fixed by the maps into the target space $M$.  This is necessary
for the validity of the variational formulas we have presented in this
paper.

Consider two elements $\alpha_{1}, \alpha_{2} \in
\Omega^{n+1}(M,x_{0})$ that represent a pair of loops in
$\Omega^{n}(M,x_{0})$, and look at their composition $\alpha_{2} \circ
\alpha_{1} \in \Omega^{n+1}(M,x_{0})$.  If $\hol(\alpha_{1})$ is the
holonomy of $\alpha_{1}$ then the first order differential equations
that defines the holonomy tells us that $\hol(\alpha_{2})
\hol(\alpha_{1}) = \hol(\alpha_{2}\circ \alpha_{1})$.  This is the
basic mechanism that leads to the concept of the holonomy group.  

The main result of this section is that the holonomy group is abelian
if $n \ge 1$.  We will demonstrate that $\hol(\alpha_{2}\circ
\alpha_{1}) = \hol(\alpha_{1}\circ \alpha_{2})$.  To show this we will
use Figure~\ref{fig:abelian} but interpret the diagram differently
using $r$-flatness instead of homotopy.
We begin at the upper left of Figure~\ref{fig:abelian} and compute 
the holonomy of $\alpha_{2} \circ \alpha_{1}$. Next what we are going 
to do is deform $\alpha_{2} \circ \alpha_{1}$ to a different element 
in $\Omega^{n+1}(M,x_{0})$ that has the same holonomy. 
\begin{figure}[tbp]
    \centering
    \includegraphics[width=0.8\textwidth]{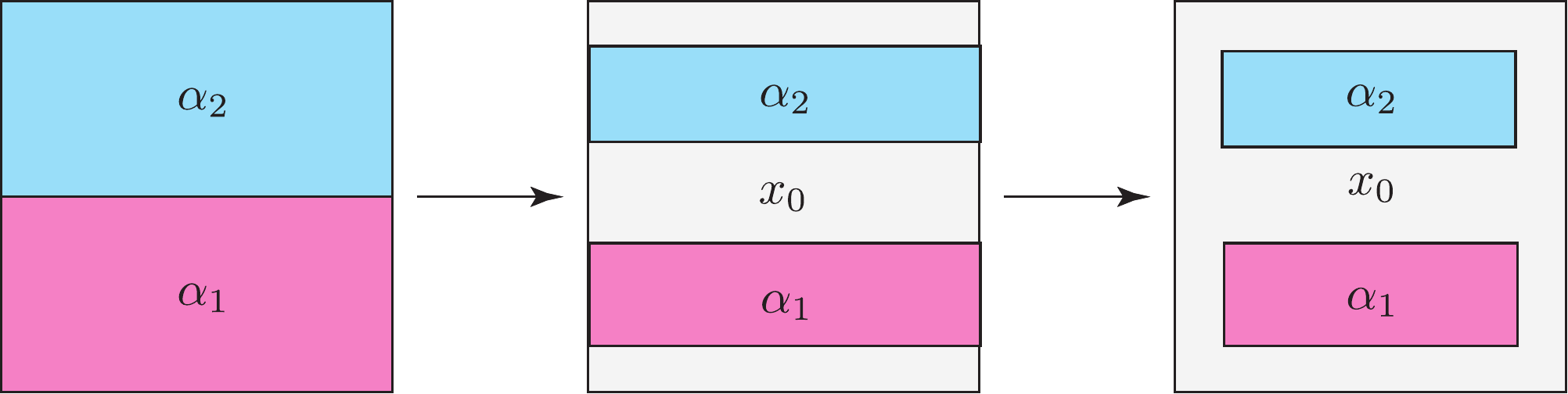}
    \begin{quotation}
	\caption[abelh]{\small Shrinking the domains of the loops in 
	such a way that the holonomy is not changed.}
	\label{fig:abelian-hol}
    \end{quotation}
\end{figure}
\begin{quote}
    \small The reader is familiar with this deformation in the case
    $n=0$.  Compute the holonomy $\hol(\gamma)$ for a loop $\gamma \in
    \Omega^{1}(M,x_{0})$.  Consider the loop $\tilde{\gamma}$ which is
    the same point set as $\gamma$ but traversed in the following way:
    you stay at $x_{0}$ for $\tau \in [0,\pi/2]$, next you go fast
    along the same point set by setting $\tilde{\gamma}(\tau) =
    \gamma(2\tau - \pi)$ for $\tau\in [\pi/2,3\pi/2]$, and finally you
    stay at $x_{0}$ until $\tau$ reaches $2\pi$.  This loop has the
    same holonomy for two reasons:(1)  in part of the loop you are not
    moving hence $\partial x/\partial \tau =0$, and (2) the
    reparametrization invariance of the holonomy in the other part of
    the path.
\end{quote}
We use the same idea as we go from the left diagram to the central one
in Figure~\ref{fig:abelian-hol}.  We shrink the respective domains of
the two loops in the $\tau$ direction exploiting the fact that at the
beginning, middle and end we are at $x_{0}$.  The holonomy is computed
using eq.~\eqref{eq:def-W} but with connection $\mathcal{A}$. The
same arguments presented in the block quote above are valid.  This new
loop has the same holonomy $\hol(\alpha_{2}\circ \alpha_{1})$.  Next
we move from the central diagram to the right diagram in
Figure~\ref{fig:abelian-hol} by shrinking the domains in the
$\sigma^{n}$ direction.  Inspecting \eqref{eq:def-calA-N} we see that
the connection is essentially unaltered because $\partial x/\partial
\sigma^{n}=0$ in the extension parts and the automatically built in
reparametrization invariance in the $\bm{\sigma}$ directions.  This
may be seen more concretely by studying the $n=1$ case of
\eqref{eq:calA} and again noticing that $\partial x/\partial \sigma$
vanishes in the extension parts and the reparametrization invariance
in $\sigma$.  This deformation does not change the holonomy.  We finish
by sequentially shrinking the domains in
$\sigma^{n-1},\sigma^{n-2},\ldots,\sigma^{1}$.  Once there we can go
to Figure~\ref{fig:abelian} and move things around using
reparametrization invariance and the fact that we have a $r$-flat
connection.  These diffeomorphisms do not change the holonomy because
of the $r$-flatness of $\mathcal{A}$.  We finish by undoing the domain
shrinking.  Throughout this entire procedure the holonomy has not
changed and thus we conclude that $\hol(\alpha_{2} \circ \alpha_{1}) =
\hol(\alpha_{1} \circ \alpha_{2})$ and we are finished with the proof.
\begin{theorem}
    \label{thm:r-flat-abel-hol}
    The holonomy group of a $r$-flat connection $\mathcal{A}$ on
    $\Omega^{n}(M,x_{0})$ is abelian.
\end{theorem}

Next we argue that there is an Ambrose-Singer type theorem in play
here by using some of the theorems from Section~\ref{sec:higher-loop}
about $r$-flat connections on $\Omega^{n}(M,x_{0})$.  We probe the local
curvature $D_{A}B$ at some point $x \in M$ of the connection
$\mathcal{A}$ in the following way.  Pick a reference path from
$x_{0}$ to $x$.  Consider a ``degenerate loop'' in
$\Omega^{n+1}(M,x_{0})$ that has collapsed to the reference path in
analogy to the top diagram in Figure~\ref{fig:AS}.  Such a loop has
trivial holonomy.  Next, at the endpoint $x$ on the path, we blow up
the loop make a small infinitesimal \emph{bulb}.  Note that the
surface of the bulb is $(n+1)$ dimensional while the ``interior'' of
the bulb is morally $(n+2)$ dimensional.  The holonomy for this loop
will be the parallel transport along the path of the $(n+2)$ form
$D_{A}B$ evaluated on the small volume.  Mimicking Ambrose and Singer
we parallel transport all the $D_{A}B$ from all points back to the
basepoint $x_{0}$.  These span some linear subspace $\mathfrak{h}$ of
$\lieg$.  This Lie algebra subspace represent the infinitesimal
holonomy.  The argument is analogous to what is done in the
Ambrose-Singer Theorem.  Take the loop and approximate it with many
bulbs.  The holonomy group is abelian and so we expect $\mathfrak{h}$
to be abelian and to also be the Lie algebra of the holonomy group. 
In other words, the curvature $D_{A}B$ is related to the Lie algebra 
of the holonomy group \emph{\`{a} la} Ambrose and Singer.
We have argued that $\mathfrak{h}$ is an abelian subalgebra of $\lieg$
but we have not found an argument for why it should be an abelian
ideal.

\subsection{Flat connections and holonomy}
\label{sec:flat-hol}

We remind the reader about a standard theorem in the theory of
connections.  Assume $X$ is a manifold with a flat connection with
structure group $G$.  The holonomy of a flat connection gives a group
homomorphism, \emph{i.e.}, a representation, $\rho: \pi_{1}(X,x_{0})
\to G$ that characterizes the flat connection in the connected
component of $X$ containing $x_{0}$.  We can apply this to our loop
space connections.  Notice that the definition of the homotopy groups
tell us that $\pi_{k}(\Omega^{n}(X,x_{0})) = \pi_{n+k}(X,x_{0})$.

Let $\mathcal{A}$ be a flat connection on $\Omega^{n}(M,x_{0})$, $n
\ge 1$.  This connection is automatically $r$-flat and therefore has
abelian holonomy.  Our loop space is not necessarily connected because
$\pi_{0}(\Omega^{n}(M,x_{0})) = \pi_{n}(M,x_{0})$.  We can now
restrict to a single connected component of $\Omega^{n}(M,x_{0})$.
After all, we are stuck in a connected component because we
continuously develop in time.  From the previous paragraph we see that
the holonomy of the flat connection gives us a group homomorphism
$\rho: \pi_{n+1}(M,x_{0}) \to H_{x_{0}}$ where $H_{x_{0}} \subset G$
is the abelian holonomy group.  Note that $\pi_{n+1}(M,x_{0})$ is
abelian and therefore its image under $\rho$ must be abelian.  This
behavior is compatible with Theorem~\ref{thm:r-flat-abel-hol}.
\begin{theorem}
    \label{thm:flat-abelian}
    A flat connection on $\Omega^{n}(M,x_{0})$ gives a representation
    $\rho: \pi_{n+1}(M,x_{0}) \to H_{x_{0}}$ where
    $H_{x_{0}} \subset G$ is the abelian holonomy group.
\end{theorem}
Why is this theorem important for us? In our method the conserved 
charges may be obtained by taking traces of the holonomy element.

In the familiar Lax-Zakharov-Shabat construction, corresponding to
$n=0$ in this notation, a flat connection gives a map $\rho:
\pi_{1}(M,x_{0}) \to G$.  For $M=S^{1}$ then we have a map
$\rho: \bbZ \to G$ that determines the conserved quantities.  The
image of $\rho$ will be abelian.

The case $n \ge 1$ corresponds to a spatial manifold with $\dim M =
n+1$ and for simplicity we take $M=S^{n+1}$.  We see that a flat
$\mathcal{A}$ connection gives a representation $\rho:
\pi_{1}(\Omega^{n}(S^{n+1},x_{0}) \to H_{x_{0}}$.  We note that
$\pi_{n+1}(S^{n+1},x_{0}) \approx \bbZ$ and therefore we have a group
homomorphism $\rho: \bbZ \to H_{x_{0}}$ just as in the
Lax-Zakharov-Shabat case.

\subsection{Nested loop space connections}
\label{sec:nested}

The loop space connection structures we have been discussing can be
generalized in the following way~\cite{Alvarez:1997ma} to a nested
construction.  First we relax the loop space definition to toroidal 
loop spaces
\begin{align}
    \Omega^{n}(M,x_{0}) &= \bigl\{ f: \mathbb{T}^{n} \to M \; | \; 
    f(0)=x_{0} \bigr\}, 
    \nonumber \\
    &= 
    \bigl\{ f: [0,2\pi]^{n} \to M \; | \; f(0,\ldots,0)=x_{0}
    \bigr\}\,.
    \label{eq:gen-loop}
\end{align}
The notation in this Section is chosen to agree with
the conventions of Section~\ref{sec:localcurv}.  Assume we are
studying integrable models in a $(d+1)$-spacetime.  We introduce a
sequence of ordinary connections $A^{(1)}, A^{(2)},\ldots,A^{(d)}$
associated with Lie algebras $\lieg^{(1)},
\lieg^{(2)},\ldots,\lieg^{(d)}$.  We also introduce a sequence of
differential forms $B^{(1)}, B^{(2)},\ldots,B^{(d)}$ where $B^{(k)}$
is a $\lieg^{(k)}$-valued $k$-form that transforms under the
respective adjoint representation.  We will use $(A^{(k)},B^{(k)})$ to
define a new type of parallel transport on $\Omega^{k-1}(M,x_{0})$
that is a twisted version of the construction in
Section~\ref{sec:higher-loop}.

There are now a variety of games you can play.  For example, you can
make all the loop space connections independent of each other. No new 
obvious phenomenon is seen here.

You can try something that is highly non-trivial.  Assume all the Lie
algebras fit inside a big Lie algebra $\lieg$.  Let $P^{(k)}$ be the
ordinary parallel transport along paths associated with the ordinary
connection $A^{(k)}$.  Introduce a ``twisted'' parallel transport
operator $\ptop^{(k+1)}$ on $\Omega^{k}(M,x_{0})$ defined by an
inductive procedure.  The parallel transport from the constant base
loop $x_{0}$ to $\gamma \in \Omega^{k}(M,x_{0})$ will be denoted by
$\ptop^{(k+1)}(\gamma)$.  We recall that our toroidal loop spaces are
given by maps on an appropriate hypercube.  This hypercube has a
natural Cartesian coordinate system that we will use in the
construction.  The inductive definition is
\begin{align}
    0 & = 
    \frac{d}{d\tau}\; \ptop^{(k+1)} 
    \nonumber \\
    &\quad + \left[ \left(\ptop^{(k)} \right)^{-1} \left(
    \int_{0}^{2\pi} d\sigma^{k}\; \left(P^{(k+1)}\right)^{-1} \left(
    \iota_{\partial/\partial\sigma^{k}} \iota_{\partial/\partial\tau}
    B^{(k+1)} \right) P^{(k+1)} \right) \left(\ptop^{(k)} \right)
    \right] \ptop^{(k+1)}\,.
    \label{eq:def-nested-pt}
\end{align}
This equation is a bit schematic and requires some explanation.  We
start at $x_{0}$ at $\tau=0$ and we evolve in time $\tau$ to a loop
$\gamma_{\tau} \in \Omega^{k}(M,x_{0})$.  The developed surface is a
$(k+1)$-submanifold $\Sigma_{\tau}$ with boundary
$\partial\Sigma_{\tau} = \gamma_{\tau} - \{x_{0}\}$.  We note that
\begin{equation*}
    \int_{0}^{2\pi} d\sigma^{k}\; \left(P^{(k+1)}\right)^{-1} \left(
    \iota_{\partial/\partial\sigma^{k}} \iota_{\partial/\partial\tau}
    B^{(k+1)} \right) P^{(k+1)}
\end{equation*}
is a $(k-1)$-form.  At time $\tau$ and at location $(\sigma^{1},
\sigma^{2}, \ldots, \sigma^{k-1}, \sigma^{k})$ the term above is sum
of the parallel transports of $B^{(k+1)}$ using $A^{(k+1)}$ along the
curves with tangent vector $\partial/\partial\sigma^{k}$ to the
``boundary'' with coordinates $(\sigma^{1}, \sigma^{2}, \ldots,
\sigma^{k-1}, 0)$.  This boundary is a loop
$\gamma_{\tau}(\sigma^{1},\sigma^{2}, \ldots, \sigma^{k-1}, 0)$ in
$\Omega^{k-1}(M,x_{0})$ and therefore the Lie algebra element we have 
just computed can be parallel transported using $\ptop^{(k-1)}$. Some 
concrete examples are given in \cite{Alvarez:1997ma}. 
\emph{N.B.} Had we used the spherical loop spaces then at the boundary 
$\gamma_{\tau}(\sigma^{1},\sigma^{2}, \ldots, \sigma^{k-1}, 0) = 
x_{0}$ and what we are trying to do collapses and we are basically 
back  in Section~\ref{sec:higher-loop}.

The upshot of the nested structure given by \eqref{eq:def-nested-pt}
is that $B^{(k)}$ at different levels can mix.  We have not studied
$r$-flatness in this case but imposing cross level vanishing of
brackets as in \eqref{eq:localzccomm} leads to conserved quantities.

\section{The local zero curvature conditions}
\label{sec:localcurv}

We now discuss the local conditions in space-time which are
sufficient for the vanishing of the curvature of the connection on loop space.  

As we have seen in the previous sections, the implementation of the 
generalized zero curvature condition in a space-time ${\cal M}$ of $d+1$
dimensions involves a nested structure of generalized loop
spaces (see section \ref{sec:nested}). In order to define the one-form
connection on loop space we 
introduced in the space-time ${\cal M}$, $d$ pairs of antisymmetric 
tensors $B_{\mu_1\ldots \mu_N}^{(N)}$ and one-form connections
$A_{\mu}^{(N)}$, with $N=1,2,\ldots d$. The connections were used to
parallel transport the tensors along curves starting and ending at a
chosen fixed point $x_0$ of ${\cal M}$. Consequently, the tensors appear
always in the conjugated form 
\be
B_{\mu_1\mu_2\ldots \mu_N}^{W_N}\equiv  W_{N}^{-1} \,
B_{\mu_1\mu_2\ldots \mu_N}^{(N)}\, W_{N} 
\lab{bwdef}
\ee
where $W_{N}$ is obtained by integrating the connection along the
curve, pa\-ra\-me\-te\-ri\-zed by $\sigma$, through the differential equation  
\be
\frac{dW_{N}}{d\sigma} + A_{\mu}^{(N)}
\frac{dx^{\mu}}{d\sigma} W_{N} =0
\ee
In the previous sections we discussed the conditions for the
connection on loop spaces to be $r$-flat. Here we shall impose in
addition that the connections $A_{\mu}^{(N)}$ are flat, i.e.
\be
F_{\mu\nu}^{(N)}\equiv
\partial_{\mu}A_{\nu}^{(N)}-\partial_{\nu}A_{\mu}^{(N)}+  
\sbr{A_{\mu}^{(N)}}{A_{\nu}^{(N)}}=0
\lab{flatan}
\ee
That
implies that the quantities $W_{N}$ are uniquely defined on every
point of ${\cal M}$, once their values at $x_0$ are chosen. The connections
are then written in the pure gauge form
\be
A_{\mu}^{(N)} = - \partial_{\mu} W_{N}\, W_{N}^{-1}
\lab{puregaugea}
\ee
As we have seen the generalized zero curvature condition does lead to
conserved quantities expressed in terms of path ordered integrals of the
connection on loop space. In $1+1$ dimensions the loop
space coincides with (or is isomorphic to) the space-time ${\cal M}$, and so
these non-locality problems disappear leading in fact to an extension of
the usual 
formulation of two dimensional integrable theories given by the equations
\be
\partial_{\mu}A_{\nu}^{(1)}-\partial_{\nu}A_{\mu}^{(1)} +
\sbr{A_{\mu}^{(1)}}{A_{\nu}^{(1)}}=0
\lab{laxeq}
\ee
and
\be
D_{\mu}^{(1)}B_{\nu}^{(1)}-D_{\nu}^{(1)}B_{\mu}^{(1)} +
\sbr{B_{\mu}^{(1)}}{B_{\nu}^{(1)}}=0
\lab{newlaxeq}
\ee
where 
\be
D_{\mu}^{(N)}\bullet\equiv \partial_{\mu} \bullet +\sbr{A_{\mu}^{(N)}}{\bullet}
\lab{dndef}
\ee
The relation \rf{laxeq} is the usual Lax-Zakharov-Shabat equation
\cite{lax} employed in two dimensional integrable field theories, and it
leads to the conserved charges which are the eigenvalues of the path
ordered integrals 
\be
W_{1}= P\,e^{\int_{\Gamma}d\sigma\, A_{\mu}^{(1)}\frac{dx^{\mu}}{d\sigma}}
\lab{conservedcharges1}
\ee
where $\Gamma$ is the one dimensional space submanifold of the two
dimensional space-time ${\cal M}$.
However, our formulation also includes another vector $B_{\mu}^{(1)}$
satisfying \rf{newlaxeq} and that leads to another set of conserved
quantities given by the eigenvalues of the operator 
\be
V_{1}= P\,e^{\int_{\Gamma}d\sigma\,
  W_{1}^{-1}\,B_{\mu}^{(1)}\,W_{1}\,\frac{dx^{\mu}}{d\sigma}} 
\ee
The consequences of the existence of that second type of charges are now
being investigated and may perhaps unify some treatments of local and
non-local charges in two dimensional field theories \cite{miramontesnew}. 
 
The question we face in dimensions higher than two is how to relate
the loop space zero curvature condition to the dynamics (equations of
motion) of 
theories defined on the space-time ${\cal M}$. The main obstacle is the 
highly non-local character of the loop space zero curvature when
expressed in terms of the tensors and connections defined in
${\cal M}$.  
That fact makes us believe that the proper formulation of integrable  theories 
in a space-time of dimension higher than two may require not just
terms involving  particles but also terms that include fluxes or other
extended objects. The implementation  of such ideas is
therefore the main challenge to our approach in the future. 

However, one can avoid the non-locality problems of the zero curvature
condition by selecting local equations in ${\cal M}$ which are sufficient
conditions for the vanishing of the loop space  curvature. We have
seen in the previous sections that the concept of $r$-flatness leads
to an improvement of such non-locality problems, since it implies the
vanishing of the commutator term separately from the term involving
the exterior covariant derivative of the tensors $B$'s (see Theorem
\ref{thm:r-flatness-cond}).  Therefore, one observes that one way (and
perhaps the only one) of imposing local conditions on ${\cal M}$ which are
sufficient for two conditions on loop space, namely the vanishing of
the loop space 
zero curvature and its independence of the scanning of the
hypersurfaces (i.e. $r$-flatness), is to have, in addition to
\rf{flatan}, the covariant exterior derivative of 
the tensors $B^{(N)}$ equal to zero, i.e.  
\be
D^{(N)}\wedge B^{(N)}=0 \qquad\qquad N=1,2 \ldots d
\lab{localzcn}
\ee
with $D^{(N)}$ defined in \rf{dndef}, 
and in addition to have the commutators of the components of the
tensors \rf{bwdef} also vanishing, i.e. 
\be
\sbr{B_{\mu_1\mu_2\ldots \mu_N}^{W_N}}{B_{\nu_1\nu_2\ldots\nu_M}^{W_M}}=0 
\qquad\qquad M,N=1,2 \ldots d
\lab{localzccomm}
\ee
The relations \rf{flatan}, \rf{localzcn} and \rf{localzccomm} are what
we call the {\em sufficient local zero curvature conditions}. They indeed
lead to conserved quantities as we now explain. 
 From \rf{puregaugea} and \rf{localzcn} one obtains 
that the ordinary exterior
derivative of $B_{\mu_1\mu_2\ldots \mu_N}^{W_N}$ vanishes, i.e 
\be
d\wedge B^{W_N}=0
\lab{dwedgebn}
\ee
It then follows that if $V_{N+1}$ is a $N+1$-dimensional volume in
${\cal M}$, and 
$\partial V_{N+1}$ is its boundary, i.e. a $N$-dimensional closed
surface,  then by \rf{dwedgebn} and  the abelian Stokes theorem 
\be
\int_{\partial V_{N+1}} B^{W_N} = \int_{V_{N+1}}d\wedge B^{W_N}=0
\lab{abelianstokes}
\ee
Notice that, if $B^{W_N}$ lives on a vector space with basis $v^a$, we
are applying the abelian Stokes theorem to each 
component $B^{W_N}_a$ separately
($B^{W_N}_{\mu_1\ldots\mu_N}=B^{W_N}_{\mu_1\ldots\mu_N;a}\, v^a)$, and
the issue if those components 
commute among themselves is not relevant here. 
Imposing appropriate boundary conditions on the pair
$\(A^{(N)},B^{(N)}\)$ can lead to conservation laws as we now
explain. 
 In a space-time ${\cal M}$ of $d+1$ dimensions, there are $d-N+1$
orthogonal directions to a $N$-dimensional surface. Let us choose one
of those directions and let us parametrize it by $\tau$. We can choose
the volume $V_{N+1}$ such that its border can be decomposed as 
\be
\partial V_{N+1}= \Sigma_{N}^{\tau_0}+\Sigma_{N}^{\tau_1}+\Gamma_{N}
\lab{surfacedecomp}
\ee
where $\Sigma_{N}^{\tau_0}$ and $\Sigma_{N}^{\tau_1}$ are $N$
dimensional surfaces perpendicular to the direction $\tau$, and 
corresponding to fixed values $\tau_0$ and $\tau_1$ respectively, of
the parameter $\tau$. $\Gamma_N$ is a $N$ dimensional surface joining
$\Sigma_{N}^{\tau_0}$ and $\Sigma_{N}^{\tau_1}$ into the closed
surface $\partial V_{N+1}$. If the boundary conditions are such that
the integral of $ B^{W_N}$ on  $\Gamma_N$ vanishes we then have 
from \rf{abelianstokes} and \rf{surfacedecomp} that
\be
\int_{\Sigma_{N}^{\tau_0}} B^{W_N}+ \int_{\Sigma_{N}^{\tau_1}}
B^{W_N}=0
\ee
If one now orients the surfaces in the same way with respect to the
$\tau$ direction one has a conserved charge in $\tau$ given by  
\be
Q^{(N)} = \int_{\Sigma_N} \, B^{W_N}
\lab{qndef}
\ee  
where $\Sigma_N$ is any surface perpendicular to the $\tau$
direction. Of course, we will be
mainly interested in quantities conserved in time and so we will be
concerned most with the case $N=d$, with $\Sigma_N$ being the spatial
sub-manifold the space-time ${\cal M}$. In any case, the number of
conserved charges will be determined by the dimension of the space
where the tensors $B^{(N)}$ live. 

We notice that the Hodge dual of $B^{W_N}$ in the $d+1$ dimensional
space-time ${\cal M}$, i.e. 
\be
J^{\rho\mu_1\ldots \mu_{d-N}}\equiv \varepsilon^{\rho\mu_1\ldots
  \mu_{d-N}\nu_1\ldots\nu_N}\,B_{\nu_1\nu_2\ldots \nu_N}^{W_N} 
\lab{conservedtensors}
\ee
is, as a consequence of \rf{dwedgebn}, a conserved antisymmetric tensor
\be
\partial_{\rho}\,J^{\rho\mu_1\ldots \mu_{d-N}}=0
\lab{conservedtensors2}
\ee  
and that is another way of expressing the conservation law we just
discussed. 

Note that associated to every pair $\(B_{\mu_1\ldots
  \mu_N}^{(N)},A_{\mu}^{(N)}\)$, we have gauge symmetries of the
sufficient local zero curvature 
conditions \rf{flatan}, \rf{localzcn}, and \rf{localzccomm}. Consider
the transformations
\br
A_{\mu}^{(N)}&\rightarrow& g\,
A_{\mu}^{(N)}\,g^{-1}-\partial_{\mu}g\,g^{-1}
\nonumber\\
B_{\mu_1\ldots\mu_N}^{(N)}&\rightarrow&
g\,B_{\mu_1\ldots\mu_N}^{(N)}\, g^{-1}
\lab{gaugetransf1}
\er
where $g$ is an element in a group with the Lie
algebra  corresponding to  where the 
connection $A_{\mu}^{(N)}$ lives, and $g$  acts on  the
tensor $B_{\mu_1\ldots\mu_N}^{(N)}$. It then follows that the
covariant derivatives of $B_{\mu_1\ldots\mu_N}^{(N)}$ transform in the
same way
\be
D_{\nu}^{(N)}B_{\mu_1\ldots\mu_N}^{(N)} \rightarrow g\,
D_{\nu}^{(N)}B_{\mu_1\ldots\mu_N}^{(N)}\, g^{-1}.
\ee
Therefore, \rf{flatan} and \rf{localzcn} are clearly invariant under
\rf{gaugetransf1}. In addition, we have that under \rf{gaugetransf1}
\be
W_N \rightarrow g\(x\)\, W_N \, g^{-1}\(x_0\)
\ee
where $x_0$ and $x$ are the initial and final points respectively, of
the curve where $W_N$ is calculated, with $x_0$ being the fixed point
of ${\cal M}$ we introduced above. Consequently, $B_{\mu_1\ldots
  \mu_N}^{W_N}$ is invariant under \rf{gaugetransf1}, and so is the
condition \rf{localzccomm}. It also follows that the conserved charges 
\rf{qndef} are invariant under \rf{gaugetransf1}. 

The covariant derivatives \rf{dndef} commute since the connections
$A_{\mu}^{(N)}$ are flat, and so most of the properties of the
ordinary exterior derivatives apply as well as to covariant exterior
derivatives, in particular $D^2=0$. Therefore, the conditions \rf{flatan} and
\rf{localzcn} are invariant under the gauge transformations
\br
A^{(N)}&\rightarrow& 
A^{(N)}
\lab{gaugetransf2}\\
B^{(N)}&\rightarrow& B^{(N)} + D^{(N)}\wedge \alpha^{(N-1)}
\nonumber
\er
where $\alpha^{(N-1)}$ is an antisymmetric tensor of rank $N-1$. The
invariance of the condition \rf{localzccomm} under \rf{gaugetransf2}
needs some more refined structures which we discuss below. 

Basically there are two ways of implementing the conditions
\rf{localzcn} and \rf{localzccomm}. The first one is as follows. 
 Given the reference point $x_0$ of ${\cal M}$ we
take the components of the tensors $B^{(N)}$ on that point to commute,
i.e.
\be
\sbr{B^{(N)}_{\mu_1\ldots\mu_N}\(x_0\)}{B^{(N)}_{\nu_1\ldots\nu_N}\(x_0\)}=0
\lab{commutebatx0}
\ee
Then we use the fact that the connections
$A_{\mu}^{(N)}$ are flat (see \rf{puregaugea}) and construct the
tensors $B^{(N)}$ on any point $x$ of ${\cal M}$, by parallel transport with
the connection $A_{\mu}^{(N)}$ along a given curve from the fixed
point $x_0$ to $x$. Notice that since the connections are flat it does
not matter the curve we choose to link $x_0$ to $x$. We then have that
\be
B^{(N)}_{\mu_1\ldots \mu_N}= W_N\, B^{(N)}_{\mu_1\ldots
  \mu_N}\(x_0\)\, W_N^{-1}
\lab{bforsecondcase}
\ee
Consequently, from \rf{bwdef} one has 
\be
B_{\mu_1\mu_2\ldots \mu_N}^{W_N} = B^{(N)}_{\mu_1\ldots\mu_N}\(x_0\)
\lab{bwtopological}
\ee
and so \rf{localzccomm} is satisfied. 
Of course,  such tensors are covariantly constant 
\be
D^{(N)}_{\nu}  B^{(N)}_{\mu_1\ldots \mu_N} =0
\lab{covconstbn}
\ee
and so they trivially satisfy \rf{localzcn}. Notice that
 \rf{commutebatx0} and \rf{bforsecondcase} imply that the components
 of the tensors $B^{(N)}_{\mu_1\ldots \mu_N}$ commute on every point
 on ${\cal M}$. However, those components at different points do not have to do
 so. 

The invariance of the condition \rf{localzccomm} under
\rf{gaugetransf2} can be established by taking $\alpha^{(N-1)}$ 
as the parallel transport of a tensor $\alpha^{(N-1)}_0$, in a way similar to
\rf{bforsecondcase}, i.e. 
\be
\alpha^{(N-1)} = W_N\,\alpha^{(N-1)}_0\, W_N^{-1} 
\ee
We then have $D^{(N)}\wedge \alpha^{(N-1)} = W_N\,d\wedge \alpha^{(N-1)}_0\,
W_N^{-1}$, and if we impose that $d\wedge \alpha^{(N-1)}_0$ live in
the same abelian algebra as the components of
$B^{(N)}_{\mu_1\ldots\mu_N}\(x_0\)$ (see  \rf{commutebatx0}), we have
the invariance of \rf{localzccomm} under \rf{gaugetransf2}. 

The conserved charges \rf{qndef} in such case have a geometrical
meaning  and 
correspond to projections of hypersurfaces in the directions defined
by those constant tensors. Indeed, from \rf{qndef} and
\rf{bwtopological} one has
\be
Q^{N}= \int_{\Sigma_N} \, B^{W_N} =
B^{(N)}_{\mu_1\ldots\mu_N}\(x_0\)\,\int_{\Sigma_N}\,d\Sigma^{\mu_1\ldots\mu_N} 
\lab{topologicalcharges}
\ee

The examples we found that fit in the first 
 type of local zero curvature condition are topological field
 theories like Chern-Simons and BF theories. Indeed, the BF theory in
 $2+1$ dimensions \cite{bf} involves an antisymmetric tensor
 $B_{\mu\nu}$ and a flat connection $A_{\mu}$, and its equations of
 motion  are given by \rf{flatan} and \rf{covconstbn} for
 $N=2$. Obviously, the Chern-Simons theory also fits into the scheme
 since it involves just a flat connection. We believe however that the
 important applications of our methods to  topological theories
 will appear when we consider our loop spaces defined on space-times
 ${\cal M}$ with non-trivial topological structures like holes, handles,
 etc. We will then have to use  modifications of the non-abelian
 Stoke's theorem on loop space on the lines of \cite{hirayama}. In the
 case of Chern-Simons and BF theories we may perhaps relate the
 modification of our
 conserved quantities \rf{topologicalcharges} to the knot theory
 invariants which are known to appear in those models. 

A second way of implementing the local conditions \rf{localzcn} and
\rf{localzccomm} involves taking the pairs $\(B_{\mu_1\ldots \mu_N}^{(N)}, 
A_{\mu}^{(N)}\)$ to live in a non-semisimple Lie algebra ${\cal
  G}^{(N)}$,  such  that $B_{\mu_1\ldots \mu_N}^{(N)}$ has
components only in the direction of the abelian ideal
${\cal P}^{(N)}$ of ${\cal G}^{(N)}$. It then follows that $ W_{N}$
belongs to the group whose Lie algebra is ${\cal G}^{(N)}$ and
therefore $B_{\mu_1\mu_2\ldots \mu_N}^{W_N}$, defined in \rf{bwdef},
also belongs to the abelian ideal ${\cal P}^{(N)}$. If we now impose
that the different abelian ideals commute 
\be
\sbr{{\cal P}^{(M)}}{{\cal P}^{(N)}}=0 \qquad\qquad M,N=1,2 \ldots d
\ee
then we satisfy condition \rf{localzccomm}. The equation 
\rf{localzcn} is then the only condition to be imposed on the tensors $
B^{(N)}$ and it will therefore define the dynamics of the generalized 
integrable theory as specified below. Notice that such formulation
does not need to specify the commutation relations among the
complements of the ${\cal P}^{(N)}$s in the algebras ${\cal
  G}^{(N)}$s. The number of conserved charges coming from
\rf{qndef} is of course given by the sum of the 
dimensions of the abelian ideals ${\cal P}^{(N)}$, where the 
tensors $B^{(N)}$ live. 

The invariance of the condition \rf{localzccomm} under
\rf{gaugetransf2} is guaranteed by taking the tensors $\alpha^{(N-1)}$
to live in the abelian ideals ${\cal P}^{(N)}$. 

We then observe that the algebraic structure underlying that second
type of local integrable theories is that of non-semisimple Lie
algebras. Most of those algebras can be cast in terms of a semi-simple
Lie algebra ${\cal G}$ and a representation $R$ of it, with the
commutation relations being given by
\br
\sbr{T_a}{T_b} &=& f_{ab}^c\, T_c\nonumber\\
\sbr{T_a}{P_i}&=& P_j\, R_{ji}\(T_a\)
\lab{poincarelike}\\
\sbr{P_i}{P_j}&=&0\nonumber
\er
with $R$ being a matrix representation of ${\cal G}$, i.e. 
\be
\sbr{R\(T_a\)}{R\(T_b\)}=R\(\sbr{T_a}{T_b}\)
\ee 
Therefore, since the number of conserved currents is given by the
dimension of the abelian ideal ${\cal P}$, it follows that the
integrability concepts will be related  to
infinite dimensional representations. As we will see in the
applications, those representations will be given in general by 
infinite  direct
products 
of finite representations, i.e. $R=\otimes_k R_k$. This
differs in a crucial way 
from the algebraic structures we find in $1+1$ dimensional integrable
theories where we have infinite algebras like the Kac-Moody algebras.
Such algebras can in fact be graded as
\be
{\hat {\cal G}} = \otimes_{n=-\infty}^{\infty} {\hat {\cal G}}_n
\qquad \qquad \sbr{{\hat {\cal G}}_m}{{\hat {\cal G}}_n} \subset {\hat
  {\cal G}}_{m+n} 
\lab{gradation}
\ee
The subspace ${\hat {\cal G}}_0$ is a finite subalgebra and the other
subspaces transform under a given representation of it $\sbr{{\hat
    {\cal G}}_0}{{\hat {\cal G}}_n} \subset {\hat{\cal G}}_{n}$,
similarly to the ideals ${\cal P}$ under ${\cal G}$. However, the
crucial difference with our formulation is that the generators in
those representations do 
not have to commute. The requirement of locality is what has driven us
to the abelian character of those representations. In 
order to have the full algebraic structures of the zero curvature
condition on loop spaces, we believe we have to deal with theories
where the fundamental objects are not just particles but perhaps fluxes. 

The second way of implementing local conditions that imply the
vanishing of the loop space zero curvature shows that the relation
\rf{localzccomm} is satisfied by an algebraic procedure and so it does
not lead to conditions on the dynamics of the theory. Such conditions
have to come  from the relations  \rf{flatan} and
\rf{localzcn}. In the applications of such formulations for Lorentz
invariant theories that have appeared in
the literature so far, only the pair $\( A^{(N)},B^{(N)}\)$ for $N=d$
has been used. In such cases the Hodge dual of $B^{(N)}$ is a vector,
\emph{i.e.}
\be
{\tilde B}^{\mu}\equiv \varepsilon^{\mu\nu_1\ldots\nu_d}\, 
B^{(N=d)}_{\nu_1\ldots\nu_d}.
\lab{dualbn=d}
\ee
Therefore, the condition \rf{localzcn} becomes 
\be
D_{\mu}  {\tilde B}^{\mu}=0
\lab{truezc1}
\ee
We then have from \rf{conservedtensors} the conserved currents
\be
J^{\mu}= W^{-1}_{N=d}\, {\tilde B}^{\mu}\, W_{N=d} \qquad\qquad \qquad 
\partial_{\mu}J^{\mu}=0
\lab{currentsn=d}
\ee

In the examples constructed in the literature thus far, the
equations of motion were found to be 
equivalent to the condition \rf{truezc1} (or \rf{localzcn}), whilst
the condition 
\rf{flatan} was trivially satisfied, i.e. involved a connection that
was flat for any field configuration. For theories that are not
Lorentz invariant there are examples where the equations of motion come
from \rf{flatan} and where \rf{localzcn} was trivially satisfied,
i.e. the tensors $B^{(N=d)}$ were the exterior covariant derivative of a
lower rank tensor, i.e. $B^{(N=d)}=D^{(N=d)}\wedge \alpha^{(d-1)}$. We now
discuss some examples where this formulation was implemented.

\section{Examples}
\label{sec:examples}

\subsection{Models on the sphere $S^2$}
\label{sec:s2}

A class of models that has been well explored using the formulation
described in Section~\ref{sec:localcurv} is one where the fields take
values on the two dimensional sphere $S^2$.  The fields may be taken
to be a triplet of real scalar fields ${\vec n}$ subject to the
constraint ${\vec n}^2=1$, or alternatively a complex scalar field $u$
parametrizing the plane that corresponds to the stereographic
projection of $S^2$.  The two descriptions are related by 
\be 
{\vec n}
= \frac{1}{1+\mid u\mid^2} \, \( u+u^* , -i \( u-u^* \) , \u2 -1 \)
\qquad \quad u=\frac{n_1+i\,n_2}{1-n_3} \lab{stereo} 
\ee 
In the
examples discussed in the literature thus far only the pair
$\(A^{(N)},B^{(N)}\)$, for $N=d$ has been used (with $d$ being the
number of space dimensions).  The flat connection, satisfying
\rf{flatan}, is taken to live in the algebra of $sl(2)$ and given by
\cite{Alvarez:1997ma} 
\be A_{\mu}^{(d)} \equiv A_{\mu}
=\frac{1}{1+\u2}\left[ -i \pa_{\mu} u T_{+} -i \pa_{\mu} u^* T_{-}
+\left(u\,\pa_{\mu} u^* - u^*\,\pa_{\mu} u\right) \, T_3 \right]
\lab{s2pota} 
\ee
with the generators satisfying the $sl(2)$
commutation relations
\begin{equation}
[T_{3},T_{\pm}] = \pm T_{\pm} \;  \qquad\qquad
[T_{+},T_{-}] = 2 T_{3}
\lab{sl2commrel}
\end{equation}
Notice that such a connection is flat and satisfies \rf{flatan} for
any configuration of the complex field $u$.  In fact, you can write
the connection in the pure gauge form \rf{puregaugea} with
$W_{N=d}\equiv W_i$, $i=1,2$, and
\begin{equation}
W_1 =  e^{iuT_+}\, e^{\varphi T_3}\, e^{iu^*T_-} \, ; \qquad
 \qquad \qquad
W_2 = e^{iu^*T_-}\, e^{-\varphi T_3}\, e^{iuT_+}
\lab{w1w2}
\end{equation}
where $\varphi =  \ln (1 + \u2 )$. Notice that $W_1$ and $W_2$ are
elements of the group $SL(2, \IC )$, and not of $SU(2)$, but $i
A_{\mu}$ does belong to the algebra of $SU(2)$. The commutation
relations \rf{sl2commrel} are compatible with the hermiticity
conditions, $T_3^{\dagger} = T_3$, $T_{\pm}^{\dagger} = T_{\mp}$, and
so $W_1^{\dagger} = W_2^{-1}$. In the defining (spinor) representation
$R^{(1/2)}$ of
$SL(2, \IC )$ one has that the elements $W_1$ and $W_2$
coincide, i.e. 
\br
W\equiv R^{(1/2)}\( W_1\)= R^{(1/2)}\( W_2\)
= {1\over {\sqrt{1 + \mid u\mid^2}}}\, \( 
\begin{array}{cc}
1 & i u \\
i u^* & 1
\end{array} \)
\lab{wdef}
\er
Therefore, they are unitary two by two matrices of unity determinant and
so elements of $SU(2)$.  

Another interesting point is that the sphere $S^2$ can be mapped 
isometrically into the symmetric space $SU(2)/U(1)$ that may be 
identified with the
complex projective space $CP^1$. The $U(1)$
subgroup is invariant under the 
involutive automorphism of the  algebra \rf{sl2commrel}
\be
\sigma \( T_3\) = T_3 \qquad \qquad \sigma\( T_{\pm} \) = - T_{\pm}
\lab{autom}
\ee
The automorphism \rf{autom} is inner and given by 
\be
\sigma \( T\) \equiv e^{i\pi T_3} \, T \, e^{-i\pi T_3}
\lab{inneraut}
\ee
The elements of $SU(2)/U(1)$ can be parametrized by the variable
$x\( g \) \equiv g \sigma \( g\)^{-1}$,  $g \in SU(2)$, 
since $x\( g \)=x\( g k\)$ with $k\in U(1)$. In addition one has that 
$\sigma \( x\) = x^{-1}$. In the spinor representation one has
\br
R^{(1/2)}\( e^{i\pi T_3}\) =  
\( 
\begin{array}{cc}
i & 0 \\
0  & -i
\end{array} \)
\er
and so $W$ defined in \rf{wdef} satisfies.
\be
\sigma \( W\) = W^{-1}
\ee
Therefore, $W$ takes the place of the variable $x\( g \)$, and so
parametrizes the elements of the symmetric space 
$SU(2)/U(1)$, or equivalently of the sphere $S^2$ \cite{afz1}. 

The other element in the pair, namely $B^{(d)}$, has to live in an
abelian ideal, and so transform under some representation of $sl(2)$
(see discussion leading to \rf{poincarelike}). Since we want models
with an infinite number of conserved charges we must work with
infinite dimensional representations. One way of doing that is to use
the Schwinger's construction. Let $R\(T\)$ be a
(finite) matrix 
representation of the algebra $T$, i.e.
\be
\sbr{R\(T\)}{R\(T^{\prime}\)} = R\(\sbr{T}{T^{\prime}}\)
\ee
Consider oscillators in equal numbers to the dimension of $R$, 
\be
\sbr{a_i}{a_j}=0 \qquad \sbr{a_i^{\dagger}}{a_j^{\dagger}}=0 \qquad
\sbr{a_i}{a_j^{\dagger}}= \delta_{ij} \qquad \quad i,j=1,\ldots {\rm
  dim}\; R
\ee
It follows that the operators 
\be
S\( T\) \equiv \sum_{i,j} a^{\dagger}_i R_{ij}\(T\) a_j
\ee
constitute a representation of $T$ 
\be
\sbr{S\(T\)}{S\(T^{\prime}\)} = S\(\sbr{T}{T^{\prime}}\)
\ee 
The oscillators can be realized in terms of differential operators on
some variables $\lambda_i$, as
$a_i \equiv \frac{\pa\;}{\pa \lambda_i}$ and $a_i^{\dagger} = \lambda_i$.
In the case of the  algebra \rf{sl2commrel}, 
one gets that its two dimensional matrix representation leads to the
following realization in terms of differential operators (with the
parameters $\lambda_i$, $i=1,2$, being denoted $\lambda$ and
$\bar{\lambda}$) \cite{babelon} 
\begin{equation}
S\(T_{+}\)\equiv \lambda \, \frac{d\;}{d\bar{\lambda}} \; ; \qquad
S\(T_{-}\)\equiv \bar{\lambda} \, \frac{d\;}{d\lambda} \; ; \qquad
S\(T_{3}\)\equiv \frac{1}{2}\left( \lambda \, \frac{d\;}{d\lambda} -
  \bar{\lambda} 
\, \frac{d\;}{d\bar{\lambda}}\right) 
\lab{twoparrep}
\end{equation}
The states of the representations corresponding to such realization are
functions of $\lambda$ and $\bar{\lambda}$. 
The action of the operators are given by
\begin{eqnarray}
S\(T_{3}\) \lambda^p\, \bar{\lambda}^q &=& \frac{p-q}{2} \,  \lambda^p\,
\bar{\lambda}^q \nonumber\\ 
S\(T_{+}\) \lambda^p\, \bar{\lambda}^q &=& q \,  \lambda^{p+1}\,
\bar{\lambda}^{q-1} \nonumber\\ 
S\(T_{-}\) \lambda^p\, \bar{\lambda}^q &=& p \,  \lambda^{p-1}\,
\bar{\lambda}^{q+1} 
\lab{actiononl}
\end{eqnarray}
Notice that  from \rf{actiononl} that the action of $S\(T_3\)$ and
$S\(T_\pm\)$ leaves the 
sum of the powers of $\lambda$ and ${\bar \lambda}$ invariant.
Therefore, one can construct irreducible representations by
considering the states  
\begin{equation}
\ket{\left(p,q\right),m}\equiv \lambda^{p+m}\, \bar{\lambda}^{q-m}
\end{equation}
with $ m\in \IZ$ and $\left(p,q\right)$ being any pair of numbers
(real or even complex). 
Then
\begin{eqnarray}
S\(T_{3}\) \ket{\left(p,q\right),m} &=& \left( \frac{p-q}{2} +m\right) \,
\ket{\left(p,q\right),m}\nonumber\\  
S\(T_{+}\) \ket{\left(p,q\right),m} &=& \left( q-m\right) \,
\ket{\left(p,q\right),m+1}\nonumber\\ 
S\(T_{-}\) \ket{\left(p,q\right),m} &=& \left( p+m\right) \,
\ket{\left(p,q\right),m-1} 
\lab{actiononket}
\end{eqnarray}
On the subspace with fixed $(p+q)$, the Casimir operator acts as:
$$
\left[S\(T_3\)^2 +{1\over 2} \bigl( S\(T_+\) S\(T_-\) + S\(T_-\) 
S\(T_+\) \bigr) \right]
\ket{\left(p,q\right),m} =  
s(s+1) \ket{\left(p,q\right),m}
$$
with $s = {1\over 2}(p+q)$. 
The parameter $s$ is the spin of the representation.

From the relations \rf{actiononket} one notices that if $p$ is an
integer, then $\ket{\left(p,q\right),-p}$ is a lowest weight state, since it
is annihilated by $S\(T_{-}\)$. Analogously, if $q$ is an integer, then
$\ket{\left(p,q\right),q}$ is a highest weight state. If $p$ and $q$
are integers and $q>-p$, then the irrep. is finite dimensional. In
order to have integer spin representations we need $\frac{p-q}{2} \in
\IZ$. The spin zero state is $\ket{\left(p,q\right),-\frac{p-q}{2}} =
\left(\lambda\, \bar{\lambda}\right)^{\frac{p+q}{2}}$.  
Notice however, that not all irreps. for integer spin will have the zero spin
state. The reason is that if $p$ is a negative integer, or $q$ is a positive
integer, then  the
representation will truncate before reaching the zero spin state.

In the $\lambda-{\bar \lambda}$ representations  \rf{twoparrep} the
potential \rf{s2pota} becomes 
\be
A_{\mu} \equiv \frac{1}{1+\u2}\left( -i \pa_{\mu} u \lambda 
\frac{d\;}{d\bar{\lambda}} 
-i \pa_{\mu} u^* \bar{\lambda} \frac{d\;}{d\lambda} 
+\left(u\pa_{\mu} u^* - u^*\pa_{\mu} u\right) 
\frac{1}{2}\left( \lambda \, \frac{d\;}{d\lambda} - \bar{\lambda} \,
\frac{d\;}{d\bar{\lambda}}\right) \right) 
\lab{s2potalambda}
\ee
In a space-time of $d+1$ dimensions the Hodge dual of the rank $d$
antisymmetric tensor  $B^{(N=d)}_{\mu_1\ldots\mu_{d}}$ is a vector, as
given in \rf{dualbn=d}. We
  shall introduce such dual vector in a spin $s$ representation of the
  algebra \rf{sl2commrel} as 
\be
{\tilde B}_{\mu}^{(s)} \equiv \frac{1}{1+\u2}\left(  \ck_{\mu} 
\lambda^{s+1} \bar{\lambda}^{s-1} -
 \ck_{\mu}^* \lambda^{s-1} \bar{\lambda}^{s+1}\right)
\lab{s2potblambda}
\ee
where the vector ${\cal K}_{\mu}$ can be  a priori any functional of
$u$, $u^*$ and 
their derivatives. Notice that we have chosen
$B_{\mu}^{(s)}$ to live in 
a representation where  $p=q=s$, and it has components only in the
direction of the states with eigenvalues $\pm 1$ of $T_3$. 

As we have seen, the connection \rf{s2potalambda} (or \rf{s2pota}) is
flat for any configuration of the field $u$. In addition, the
condition \rf{localzccomm} is satisfied because the tensors $B^{W_d}$
live in an abelian ideal, namely that generated by functions of the
parameters $\lambda$ and ${\bar \lambda}$ (see
\rf{s2potblambda}). Therefore, the only local condition to be
satisfied is \rf{truezc1} (or \rf{localzcn}), i.e. 
\be
D^{\mu}{\tilde B}_{\mu}^{(s)}= \pa^{\mu} {\tilde B}_{\mu}^{(s)} +
\sbr{A^{\mu}}{{\tilde B}_{\mu}^{(s)}}=0 
\lab{s2beqmot}
\ee
Therefore the equations of motion for the field $u$ come from the
condition \rf{s2beqmot}, and they depend of course on the choice of the
spin $s$ of the representation. 

For $s=1$, one gets that \rf{s2beqmot} implies the equations
\be
\(1+\mid u\mid^2\)\,\pa^{\mu}{\cal K}_{\mu}-2\,u^*\,\pa^{\mu}u\,{\cal
  K}_{\mu}=0 \qquad \qquad 
\pa^{\mu}u\,{\cal K}_{\mu}^*-\pa^{\mu}u^*\,{\cal K}_{\mu} =0
\lab{s=1model}
\ee
together with the complex conjugate of the first one. 

For $s=-1$, \rf{s2beqmot} implies instead
\be
\pa^{\mu}{\cal K}_{\mu} = 0 \qquad \qquad 
\pa^{\mu}u\,{\cal K}_{\mu}=0
\lab{s=-1model}
\ee
together with their complex conjugates.

For $s\neq \pm 1$, one gets from \rf{s2beqmot} 
\be
\pa^{\mu}{\cal K}_{\mu} = 0 \qquad \qquad 
\pa^{\mu}u\,{\cal K}_{\mu}=0\qquad \qquad 
\pa^{\mu}u\,{\cal K}_{\mu}^*-\pa^{\mu}u^*\,{\cal K}_{\mu} =0
\lab{anysmodel}
\ee
together with the complex conjugates of the first two equations.

The conserved charges following from \rf{s2beqmot} are those given in
\rf{qndef} for the case $N=d$, and with $\Sigma_N$ corresponding to
the $d$ dimensional subspace of the space-time ${\cal M}$, orthogonal
to the time direction. These  charges are conserved in time and the
corresponding conserved currents are given by \rf{currentsn=d}, 
\emph{i.e.}
\be
J_{\mu}^{\(s\)}= W^{-1}_j\, {\tilde B}_{\mu}^{(s)}\, W_j
\lab{currentanys}
\ee
with $W_j$, $j=1,2$, being given in \rf{w1w2}. The fact that the two
group elements $W_1$ and $W_2$ give the same currents can be checked
by explicit calculations. Indeed, using \rf{twoparrep} one has that,
for an arbitrary function $f\(\lambda,{\bar \lambda}\)$,
\begin{eqnarray}
e^{\alpha S\(T_+\)} f\left(\lambda ,\bar{\lambda}\right) e^{-\alpha S\(T_+\)} &=&
f\left(\lambda ,\bar{\lambda} + \alpha \lambda\right)\nonumber\\
e^{\beta S\(T_-\)} f\left(\lambda ,\bar{\lambda}\right) e^{-\beta S\(T_-\)} &=&
f\left(\lambda + \beta \bar{\lambda} ,\bar{\lambda} \right)\nonumber\\
e^{\gamma S\(T_3\)} f\left(\lambda ,\bar{\lambda}\right) e^{-\gamma S\(T_3\)} &=&
f\left(e^{\gamma/2}\lambda 
,e^{-\gamma/2}\bar{\lambda} \right)
\end{eqnarray} 
Then you can check that
\begin{eqnarray}
W_j^{-1} f\left( \lambda , \bar{\lambda}\right) W_j = f\left( 
\frac{\lambda - i u^* \bar{\lambda}}{\sqrt{1+\u2}},
\frac{\bar{\lambda} - i u \lambda}{\sqrt{1+\u2}}\right) \qquad j=1,2
\label{conjw1w2}
\end{eqnarray}
So, the two group elements give the same rotation on the parameters. Notice
they look like a Lorentz boost on the space $\left( \lambda , 
\bar{\lambda}\right)$ and with complex
velocity $iu$. 

The models defined by the equations \rf{s=1model}, corresponding to
the case $s=1$, have only three conserved currents \rf{currentanys}
and they are given by the three components of 
\be
J_{\mu}^{\(1\)}= \frac{{\cal K}_{\mu}+u^2\,{\cal K}_{\mu}^*}{\(1+\mid
  u\mid^2\)^2} \; \lambda^2 -
2\,i\,\frac{u^*\,{\cal K}_{\mu}-u\,{\cal K}_{\mu}^*}{\(1+\mid
  u\mid^2\)^2} \; \lambda\,{\bar\lambda} -
\frac{{\cal K}_{\mu}^*+{u^*}^2\,{\cal K}_{\mu}}{\(1+\mid
  u\mid^2\)^2} \; {\bar \lambda}^2 
\lab{conscurrs=1}
\ee
Now, the models defined by the equations \rf{s=-1model} corresponding
to $s=-1$, have instead an infininte number of conserved
currents. Indeed, one can check that
\be
J_{\mu}^{\(-1\)}= \frac{{\cal K}_{\mu}}{\({\bar
    \lambda}-i\,u\,\lambda\)^2} - 
\frac{{\cal K}_{\mu}^*}{\(
    \lambda-i\,u^*\,{\bar \lambda}\)^2} 
\lab{conscurrs=-1}
\ee
So, expanding in powers of $\lambda$ and ${\bar \lambda}$ one gets an
infinite number of currents.

The models given by equations \rf{anysmodel} have a much larger set of
conserved currents since they admit a zero curvature representation for
any spin $s$. Looking at eq. \rf{s2potblambda}, we see that
$$
{\tilde B}_\mu^{\(s\)} = (\lambda \bar{\lambda} )^{(s+1)} {\tilde B}_\mu^{\(-1\)}
$$
If we consider a general ${\tilde B}_\mu = \sum_s \beta_s {\tilde
  B}_\mu^{\(s\)}$, we have
$$
{\tilde B}_\mu = b(\lambda \bar{\lambda}){\tilde B}_\mu^{\(-1\)}
$$
where $b(z)= \sum_s \beta_s z^{s+1}$ is essentially an arbitrary function.

At the level of currents, this means
$$
J_\mu = b\left({(\lambda - iu^* \bar{\lambda} ) (\bar{\lambda} - iu \lambda ) 
\over 1 + uu^* } \right) J_\mu^{\(-1\)}
$$
We can write this in the nice form \cite{afz1,afz2,fujii2}:
\be
J_\mu =  {\cal K}_\mu {\delta G \over \delta u} - 
{\cal K}^*_\mu {\delta G \over \delta u*} 
\lab{niceformcurr}
\ee
where
$$
G = i \int^{v(u,u^*)} {dv \over v^2} \, b(v),\qquad\qquad
v(u,u^*) = {(\lambda - iu^* \bar{\lambda} ) (\bar{\lambda} - iu \lambda ) 
\over 1 + uu^* }
$$
Essentially $G$ is an arbitrary functional of $u$ and $u^*$, but not
of its derivatives, and we have a conserved current for every $G$. 

A large number of theories have been studied in the literature using
the local zero curvature formulation we have just presented.  We list
here some examples.

\subsubsection{The $CP^1$ model}

The equations of motion are given by
\be
\(1+\mid u\mid^2\)\partial^2 u - 2 \, u^*\,
\partial^{\mu}u\,\partial_{\mu}u =0
\lab{cp1eq}
\ee
together with its complex conjugate. In such case we have ${\cal
  K}_{\mu}\equiv \partial_{\mu} u$, and so the first equation in
\rf{s=1model} corresponds to \rf{cp1eq}, and the second is trivially
satisfied.

There exists a very interesting submodel of the $CP^1$ theory defined by
the equations \cite{ward,Alvarez:1997ma}
\be
\partial^2 u=0 \qquad\qquad\qquad \partial^{\mu}u\,\partial_{\mu}u =0
\lab{cp1submodel}
\ee
Such equations correspond to \rf{anysmodel}, again with ${\cal
  K}_{\mu}\equiv \partial_{\mu} u$. Therefore it has an infinite set
of conserved currents.

\subsubsection{The Skyrme-Faddeev model and its extension}

The extended  Skyrme-Faddeev model is a theory defined on $3+1$
dimensions and given by the Lagrangian \cite{gies,s35}
\be
{\cal L} = M^2\, \partial_{\mu} {\vec n}\cdot\partial^{\mu} {\vec n}
 -\frac{1}{e^2} \, \(\partial_{\mu}{\vec n} \wedge 
\partial_{\nu}{\vec n}\)^2 + \frac{\beta}{2}\,
\left(\partial_{\mu} {\vec n}\cdot\partial^{\mu} {\vec n}\right)^2
\lab{sfextendedaction}
\ee 
where ${\vec n}$ is a triplet of real scalar fields taking values on
the sphere $S^2$, $M$ is a coupling constant with dimension of mass,
 $e^2$ and $\beta$ are dimensionless coupling constants. The usual
Skyrme-Faddeev model \cite{sf} corresponds to the case $\beta=0$.

If one makes the stereographic projection of the sphere $S^2$ on the
plane and works with the complex $u$ field as defined in \rf{stereo}
one gets 
\br
{\vec n}\cdot\(\partial_{\mu}{\vec n} \wedge 
\partial_{\nu}{\vec n}\)
&=&-2i\frac{\(\partial_{\mu} u\partial_{\nu} u^* - 
 \partial_{\nu} u \partial_{\mu} u^*\)}{\(1+\u2\)^2} \equiv H_{\mu\nu}
\lab{hmunudef}\\ 
\left(\partial_{\mu} {\vec
  n}\cdot \partial^{\mu} {\vec
  n}\right)&=& 4\,\frac{\partial_{\mu}u\;\partial^{\mu}u^*}{\(1+\u2\)^2}
\lab{dn2}
\er
The Euler-Lagrange equations following from \rf{sfextendedaction}  read 
\be
\(1+\u2\)\, \partial^{\mu}{\cal K}_{\mu}-2\,u^{*}\,{\cal K}_{\mu}\,
\partial^{\mu} u=0\,,
\lab{sfextendedeqmot}
\ee
together with its complex conjugate, and where
\be
{\cal K}_{\mu}\equiv M^2\, \partial_{\mu}u-\frac{4}{e^2}\,\frac{ 
\left[\(1-\beta\,e^2\)\,\(\partial_{\nu}u\,\partial^{\nu}u^{*}\)\,
\partial_{\mu} u-
\(\partial_{\nu}u\partial^{\nu} u\) \partial_{\mu}u^{*}\right]}{\(1+\u2\)^2}
\lab{kdefsf}
\ee
So \rf{sfextendedeqmot} corresponds to the first equation in \rf{s=1model},
and \rf{kdefsf} trivially satisfies the second equation in
\rf{s=1model}. Therefore, such theory has the three conserved currents
given in \rf{conscurrs=1}, and they correspond in fact to the Noether
currents associated to the global $SO(3)$ symmetry of
\rf{sfextendedaction}. 

However if one imposes the constraint 
\be
\partial^{\mu}u\,\partial_{\mu}u =0
\ee
one observes that \rf{kdefsf} satisfies the
eqs. \rf{anysmodel}. Therefore, such a submodel has an infinite number of
conserved currents given by \rf{niceformcurr}.

\subsubsection{Models with exact Hopfion solutions}

An interesting class of models is given by the actions  
\be
S = \int d^{n}x\; \(H_{\mu\nu}^2\)^{d/4}
\lab{hopfionaction}
\ee
where $H_{\mu\nu}$ is the pull back of the  area form on the sphere
$S^2$, given in \rf{hmunudef}. 

In a Minkowski space-time of $n=d+1$ dimensions, the power $d/4$ is
chosen to comply with the requirements of Derrick's 
theorem. In fact, it implies that the static solitons have an energy which is
invariant under rescaling of the space variables. However, one can
have time dependent solutions for $n=d$, or  solutions in an
Euclidean space of $n=d$ dimensions. 

The Euler-Lagrange equations following from \rf{hopfionaction} are
given by
\be
\partial^{\mu}{\cal K}_{\mu}=0
\lab{hopfioneq}
\ee
with
\be
{\cal K}_{\mu}=
\(H_{\rho\sigma}^2\)^{\frac{\(d-4\)}{4}}H_{\mu\nu}\partial^{\nu}u
\lab{hopfionkmu}
\ee
Notice that \rf{hopfionkmu} trivially satisfy the last two equation in
\rf{anysmodel}, and \rf{hopfioneq} corresponds to the first
one. Therefore, these models have an infinite set of conserved currents
given by \rf{niceformcurr}.

\subsection{The multidimensional Toda systems}
\label{sec:multitoda}

The multidimensional Toda systems were introduced by Saveliev and
Razumov \cite{multitoda} as a generalization of the two dimensional
Toda models to a space-time of even dimension and with a metric that
has an equal number of eigenvalues $+1$ and $-1$. The scalar
product is invariant under the group $SO(p,p)$. We shall use light
cone coordinates $z_{\pm i}=t_i\pm x_i$, $i=1,2\ldots p$, with $t_i$
and $x_i$ being the time and space coordinates associated to the
eigenvalues $-1$ and $+1$ respectively. The model is introduced
through a Kac-Moody algebra ${\hat {\cal G}}$ furnished with an
integer gradation  
\be
{\hat {\cal G}} = \otimes_{n=-\infty}^{\infty} {\hat {\cal G}}_n
\qquad \qquad 
\sbr{{\hat {\cal G}}_m}{{\hat {\cal G}}_n}\subset {\hat {\cal G}}_{m+n}
\ee
The fields of the model are  the elements $\gamma$ of the grade zero
subgroup, i.e. that group obtained by exponentiating the subalgebra
${\hat {\cal G}}_0$.  In addition, there are $p$ elements $E_{+i}$ and
$E_{-i}$, $i=1,\ldots p$, with grades $+1$ and $-1$ respectively and satisfying
\be
\sbr{E_{+i}}{E_{+j}}=0 \qquad \qquad \sbr{E_{-i}}{E_{-j}}=0 \qquad
\qquad i,j=1,\ldots p
\lab{econd}
\ee
The equations of motion constitute an overdetermined system and are
given by the 
following three sets of equations
\be
\partial_{+i}\(\partial_{-j}\gamma\,\gamma^{-1}\)+
\sbr{\gamma\,E_{+i}\,\gamma^{-1}}{E_{-j}}=0
\lab{multitodaeq1}
\ee
and
\br
\partial_{+i}\(\gamma\,E_{+j}\,\gamma^{-1}\)&=&
\partial_{+j}\(\gamma\,E_{+i}\,\gamma^{-1}\)\nonumber\\
\partial_{-i}\(\gamma^{-1}\,E_{-j}\,\gamma\)&=&
\partial_{-j}\(\gamma^{-1}\,E_{-i}\,\gamma\)
\lab{multitodaeq2}
\er
For the two dimensional case, i.e. $p=1$, the equations
\rf{multitodaeq2}, as well as the conditions \rf{econd}, become
trivial, and \rf{multitodaeq1} becomes the usual two dimensional Toda
equations.  

The model admits a zero curvature representation 
\be 
F_{\mu\nu}=\partial_{\mu}A_{\nu}-\partial_{\nu}A_{\mu}+\sbr{A_{\mu}}{A_{\nu}}=0
\qquad\qquad \mu ,\nu=\pm 1,\pm 2\ldots \pm p
\lab{zcmultitoda}
\ee
with the connection $A_{\mu}$ given by 
\be
A_{+i}= \gamma \,E_{+i}\,\gamma^{-1} \qquad \qquad 
A_{-i}=-\partial_{-i}\gamma\,\gamma^{-1}-E_{-i} \qquad \qquad
i=1,\ldots p
\lab{potmultitoda}
\ee
One can easily check that \rf{zcmultitoda} with the connection
\rf{potmultitoda}, are equivalent to the equations \rf{multitodaeq1}
and \rf{multitodaeq2}.  

The solutions of such model were constructed in \cite{multitoda} using
a generalization of the so-called Leznov-Saveliev method
\cite{leznovsaveliev1,leznovsaveliev2}, and we will not discuss here
the details of that construction. 

We can use our formulation to try to construct conserved charges for
such model. The pairs $\(A^{(N)},B^{(N)}\)$ can be constructed by
taking any of the connection $A^{(N)}$ as that given in
\rf{potmultitoda}, and the tensors $B^{(N)}$ as
\be
B^{(N)}= D^{(N)}\wedge \alpha^{(N-1)}
\lab{bnmltitoda}
\ee
with $\alpha^{(N-1)}$ being an arbitrary antisymmetric tensor of rank
$N-1$, and living on the abelian ideal ${\cal P}$ of the non semisimple
algebra ${\cal G}={\hat {\cal G}}+{\cal P}$, with ${\hat {\cal G}}$
being the Kac-Moody algebra introduced above. One way of constructing
${\cal P}$ is to take it to transform under the adjoint representation
of ${\hat {\cal G}}$, i.e. if $T$ is a generator of
${\hat {\cal G}}$, we introduce $P\(T\)\in {\cal P}$ such that 
\be
\sbr{T}{P\(T^{\prime}\)} = P\(\sbr{T}{T^{\prime}}\) \qquad \qquad 
\sbr{P\(T\)}{P\(T^{\prime}\)} = 0
\ee
Clearly the  pairs $\(A^{(N)},B^{(N)}\)$ satisfy the sufficient local
curvature conditions \rf{flatan}, \rf{localzcn} and \rf{localzccomm}
and we can use them to try to construct the charges \rf{qndef}. Notice
that the equations of motion are equivalent to the condition
\rf{flatan}. However, in order to show that  \rf{bnmltitoda} satisfy
\rf{localzcn} we need the equations of motion, since we use the fact
that the covariant derivatives commute. 

It follows that the tensors \rf{bwdef} are given by
\be
B^{W_N}= d\wedge \(W^{-1}\,\alpha^{(N-1)}\, W\)
\lab{bwnsimple}
\ee
with $W$ being the group element such that the connection
\rf{potmultitoda} can be written as
$A_{\mu}=-\partial_{\mu}W\,W^{-1}$. 

One can construct conserved quantities following the reasoning
explained in \rf{dwedgebn}-\rf{qndef}. By decomposing the border of a
$N+1$ dimensional volume $V_{N+1}$ as in \rf{surfacedecomp}, we need to
impose boundary conditions such that 
\be
\int_{\Gamma_N} \, B^{W_N} = \int_{\partial \Gamma_N} \, 
W^{-1}\,\alpha^{(N-1)}\, W =0
\lab{boundarygamma}
\ee
where we have used \rf{bwnsimple} and the abelian Stokes theorem. 
The conserved quantities are given by \rf{qndef}, i.e.  
\be
Q^{(N)} = \int_{\Sigma_N} \, B^{W_N} = \int_{\partial \Sigma_N} \, 
W^{-1}\,\alpha^{(N-1)}\, W
\lab{conschargemultitoda}
\ee
where again we have used \rf{bwnsimple} and the abelian Stokes
theorem. The surface $\Sigma_N$ corresponds to a fixed value of the
parameter $\tau$, and the charge is conserved in $\tau$ (see
discussion in \rf{dwedgebn}-\rf{qndef}). 

Consider solutions of the multidimensional Toda model such
that the group valued fields $\gamma$ satisfy the boundary condition 
\be 
\gamma \rightarrow 1 \qquad \qquad \mbox{\rm as any $x_i\rightarrow \pm
  \infty$}
\ee
Then the connections satisfy 
\be
A_{t_i}\equiv A_{+i} + A_{-i} \rightarrow  E_{+i}-E_{-i}
  \qquad \qquad \mbox{\rm as any $x_i\rightarrow \pm
  \infty$}
\ee
As an example, and for simplicity, let us take $N=1$ and choose the
surfaces $\Sigma_1^{\tau_0}$ and $\Sigma_1^{\tau_1}$ in
\rf{surfacedecomp}, to be two infinite straight lines parallel to one
of the axis $x_i$,  at different values of the times
$t_i$.  The surface 
$\Gamma_1$ will be disjoint and made of two  lines parallel to a given
axis $t_i$ and joining the
ends of $\Sigma_1^{\tau_0}$ and $\Sigma_1^{\tau_1}$, and so they lie
at spatial infinity. Since $A_{\mu}$ is flat it follows that $W$ is path
independent. We choose  $W=1$ at the point
$\(x_i,t_i\)=\(-\infty ,0\)$ for $i=1,\ldots p$. Then one can always
choose the path in such way that $W$ at any point of $\Gamma_1$ is of
the form $W=\exp\(\int dt_i\(E_{+i}-E_{-i}\) \)\, W_0$, where $W_0$ is
the value of $W$ at the initial point of $\Gamma_1$. 
 If we now take $\alpha^{(0)}$ to be independent of the
times $t_i$, and to commute with   $E_{+i}-E_{-i}$, \emph{i.e.}
\be
\left[ E_{+i}-E_{-i}\, , \, \alpha^{(0)}\right] =0
\ee
then the boundary conditions \rf{boundarygamma} are
satisfied. Under such conditions the 
charges \rf{conschargemultitoda}, for $N=1$, are conserved at any time
$t_i$.   With appropriate modifications, similar procedures
can be applied to higher dimensional surfaces $\Sigma_N$. 

\section{Construction of solutions}
\label{sec:solutions}

In $1+1$ dimensional integrable fields theories most of the methods for
constructing solutions are based on the Lax-Zakharov-Shabat  equation
\cite{lax} or zero curvature condition given in
\rf{laxeq}.  One of the  most powerful
techniques to construct solutions is the so-called dressing method
\cite{dressing}. Its main ingredient is to use the gauge symmetry of
\rf{laxeq} 
to map a given solution into another. However, it has other important
ingredients. The connection $A_{\mu}$ for an integrable theory in
$1+1$ dimensions is graded according to a gradation \rf{gradation} of the
corresponding Kac-Moody algebra ${\hat {\cal G}}$. That plays a
crucial role in the development of the method. 

The local zero curvature conditions \rf{flatan}, \rf{localzcn} and
\rf{localzccomm} have a quite large gauge symmetry given by the
transformations \rf{gaugetransf1} and \rf{gaugetransf2}.  Of course
those transformations map solutions into solutions of the zero
curvature conditions, and it could be used to design a method for
constructing solutions.  Some attempts in that direction were done but
the results did not lead to a concrete and effective method for
obtaining solutions.  It was possible to show however
\cite{luizbabelon} that for the theories, discussed in section
\ref{sec:s2}, with target space $S^2$ and possessing an infinite
number of conserved currents we can use the local zero curvature 
condition in
the $\lambda$-${\bar \lambda}$ representation (see
\rf{s2potalambda}-\rf{s2potblambda}) to perform one integration of
the equations of motion.  The second integration has to be done by
direct methods.  In particular, Ward's solution \cite{ward} of the
$CP^1$ submodel \rf{cp1submodel} was re-obtained in this way.  For
theories which are not Lorentz invariant the situation is different.
Indeed, for the multidimensional Toda systems \cite{multitoda},
discussed in section \ref{sec:multitoda}, it is possible to construct
a quite wide class of solutions using the zero curvature construction
as shown in \cite{multitoda}.  The development of a concrete method
for constructing solutions is currently not possible because we
believe that there are some important undiscovered structures missing
in the current formulation.  Perhaps these can be discovered by
investigating more deeply the role of zero curvature in loop space.

The solutions constructed so far in the literature for the models
admitting the local  zero curvature conditions \rf{flatan}, \rf{localzcn} and
\rf{localzccomm}, were obtained through direct methods. We discuss
below some of those cases.

\subsection{Exact Hopfion solutions}

The theories introduced in \rf{hopfionaction} have a large symmetry
group. The two form $H_{\mu\nu}$ is the pull back of the area form on
the sphere $S^2$, and so the action and equations of motion are
invariant under the area preserving diffeomorphisms of the sphere
which is an infinite dimensional group \cite{diffeo1}. The same is true for other
two dimensional target spaces like the Euclidean 
plane, the Poincar\'e hyperbolic disc, etc. In those cases one has to
take $H_{\mu\nu}$ as the pull back of the corresponding area forms
\cite{babelon,s31}. On the other hand, the 
theories \rf{hopfionaction} are conformally invariant when the
dimension of space-time equals the integer $d$ appearing in the exponent of
$H_{\mu\nu}$. Here one clarification is necessary: the dimension $n$
of the space-time has in fact to be the dimension of the subspace
where the solution in being constructed. For instance, if we
consider the theory \rf{hopfionaction} in a space-time of $3+1$
dimensions but want to construct static solutions then the conformal 
invariance has to be present in the three dimensional spatial submanifold
and we have to take $d=3$. That is exactly the case considered in
 \cite{afz2}, where the theory was not conformally invariant in
$3+1$ dimensions but only in the three dimensional Euclidean spatial 
submanifold.  

Notice that the theory \rf{hopfionaction} in two dimensions and for
$d=2$, is trivial \cite{babelon}. Indeed,  in such case $H_{\mu\nu}$ has only one
component and the equation \rf{hopfioneq} is satisfied for any
configuration of the field $u$. In addition, that theory is invariant
under the infinite dimensional conformal group in two dimensions. 

The idea for constructing the solutions is to use Lie's  good  old
method for exploring the symmetries of the equations of motion
and set up an ansatz based on that symmetry. We are interested in
finite energy solutions (or finite action in the Euclidean case) with
a non trivial topology. In the case of the models discussed in section
\ref{sec:s2} one of the relevant topological charges is the Hopf
invariant classifying the homotopy classes of maps $S^3\rightarrow
S^2$. It turns out that the conformal symmetry has an important
relation with the Hopf charge. The solutions with non-trivial Hopf
charge are invariant under the composition of some specific conformal
and target space transformations \cite{babelon}. The relevant (target
space) area preserving diffeomorphism is the global phase transformation 
\be
u\rightarrow e^{i\,\alpha}  \, u
\lab{phasetransf}
\ee
with the complex scalar field $u$ defined in \rf{stereo}. One now
builds an ansatz for configurations that are invariant under the
combined action of the phase
transformation \rf{phasetransf} and some commuting conformal
transformations.  

\subsubsection{3d solutions}

To make things concrete consider the case of a three dimensional
Euclidean space with Cartesian coordinates $x_i$, $i=1,2,3$. The
conformal group on that space is $SO(4,1)$, which has rank two. The
two commuting conformal transformations we choose correspond to
rotations on the $x_1\,x_2$ plane, and a combination of the translation
and the special conformal transformation associated to $x_3$
direction.  They are generated by the vector fields \cite{babelon} 
\br
\partial_{\varphi} &\equiv& x_2\,\partial_{x_1} - x_1\,\partial_{x_2}  
\lab{transfd=3}\\
\partial_{\xi} &\equiv& 
\frac{1}{2a}\,\left[2\, x_3\,\sum_{i=1}^3 x_i\,\partial_{x_i}
  -\sum_{i=1}^3 x_i^2 \,\partial_{x_3} + a^2\,\partial_{x_3}\right]
\nonumber
\er 
where $a$ is an arbitrary parameter with the dimension of length. 
Through the above formulas we have introduced the coordinates
$\varphi$ and $\xi$ parametrizing the curves generated by the one
parameter subgroups defined by the above commuting conformal
transformations. The third coordinate $z$ is chosen to be orthogonal to
$\varphi$ and $\xi$, to satisfy $\partial_{\varphi}z=\partial_{\xi}z=0$, and
is  given by
\be
z\equiv 4\,a^2\,\frac{x_1^2+x_2^2}{\(a^2+x_1^2+x_2^2+x_3^2\)^2}
\ee
Inverting the coordinates we have\footnote{Those correspond to the
  toroidal coordinates used in \cite{afz2} with $z\rightarrow \tanh^2
  \eta$, and $\eta>0$}
\br
x_1&=&\frac{a}{p}\,\sqrt{z}\,\cos\varphi\nonumber\\
x_2&=&\frac{a}{p}\,\sqrt{z}\,\sin\varphi\nonumber\\
x_3&=&\frac{a}{p}\,\sqrt{1-z}\,\sin\xi
\lab{3dcoord}
\er
with $p=1-\sqrt{1-z}\,\cos\xi$, and $0\leq \varphi\, ,\,\xi\leq2\,\pi$,
$0\leq z\leq 1$. The metric is
\be
d\, s^2= \(\frac{a}{p}\)^2\left[\frac{d\,z^2}{4\,
    z\(1-z\)}+\(1-z\)\,d\,\xi^2+z\, d\,\varphi^2\right]
\ee
The ansatz invariant under the combined action of \rf{phasetransf} and
\rf{transfd=3} is given by \cite{afz2}
\be
u=\sqrt{\frac{1-g}{g}}\,e^{i\,\(m\,\xi+n\,\varphi\)}
\lab{3dansatz}
\ee
with $m$ and $n$ being integers, $g\equiv g\(z\)$, and $0\leq g\leq
1$. We have chosen to parametrize the modulus of $u$ with the profile
function $g$, taking values between $0$ and $1$, because it constitutes
some sort of Darboux variable for the two form
$H_{\mu\nu}$ defined in \rf{hmunudef} . Indeed 
\be
H_{\mu\nu}= \partial_{\mu}g\,\partial_{\nu}\theta-
\partial_{\nu}g\,\partial_{\mu}\theta \qquad\qquad {\rm with} \qquad\qquad
u=\sqrt{\frac{1-g}{g}}\,e^{i\,\frac{\theta}{2}}
\ee
Replacing the ansatz \rf{3dansatz} into \rf{hopfioneq} we get that the
equation of motion for the profile function $g$ is
\be
\partial_z\left[\Lambda^{3/4}\,\(\partial_z g\)^{1/2}\right]=0 \qquad
\qquad \qquad  
\Lambda = m^2\, z+n^2\,\(1-z\)
\lab{eqgd=3}
\ee
Therefore, we need to have $g$ as a monotonic function on the interval
$0\leq z\leq 1$ in order to obtain real solutions. Notice that the
equation \rf{eqgd=3} is invariant under the transformations
\be
g \leftrightarrow 1-g
\ee
and
\be
z \leftrightarrow 1-z \qquad \qquad {\rm and}\qquad\qquad
m^2\leftrightarrow n^2
\ee
We look for solutions satisfying the boundary conditions
\be
g\(0\)=0\qquad\qquad\qquad\qquad g\(1\)=1 
\ee
which implies that (see \rf{stereo}, \rf{3dcoord} and \rf{3dansatz})
${\vec n}\rightarrow \(0,0,1\)$ on $x_3$ axis and at spatial infinity,
and ${\vec n}\rightarrow \(0,0,-1\)$ on a circle of radius $a$ on the
$x_1\,x_2$ plane and center at the origin. Then the solutions are given by
\br
g\(z\)&=& z \qquad\qquad \qquad \qquad\qquad \qquad \qquad \quad \;
 {\rm for} \qquad \mid n\mid = \mid m\mid \\
g\(z\)&=& \frac{\mid m\mid\,\mid n\mid\,
\Lambda^{-1/2} - \mid m\mid}{\mid n\mid - \mid m\mid}
\qquad\qquad \qquad {\rm for} \qquad  \mid n\mid \neq \mid m\mid
\nonumber
\er
Those are the solutions discussed in \cite{afz2} where it was shown 
that
the energy $E$ and the Hopf topological charge $Q_H$ of such solutions are  
\be
E \sim \sqrt{\mid m\mid\,\mid n\mid\(\mid m\mid+\mid n\mid\)} 
\qquad \qquad \qquad Q_H=-n\, m.
\ee

\subsubsection{4d solutions}

We now consider the theory \rf{hopfionaction} for $d=4$ in a four
dimensional Minkowski space-time with Cartesian coordinates $x_{\mu}$,
$\mu =0,1,2,3$. The conformal group in this case is $SO(4,2)$ which
has rank three. In order to construct the ansatz we choose the
following three commuting conformal 
transformations defined by the vector fields
\br
\partial_{\varphi} &\equiv& x_2\,\partial_{x_1} - x_1\,\partial_{x_2}  
\lab{transfd=4}\\
\partial_{\xi} &\equiv& 
\frac{1}{2a}\,\left[2\, x_3\,\sum_{\mu=0}^3 x_{\mu}\,\partial_{x_{\mu}}
  -\sum_{\mu=0}^3 x_{\mu}^2 \,\partial_{x_3} + a^2\,\partial_{x_3}\right]
\nonumber\\
\partial_{\zeta} &\equiv& 
\frac{1}{2a}\,\left[2\, x_0\,\sum_{\mu=0}^3 x_{\mu}\,\partial_{x_{\mu}}
  -\sum_{\mu=0}^3 x_{\mu}^2 \,\partial_{x_0} - a^2\,\partial_{x_0}\right]
\nonumber
\er 
Again $a$ is an arbitrary parameter with dimension of length, and
we have introduced the coordinates $\xi$, $\varphi$ and $\zeta$
parametrizing the curves generated by the three commuting conformal
transformations. The fourth coordinate $z$ is chosen to be orthogonal
to them, to satisfy
$\partial_{\varphi}z=\partial_{\xi}z=\partial_{\zeta}z=0$, and it
is given by
\be
z=4\,a^2\,\frac{x_1^2+x_2^2}{\(a^2+R^2\)^2-4\,x_0^2\,r^2}
\ee
with $r^2=x_1^2+x_2^2+x_3^2$, and $R^2=x_0^2+r^2$. 
In terms of these new coordinates, the Cartesian coordinates are given
by\footnote{Those are the coordinates used in 
  \cite{s32} with the change $z\rightarrow 1/\(1+y\)$, $y>0$.}
\br
x_0&=&\frac{a}{q}\,\sin\zeta\nonumber\\
x_1&=&\frac{a}{q}\,\sqrt{z}\,\cos\varphi\nonumber\\
x_2&=&\frac{a}{q}\,\sqrt{z}\,\sin\varphi\nonumber\\
x_3&=&\frac{a}{q}\,\sqrt{1-z}\,\sin\xi
\lab{4dcoord}
\er
with $q=\cos \zeta-\sqrt{1-z}\,\cos\xi$, and $0\leq \varphi\, ,\,\xi\leq2\,\pi$,
$0\leq z\leq 1$, and $0\leq \zeta \leq \pi$. The metric is
\be
d\, s^2= \(\frac{a}{q}\)^2\left[d\zeta^2 -\frac{d\,z^2}{4\,
    z\(1-z\)}-\(1-z\)\,d\,\xi^2-z\, d\,\varphi^2\right]
\ee
The ansatz leading to configurations invariant under the combined
action of the transformations \rf{phasetransf} and \rf{transfd=4} is
given by 
\be
u=\sqrt{\frac{1-g}{g}}\,e^{i\,\(m_1\,\xi+m_2\,\varphi+m_3\,\zeta\)}
\lab{4dansatz}
\ee
with $g = g\(z\)$, and $0\leq g \leq 1$. In order for $u$ to be single
valued we need $m_1$ and $m_2$ to be integers. In addition, $\( \zeta=0
, z , \xi , \varphi\)$ and $\( \zeta=\pi, z , \xi+\pi , \varphi+\pi\)$
correspond to the same point $\(x_0=0, x_1,x_2,x_3\)$. Therefore, we
also need $m_1+m_2+m_3= 2 N$, with $N$ being an integer, in order for $u$ to be
single valued. 

Replacing \rf{4dansatz} into \rf{hopfioneq} one gets that the profile
function $g$ has to satisfy the linear ordinary differential equation
\be
\partial_z\(\Omega\, \partial_z g\)=0\qquad\qquad \qquad 
\Omega= m_1^2\, z+m_2^2\,\(1-z\)-m_3^2\, z\,\(1-z\)
\lab{eqgd=4}
\ee
Similarly to \rf{eqgd=3} one notices that \rf{eqgd=4} has the symmetries
$g\leftrightarrow 1-g$, as well as $z\leftrightarrow 1-z$ and
$m_1^2\leftrightarrow m_2^2$. Again we solve \rf{eqgd=4} with the same
boundary conditions we solved \rf{eqgd=3}, i.e. $g\(0\)=0$ and
$g\(1\)=1$. Notices that \rf{eqgd=4} implies $\partial_z g \sim
\Omega^{-1}$, and so we have to avoid the zeros of $\Omega$ on the
interval $0\leq z\leq 1$. One observes that for $m_1=0$, $\Omega$ has
a zero on $z=1$, and for $m_2=0$, $\Omega$ has a zero on
$z=0$. Therefore, we shall work with both, $m_1$ and $m_2$, different
from zero. However, we can have vanishing $m_3$ because in that case
$\Omega$ is positive on the interval  $0\leq z\leq 1$.

The solutions were constructed in \cite{s32} and are given by
\br
g&=& \frac{b\, z}{1-z+b\,z} \; ; \qquad {\rm for} \;\; \Delta=0 \; ; \; \; \; 
b>0
\lab{normsolg}\\
g&=& \frac{{\rm ArcTan} \(z\,\sqrt{-\Delta}/\(1-z+b\,z\)\)}{{\rm ArcTan}
  \(\sqrt{-\Delta}/b\)} \; ; \quad {\rm for} \;\; \Delta<0 \; ; \; \; \; 
{\rm any} \;\; b\nonumber\\
g&=&
\frac{\ln\left[\(1-z+\(b+\sqrt{\Delta}\)\,z\)/\(1-z+\(b-\sqrt{\Delta}\)\,
    z\)\right]}{
\ln\left[\(b+\sqrt{\Delta}\)/\(b-\sqrt{\Delta}\)\right]}
\; ; \;\; {\rm for} \;\; \Delta,b>0 \nonumber
\er
where
\br
b&=&\left[\(m_1+m_3\)\(m_1-m_3\)+m_2^2\right]/2m_2^2 
\lab{bdeltadef}\\
\Delta&=&
\left[\(m_1+m_3\)^2-m_2^2\right]\left[\(m_1-m_3\)^2-m_2^2\right]/4m_2^4
\nonumber 
\er
We can not have $b<0$ and $\Delta \geq 0$, which happen whenever
  $\(m_1+m_3\)/m_2 \geq 1$ and $\(m_1-m_3\)/m_2 
\leq -1$ or $\(m_1-m_3\)/m_2 \geq 1$ and $\(m_1+m_3\)/m_2
\leq -1$. 

Those solutions are time dependent and due to Derrick's scaling
argument they can not be put at rest. Due to the arbitrariness of the
parameter $a$, we can scale the size of the solution as well as the 
rate in which it evolves in time. In fact it is better to work with a
dimensionless variable given by $\tau = x_0/a$.

\begin{figure}[tbp]
    \centering
    \includegraphics[width=0.6\textwidth]{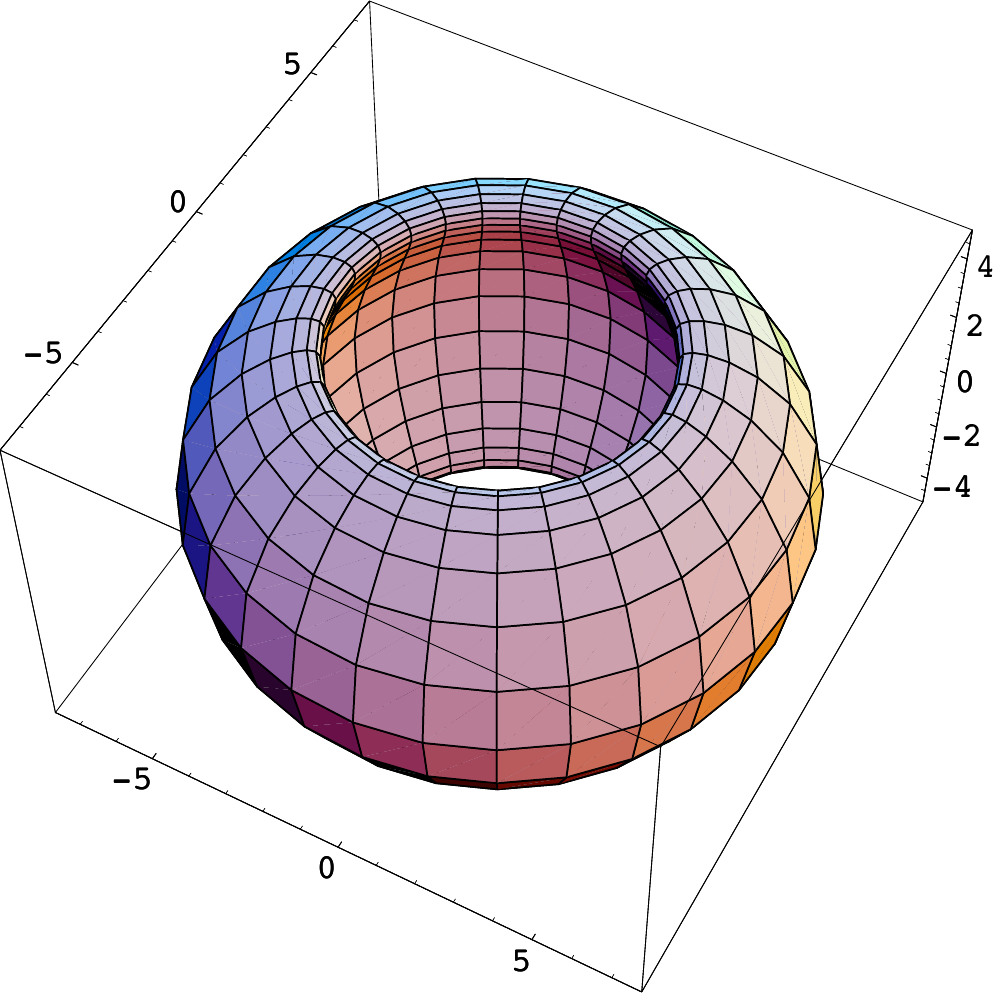}
    \begin{quotation}
    \caption[ed]{\small Surface of constant $n_3$ ($z=0.42$),
     for $\xi,\varphi=0\rightarrow 2\pi$,  
     and for the time $x_0/a = 6$. The $x_3$ axis passes through the
     center of the torus.}
     \label{fig:3dchi1t6}
     \end{quotation}
\end{figure}

A good way of visualizing the solutions is to plot the surfaces of
constant $n_3$ (the third component of the field vector introduced in
\rf{stereo}). However, $n_3$ depends only on $\mid u\mid$ and that in
its turn depends only $g$ (see \rf{4dansatz}). Now $g$ is a monotonic
function of $z$ and  therefore by fixing $z$ we fix $n_3$. The surfaces
of constant $n_3$ are of toroidal form. In Figure~\ref{fig:3dchi1t6}
we show the surface corresponding to $z=0.42$ and $\tau
=6$. Notice that such a plot applies to any of the solutions.  What
changes from solution to solution is the correspondence between the
values $z$ and $n_3$. 

\begin{figure}
    \centering
    \includegraphics[width=0.75\textwidth]{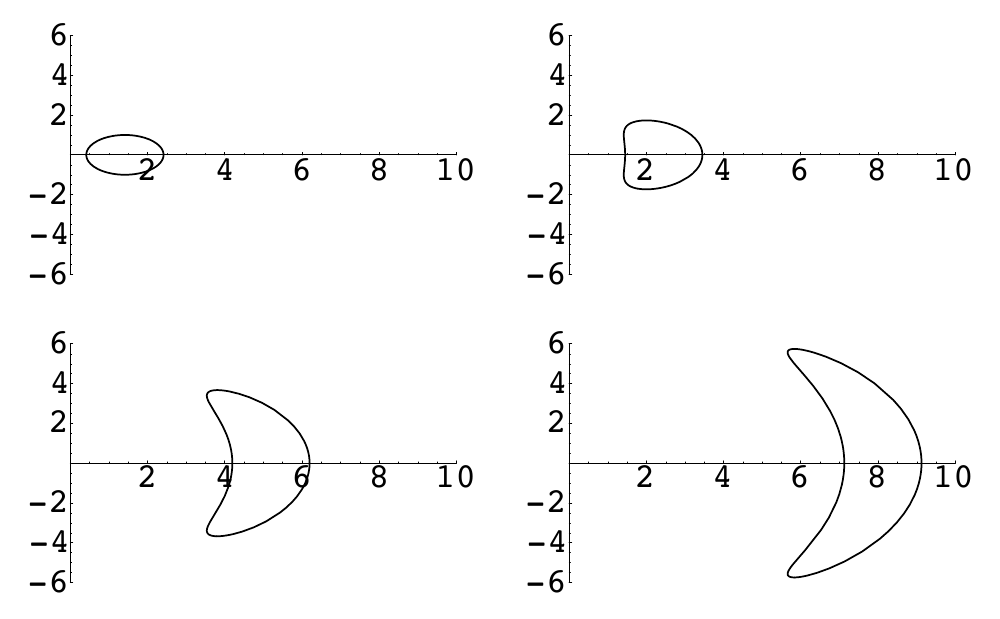}
    \begin{quotation}
    \caption[xy]{\small \label{fig:crosssec} Cross sections of the surfaces of
	constant $n_3$ for 
	$n_3=1-2g\(z=1/2\)$, and at the times $\tau =x_0/a=0,2,5,8$. The
	vertical and horizontal axis correspond to $x_3/a$ and
	$\sqrt{x_1^2+x_2^2\;}/a$ respectively. The surfaces are
	invariant under $\tau\rightarrow -\tau$}
    \end{quotation}
\end{figure}

Some general
properties of the solutions are: {\em i)} The surface for $n_3=-1$,
which implies $g=1$ and  $z=1$, is a circle on the plane $x_3=0$
with center at the origin and radius $a\sqrt{1+\tau^2}$; {\em ii)} The
surface for $n_3=1$, which implies $g=0$ and  $z=0$,
corresponds to the $x_3$-axis plus  spatial infinity for any time
$\tau$;
{\em iii)} For $\tau =0$ the surfaces of constant $n_3$ with $-1 < n_3
< 1$ are torii centered around the origin with a thickness that grows
as $n_3$ varies from $-1$ to $1$. As $\tau$ flows towards the future
or the past, those torii get thicker and their cross section deforms 
from a circle to the quarter moon shape  shown in Figure~\ref{fig:crosssec};
{\em iv)} The 
solution performs one single oscillation as $\tau$ varies from
$-\infty$ to $\infty$.  The surfaces of constant $n_3$ are
symmetrical under  the interchange $\tau
\leftrightarrow -\tau$.

For every value of time, the solutions define a map from the three
dimensional space to the target space $S^2$. However, at spatial
infinity the solution goes to a constant value of ${\vec n}$. Then one
can consider $\IR^3$ compactified into $S^3$, and one has the map
$S^3\rightarrow S^2$. The classes of homotopy are labeled by the  Hopf
invariant. This defines the Hopf topological charge of the
solution. Evaluating it you get \cite{s32} which is time
independent and equal to $m_1\,m_2$.

\subsection{Exact vortex solutions}

The extended Skyrme-Faddeed model defined by the action
\rf{sfextendedaction} has an interesting class of solutions that
belongs to an integrable submodel \cite{s35}. As we have seen, if we
impose the constraint
\be
\partial^{\mu}u\,\partial_{\mu}u =0
\lab{sfeq1}
\ee
the resulting submodel has an infinite number of conserved currents
given by \rf{niceformcurr}. However, if in addition we impose the
relation 
\be
\beta\,e^2=1
\lab{bogo3}
\ee
between the coupling constants you get
that the equation of motion \rf{sfextendedeqmot} reduces to 
\be
\partial^{\mu}\partial_{\mu} u =0
\lab{sfeq2}
\ee
Notice that the constraint \rf{sfeq1} and the equation of motion
\rf{sfeq2}, in Minkowski space-time (with Cartesian coordinates
$x^{\mu}$, $\mu =0,1,2,3$)
can be cast, respectively, into the form  
\be
\left[\(\partial_1+i\,\partial_2\) u\right]\;
\left[\(\partial_1-i\,\partial_2\) u\right] = -
\left[\(\partial_3+\partial_0\)u\right]\; 
\left[\(\partial_3-\partial_0\)u\right] 
\lab{bogo1}
\ee
and
\be
\partial^2u=0\qquad \quad{\rm or} \qquad\quad  \(\partial_1+i\,\partial_2\)
\(\partial_1-i\,\partial_2\) u = - \(\partial_3+\partial_0\)
\(\partial_3-\partial_0\)u 
\lab{bogo2} 
\ee
Of course the equations \rf{bogo1} and \rf{bogo2} are solved by field
configurations satisfying
\be
\(\partial_1+i\,\varepsilon_1\,\partial_2\)u=0 \qquad\qquad {\rm and}
\qquad \qquad 
\(\partial_3+\varepsilon_2\,\partial_0\)u=0
\lab{bogo4}
\ee
wherethe signs $\varepsilon_i=\pm 1$ are chosen independently. In
addition, equations \rf{bogo4} are satisfied by configurations of the
form \cite{s35}
\be
u=v\(z\)\, w\(y\)
\lab{ansatzu}
\ee
where $z=x^1+i\,\varepsilon_1\,x^2$ and $y=x^3-\varepsilon_2\,x^0$,
with $v\(z\)$ and $w\(y\)$ being arbitrary regular functions of their
arguments. Notice that if $u$ satisfies \rf{bogo4}, so does any
regular functional of it, ${\cal F}\(u\)$. Indeed, by taking ${\cal
  F}$ to be the logarithm one observes that the ansatz \rf{ansatzu} is 
mapped into $u=v\(z\) + w\(y\)$. See \cite{ward} for similar
discussions in $2+1$ dimensions.

We see that the extended Skyrme-Faddeev theory has a large class of exact
solutions. Among those there are interesting vortex solutions. For
instance, one can take $v\sim z^n$, and $w\(y\)=1$ in \rf{ansatzu}. Using polar
coordinates on the $x^1\, x^2$ plane, i.e.  
$x^1 + i\, \varepsilon_1\, x^2=\rho\, e^{ i\,\varepsilon_1\, \vp}$, one
obtains the static vortex 
\be
u= \(\frac{\rho}{a}\)^{n}\; e^{i\, \varepsilon_1\, n\, \vp}
\lab{staticvortex}
\ee
where $n$ is an integer, and $a$ an arbitrary parameter with dimension
of length. 

We can dress such vortices with waves traveling along the $x^3$
axis. There are several ways to do it. One way of keeping the energy
per unit of length finite is to take $w\(y\)$ in \rf{ansatzu} of the
plane wave form, leading to the vortex 
\be
u= \(\frac{\rho}{a}\)^{n}\; e^{i\left[ \varepsilon_1\, n\, \vp+
  k\,\(x^3-\varepsilon_2 \, x^0\)\right]}
\lab{vortexwave}
\ee
where $k$ is an arbitrary parameter with dimension of (length)$^{-1}$. 
These vortex solutions are of the Bogomolny type in the sense that
they saturate an energy bound related to the topological charge (see
\cite{s35} for more details). 

The energy per unit length for the static vortex \rf{staticvortex} and
the vortex with waves \rf{vortexwave} are given respectively by
\be
{\cal E}_{\rm stat. vortex} = 8\,\pi\,M^2\,\mid n\mid
\ee
and\footnote{For $n=\pm 1$, The energy per unit length for the vortex
  \rf{vortexwave} diverges.} 
\be
{\cal E}_{\rm vortex/wave} = 8\,\pi\,M^2\,\left[ \mid
  n\mid+k^2\,a^2\,I\(\mid n\mid\)\right]\qquad\qquad \mid n\mid >1
\ee
where
\be
I\(n\) = \frac{1}{n}\,
\Gamma\(\frac{n+1}{n}\)\,\Gamma\(\frac{n-1}{n}\)
\lab{indef}
\ee
The integer $n$ is the topological charge associated to the vortex,
and defined as the winding number of the map from any circle on the
$x^1x^2$ plane, centered
on the $x^3$-axis, to the circle $u/\mid u\mid$ on target
space. 

Gies \cite{gies} has calculated the Wilsonian low energy effective
action for the 
pure  (without matter) $SU(2)$ Yang-Mills theory, using the
Cho-Faddeev-Niemi decomposition \cite{cho1,niemi} of the gauge field
and has found that it corresponds to the action
\rf{sfextendedaction}. The condition \rf{bogo3} on the coupling
constants is compatible with the expression that Gies obtains in
terms of the gauge coupling constant, the infrared and the ultraviolet
cutoffs. It would be very interesting to investigate if such
vortex solutions can play a role in the low energy limit of 
Yang-Mills theory.  

\section{Applications and developments of the method}
\label{sec:app}

The previous sections presented selected illustrative examples.  We
now present a general review of the applications, with a special
emphasis on generalizations, on new results and on the physical relevance of
the models.

\subsection{Overview}

As was previously explained, the simplest way to implement the
generalized zero curvature formulation of integrability in $d+1$
dimensions with local field equations is to specify a Lie algebra
${\cal G}$ with an abelian ideal ${\cal P}$, and a $d$-form $B$ with
values in the ideal such that the vanishing of the covariant
differential $D\wedge B=0$, with respect to a flat connection $A_\mu$,
leads to generalized zero curvature equations.

To test this formulation, the first applications were for simple well
studied systems in $2+1$ dimensions like $CP^1$ and the principal
chiral model, with two and three dimensional target spaces
respectively.  As we have seen, the main results were the possibility
of integrability conditions for the complex field $u$ of the eikonal
type $(\partial_{\mu} u)^{2}=0$, producing sectors with infinitely
many conserved currents explicitly given by the construction.  In this
generalized sense these sectors are called integrable.

These first applications were accomplished by ordinary algebraic
methods with different techniques, either directly \cite {gmsg} or by
coset methods \cite {lf}, and they exhibited the difficulty of
obtaining solutions and implementing dressing methods which led later
to more involved realizations \cite {babelon} as explained in
Section~\ref{sec:s2}.  An appealing result was the generalization of
BPS first order equations in a strong form and a weaker form, and to a
systematic procedure to obtain an infinite number of conserved
currents \cite {fujii, fujii2}.

This led to a study of the Skyrme problem in 3+1 dimensions \cite
{skyrme1} which is well known for the lack of a BPS reduction.  Using
previous experience, the formulation with target space restricted from
$ S^3 $ to $ S^2$ was first undertaken, the so called
Skyrme-Faddeev-Niemi model \cite {niemi}.  The original model consists
of two terms, the first one is the sigma model ${\cal L}_2$ and the
the second ${\cal L}_4$, the quadratic Skyrme one.

Given the difficulty, an interesting development for such target
spaces was the study by Aratyn, Ferreira and Zimerman (AFZ) of a more
simple theory {$-{\cal L}_{4}^{3/4}$}, \cite  {afz1}, \cite  {afz2}
with the power chosen to  avoid scaling  instability. The model has
similar topological properties. 

In the simplest case, the field of the theory describes a map from the
one-point compactified three-dimensional space $\IR_0^3$ to the
two-sphere $S^2$.  $\IR_0^3$ is topologically equivalent to the
three-sphere $S^3$, therefore such maps are characterized by the third
homotopy group of the target space $S^2$, which is nontrivial, $\pi_3
(S^2) = Z$.  As a consequence, fields which describe maps $\IR_0^3 \to
S^2$ fall into different homotopy classes, and a soliton is a field
configuration which minimizes a given energy functional within a fixed
homotopy class.  The topological index characterizing the homotopy
class is called the Hopf index, the corresponding map is called a Hopf
map.  The minimizing solutions are sometimes called Hopf solitons.
The model has in fact infinitely many conserved currents and
infinitely many Hopf soliton solutions, characterized by explicit
topological Hopf charges.  Therefore the AFZ model is integrable in
our generalized sense.

This model was generalized in \cite{aw1}, \cite {S-2-syms}. Other
simple models with $S^2$ target space, like the Nicole model, were
analyzed using this scheme \cite{nicole}, \cite {we-nicole}.  The 
lagrangian for this model has a term that is a fractional power of 
the traditional sigma model lagrangian density:
  \be \lab{Ni-La}
{\cal L}_{\rm Ni}= ({\cal L}_2)^\frac{3}{2}.
\ee 
This model was shown to have a soliton solution with Hopf charge 1, 
but was
not integrable.  Some integrable sectors and exact energy bounds
were obtained.

A geometric understanding of these models and their conserved currents
was developed, with the important result that the AFZ model has
infinite target space symmetries which were related to the Noether
currents of the area preserving diffeomorphisms on target space
\cite{diffeo1}.  On the other hand, the Nicole model had only the
obvious symmetries: the conformal ones of base space and the modular
symmetries of the target space.  This was sufficient to implement the
(toroidal) Ansatz in a convenient way.  These and other important
insights \cite{babelon} were further extended \cite{we-abel} and
generalized  to produce new models in higher dimensions and with
higher textures \cite{s31}, \cite{s33}, \cite{pull}.

The next model studied with a 3-dimensional target space was the zero
curvature formulation of the ordinary Skyrme model \cite{skyrme2}
that is discussed below in some detail.  The existence of a subsector
with infinitely many conserved currents was unraveled that contained the
hedgehog solution with topological charge one (the nucleons)
\cite{FSG}.  Further sectors were later found and a geometric
understanding of the integrability conditions along with a general
analysis of theories with a 3-dimensional target space followed leading
to a detailed classification of possible relevant theories 
\cite{ASGW1-S3}.
 
The direct study of gauge theories was then started \cite{s32}, along
with the abelian Higgs model \cite{ASGW2-S3}, the Yang Mills dilaton
theory \cite{YMdil}, and a fresh look at the self dual Yang Mills
equations \cite {sd}.  As we shall see, these analyses exhibited the
interplay of gauge invariance and the integrability conditions, and
the importance of base space properties, especially symmetries, for
constructing convenient ansaetze.  This led to new exact solutions
\cite {s34} and the clarification of the zero curvature integrability
conditions as a weaker generalization of BPS equations for sectors
where the latter cannot hold.  Closely related to the gauge theory,
the Skyrme Faddeev formulation has been revisited from different
viewpoints \cite{wesf}, \cite{s35}, \cite{1st}.

\subsection{Integrable models with 2-dimensional target space}
\label{sec:2d-target}

Here we discuss generic features of the generalized zero curvature
(GZC) integrability method that apply to many models: from the initial
$CP^1$ model (also called the Baby Skyrme model) to extensions of the
Skyrme-Faddeev model.  The applications are important because they
range from exact solutions \cite{afz1} to models with potential
physical relevance \cite{s35}.

First we review how the direct approach works.  We consider models
defined on a spacetime of $d+1$ dimensions and with a two-dimensional
manifold $\mathcal{M}$ as the target space.  As before we use the
complex coordinates $u$ on the target manifold.

According to the general prescription we fix the Lie algebra and an
abelian ideal.  We will be studying a Lie algebra with structure
\eqref{eq:poincarelike}.  The simple part of the algebra is
$\mathfrak{su}(2)$, the Lie algebra of $SU(2)$.  We take the abelian
ideal $\mathcal{P}$ be a representation space of $\mathfrak{su}(2)$.
The standard basis for the representation will be denoted by
$\{P^{(j)}_{m}\}$ where $j$ is the angular momentum quantum number and
$m$ is the magnetic quantum number.  In the zero curvature formulation
we will only need the use of elements of $\mathcal{P}$ with magnetic
quantum number restricted to $m=\pm 1$.  

First we choose
the triplet representation and let $\partial_{\mu} u \equiv
u_{\mu}$ etc., we see that according to \rf{s2pota} we have
\begin{equation}
A_{\mu}=-\partial_{\mu}W W^{-1}=\frac{1}{1+|u|^2} \left( -iu_{\mu}
T_+-i{u^\ast}_{\mu}T_- + (u{u^\ast}_{\mu} -{u^\ast} u_{\mu})T_3 
\right),
\end{equation}
and
\begin{equation}
\tilde{B}_{\mu}=\frac{1}{1+|u|^2} \left(
{\mathcal{K}^\ast}_{\mu}P^{(1)}_1- \mathcal{K}_{\mu} P^{(1)}_{-1}
\right),
\end{equation}
where $\mathcal{K}_{\mu}$ is so far an arbitrary
vector\footnote{Notice a difference (conjugation) in the convention for
the basis of the abelian ideal.}  depending on the fields as well as
their derivatives, $W$ is an element of $SU(2)/U(1)$ given by
\rf{wdef}.
Note that $T_{3}$,  $T_{\pm}=(T_1\pm
T_2)/2$ and $P^{(j)}_m$ constitute a basis of the Lie algebra 
including
the Abelian ideal.  $T_1,T_2,T_3$ may be taken to be Pauli matrices.
The commutators are $[T_3,T_{\pm}]=\pm T_{\pm}, \;\; [T_+,T_-]=2T_3$,
$[T_3,P_m^{(j)}]=m P_m^{(j)}$, $[T_{\pm},P_m^{(j)}] =
\sqrt{j(j+1)-m(m\pm 1)} P_{m \pm 1}^{(j)}$,
$[P^{(j)}_m,P^{(j')}_{m'}]=0$.  The connection $A_{\mu}$ is flat by
construction.  Thus, the only nontrivial condition in the GZC
formulation is that the covariant divergence of the $\tilde{B}_{\mu}$ 
field vanishes.
In the triplet representation the results is
\begin{equation}
(1+|u|^2)\partial^{\mu} \mathcal{K}_{\mu}-2u \mathcal{K}_{\mu}
{u^\ast}^{\mu}=0 \quad {\rm and} \quad \Im (u^\mu \mathcal{K}_\mu )=0.\lab{2dim int eq gen}
\end{equation}
However, in a higher spin representation you get
\rf{2dim int eq gen} and an additional constraint
\begin{equation}
\mathcal{K}_{\mu} {u^\ast}^{\mu}=0. \lab{2dim constrain}
\end{equation}
Note that we can take the abelian ideal to be infinite dimensional and
spanned by all the irreducible representations. In this way we will get
an infinite number of conservation laws.

We can conclude that a dynamical model with a two dimensional target
space is integrable if we can define a vector quantity
$\mathcal{K}_{\mu}$ such that $\mathcal{K}_{\mu}{u^\ast}^{\mu}\equiv 0
$ and the relevant equations of motion read
\begin{equation}
\partial^{\mu} \mathcal{K}_{\mu}=0. \lab{2dim int eq}
\end{equation}
Models with these properties are now known as models of the
AFZ type \cite{afz2}. They are integrable in
the GZC  sense: they have a GZC formulation with an 
infinite-dimensional Abelian ideal. They are given by the
following Lagrange density
\begin{equation}
\mathcal{L}= \omega(u{u^\ast})  H^q,
\end{equation}
where
\begin{equation}
H \equiv  u_{\mu}^2 {u^\ast}_{\nu}^2 - (u_{\mu} {u^\ast}^{\mu})^2.
\lab{H afz}
\end{equation}
and $\omega$ is any function of $u{u^\ast}$ and $q$ is a positive real
parameter.  A specially important example of such an integrable models
in four dimensional Minkowski space-time is given by the expression
\begin{equation}
\mathcal{L}_{AFZ}= \omega(u{u^\ast}) H^{\frac{3}{4}}, \lab{afz}
\end{equation}
where the value of the power is taken to avoid the Derrick's argument
for the non-existence of static solitons \cite{afz1}.  The AFZ model
describes soliton excitations of a three component unit vector field
$\bm{n}= \( n_1, n_2, n_3\)$, with $\bm{n}^2=1$, that may be related
via the standard stereographic projection with the unconstrained
complex field $u$.

The static solutions are maps from
compactified $\IR^3$ to the $S^2$ target space and carry the
corresponding topological charge, \emph{i.e.}, the Hopf index $Q \in
\pi_3(S^2) \cong \bbZ$.  The pre-images of points on the target sphere
are closed lines, which can be linked forming knots and thus provide
topological stability to the soliton solutions.  In this model such
topologically nontrivial solitons (hopfions) have been in fact derived
in an exact form \cite{afz2}.  
Moreover, you can also construct infinitely many conserved currents
\begin{equation}
j_{\mu}= G_{{u^\ast} } \mathcal{K}_{\mu}- G_{u }
{\mathcal{K}^\ast}_{\mu}, \lab{afz currents}
\end{equation}
where $G$ is an arbitrary function of $u$ and $u^{*}$, $G_{u }\equiv
\partial_{u }G $, etc.  The currents give an explicit form for the
generalized momentum ${\cal{K}}_{\mu}$ of the integrable AFZ field.
They are related on the one hand to the volume preserving
diffeomorphisms of the target space and on the other to the conformal
properties of the base space because of the product structure.  When
the model is integrable, both sets provide symmetries and conservation
laws, according to the Noether theorem.  If the model is not
integrable, those currents can be conserved currents of an integrable
subsector of the model defined by imposing the constraints \rf{2dim
constrain}, as illustrated with the $ CP^1$ and Skyrme Faddeev models.
In some cases, also the symmetries of the submodel are enhanced (this
is especially transparent in the $CP^1$ model, where the submodel has
an additional conformal symmetry on target space).  Besides, the
submodel has the Noether currents of the volume preserving
diffeomorphisms as additional conserved currents.  As both the
enhanced symmetries and the enhanced conservation laws only exist at
the level of the submodel, the Noether theorem does not apply (the
constraints \rf{2dim constrain} are not of the Euler-Lagrange type),
and there is no one-to-one correspondence between the enhanced
symmetries and the enhanced conservation laws.  For the $CP^1$,
Faddeev-Niemi and Nicole models, we present the corresponding results
in Table~\ref{tbl:models}

\begin{table}
    \centering
    \begin{tabular}{||r|c|l|c||}
    \hline
     & $\infty $ many & geometric &  solutions \\
     model & conserv. laws & symmetries &  known \\
    \hline
    Baby Skyrme & yes$^a$ & $C_2 \times SU(2) $  & yes \\
    \hline
    submodel & yes & $C_2 \times C_2 $ &  yes \\
    \hline
    Nicole  & no & $C_3 \times SU(2)$  & yes \\
    \hline
    submodel & yes & $C_3 \times SU(2)$ &  yes \\
    \hline
    Faddeev--Niemi & no & $E_3 \times SU(2)$  & yes$^b$ \\
    \hline submodel & yes & $E_3 \times SU(2)$  & no \\
    \hline 
    \end{tabular} 
    \begin{quotation}
	\caption[aaa]{\small Some results for the three soliton models and their
	submodels.  Here $C_d$ is the conformal group in $d$
	dimensions and  $E_d$ is the Euclidean group
	(translations and rotations) in $d$ dimensions.\\
	\strut$\quad{}^a$~due
	to the infinite-dimensional base space symmetries $C_2$\\
	\strut$\quad{}^b$~only known  numerically	}
	\label{tbl:models}
    \end{quotation}
\end{table}

\subsection{A weaker integrability condition}

As we have seen, the AFZ model is special because it has infinitely many
conserved currents and also infinitely many explicit solutions in an
Ansatz that realizes the generalized integrability.

Many other models we have considered do not have an infinite number of
symmetries, but we saw that the complex eikonal equation
$(\partial_{\mu} u)^2=0$ defines integrable subsectors with infinitely
many conserved currents $J^{G}_\mu$ parametrized by an arbitrary
function $G(u,{\bar {u}})$. These are the Noether currents for the
area preserving diffeomorphisms on the target space.  In \cite{we-abel}
it was demonstrated that for all theories with a two dimensional target
space there is a weaker condition than the complex eikonal equation 
given by 
\begin{equation}
{u^\ast}^2(\partial_{\mu} u)^2-{u}^2(\partial_{\mu} u^{*})^2=0,
\lab{weaker}
\end{equation}
that leads to sectors with infinitely many conserved currents.  The
infinitely many conserved currents have an additional restriction
$G=G(u{\bar u})$ (not  separately on $u,\bar u$). These are the
Noether currents for an abelian subgroup of the area preserving
diffeomorphisms on the target space.

The meaning of the new condition \rf{weaker} is better seen
reexpressing the field in terms of the modulus and the phase $u=\exp
(\Sigma + i \phi)$, where the condition becomes $\partial^{\mu} \Sigma
\partial_{\mu}\phi =0$, which is nothing but the orthogonality of
gradients of the modulus and phase in the relevant static case.  The
eikonal equation implies an additional condition that their squares
must be equal $({\partial_{\mu}\Sigma })^2 =
({\partial_{\mu}\phi})^2$.

The strong and weak integrability conditions can be generalized to the
$U(1)$ gauge case with 3-dimensional target spaces.  They are given by
``minimally replacing'' the partial derivatives $\partial_{\mu}$ with
the covariant derivatives $D_{\mu}$ \cite{ASGW2-S3}.  There is however
a difference; while the currents from the weak condition are gauge
invariant and have gauge invariant conservation equations, in the
strong case one has only the (gauge invariant) existence of a gauge
where the currents are conserved.  The geometrical meaning of the weak
condition is also the orthogonality of the covariant derivative of the
phase and modulus.  These results have been analyzed in a number of
models, from the abelian Higgs model, to abelian generalizations of $CP^1$
and to Chern Simons theory.  We see a pattern where the weak sector is a
generalization of the strong and the Bogomolny sectors, providing
infinitely many conserved currents in sectors without the Bogomolny
condition nor the eikonal constraint (its Lorentz invariant
generalization).  This typically happens outside some critical values
of the couplings.  In addition, all theories considered here with
known solutions have been shown to be described by first order
equations \cite{1st}.

\subsection{Integrable models with 3-dim target space}

The bigger complexity of the 3-dimensional target space will allow for
many more possibilities for integrable models and submodels.

The Lie algebra setup is identical to Section~\ref{sec:2d-target}
except for one important change.  In 2d target spaces the
constructions for the various fields only involved $P^{(j)}_{\pm 1}$.
In a 3d target space will will also have to introduce $P^{(j)}_{0}$.
In the spin $j$ representation, the flat connection and the Hodge
dual field are
\begin{equation}
A_{\mu} = -\partial_{\mu} W \; W^{\dagger} = \frac{1}{1+|u|^2}
\left( -iu_{\mu} T_+ - i{u^\ast}_{\mu} T_- + (u
{u^\ast}_{\mu}-{u^\ast}u_{\mu} ) T_3 \right)
\label{eq:A-dW}
\end{equation}
as before and
\begin{equation}
\tilde{B}_{\mu}^{(j)} = \frac{i}{(1+|u|^2)^2} \mathcal{H}_{\mu}
P^{(j)}_0+\frac{1}{1+|u|^2} \left({\mathcal{K}^\ast}_{\mu}
P^{(j)}_1 - \mathcal{K}_{\mu} P^{(j)}_{-1} \right).
\end{equation}

The main new ingredient is the appearance of a new field functional
along the direction $P^{(j)}_0$ denoted by
$\mathcal{H}_{\mu}$.  For Lagrangians that can be expressed as the
$q$-th power of the pullback of the volume three form $H$ times a
factor that does not contain the derivatives of the fields it is easy
to show the following important properties are obeyed by the objects
$\mathcal{K}_{\mu}$ and $\mathcal{H}_{\mu}$:
\begin{equation}
\mathcal{H}_{\mu} {u^\ast}^{\mu}=0, \;\; \mathcal{K}_{\mu}
\xi^{\mu}=0, \;\;\;\; \mathcal{H}_{\mu} u^{\mu}=0, \;\;
\mathcal{K}_{\mu} {u^\ast}^{\mu}=0.
\end{equation}

 The vanishing covariant divergence of the Hodge
dual field $\tilde{B}_{\mu}$ gives
$$
 \partial^{\mu} \mathcal{H}_{\mu} P^{(j)}_0 +
(1+|u|^2) \left(\partial^{\mu}{\mathcal{K}^\ast}_{\mu}
P^{(j)}_1 - \partial^{\mu} \mathcal{K}_{\mu} P^{(j)}_{-1} \right)
-  \left(u{u^\ast}^{\mu}
{\mathcal{K}^\ast}_{\mu} P^{(j)}_1 - {u^\ast} u^{\mu}
\mathcal{K}_{\mu} P^{(j)}_{-1} \right)-
$$
\begin{equation}
i \left( {u^\ast}^{\mu}
{\mathcal{K}^\ast}_{\mu} -u^{\mu} \mathcal{K}_{\mu}\right)
\sqrt{j(j+1)} P^{(j)}_0 + u{u^\ast}^{\mu}
{\mathcal{K}^\ast}_{\mu} P^{(j)}_1  - {u^\ast}
u^{\mu} \mathcal{K}_{\mu} P^{(j)}_{-1}=0.
\end{equation}
Moreover, if we notice that
\begin{equation}
\mathcal{K}_{\mu} u^{\mu}= {u^\ast}^{\mu} {\mathcal{K}^\ast}_{\mu}
\end{equation}
then we arrive  at the field equations 
\begin{equation}
\partial_{\mu} \mathcal{K}^{\mu}=0, \;\;\;\;\;\partial_{\mu} \mathcal{H}^{\mu}=0.
\end{equation} 
Therefore, we conclude that these models are
integrable. The Abelian ideal we used in the generalized zero
curvature is  infinite dimensional.

Observe that the group element $W\in SU(2)$ that appears in
\eqref{eq:A-dW} for $A_{\mu}$ may be identified with an element of the
coset space $SU(2)/U(1)$ as is the case for models with a two
dimensional target space.  Moreover, the dual field $\tilde{B}_{\mu}$
is defined up to an arbitrary function of $u$ and ${u^\ast}$ which
multiplies $P^{(j)}_0$.  \\
Finally, you can verify that this family of models possesses  three families
with infinitely many on-shell conserved currents
\begin{equation}
j_{\mu}^{(G)} = G_{{u^\ast}} \mathcal{K}_{\mu} -G_{u}
{\mathcal{H}}^{\ast}_{\mu} \lab{curr G n3 1},
\end{equation}
\begin{equation}
j_{\mu}^{(\tilde G)} = \tilde G_{\xi} \mathcal{K}_{\mu}- \tilde G_{u}
\mathcal{H}_{\mu} \lab{curr G n3 2},
\end{equation}
\begin{equation}
j_{\mu}^{(\tilde {\tilde G})} = \tilde {\tilde G}_{\xi}
{\mathcal{K}}^{\ast}_{\mu}
- \tilde {\tilde G}_{{u^\ast}}
\mathcal{H}_{\mu}. \lab{curr G n3 3},
\end{equation}
where
\begin{equation}
G=G(u,{u^\ast},\xi), \;\;\; \tilde G=\tilde G(u,{u^\ast},\xi),
\quad \tilde {\tilde G} = \tilde {\tilde G} (u,{u^\ast},\xi).
\lab{curr dep n3}
\end{equation}
Moreover, there is a good understanding of the geometrical origin of
the currents and you can show that the conservation laws found for the
integrable models are generated by a class of geometric target space
transformations.  As expected, they are the Noether currents that
generate the volume-preserving diffeomorphisms \cite{pull}.

This is not the case for the GZC formulation of the Skyrme model
\cite{FSG}, where you also have a field functional in the Cartan
direction.  The analogous fields do not have such a simple geometrical
formulation and the model is of course not integrable.  It admits
again integrability conditions giving sectors with infinite conserved
currents, as we will see in the next section.

\subsection{The Skyrme model}

The Lagrangian for the Skyrme model is
\begin{equation}
{\cal L} =  {f_\pi^2 \over 4}
\tr \left ( U^\dagger \pa_{\mu} U U^\dagger \pa^{\mu} U \right ) -
{1 \over 32 e^2}
\tr \left [ U^\dagger\pa_{\mu} U,U^\dagger\pa_{\nu} U \right ]^2
\end{equation}
where $f_\pi$ and $e$ are phenomenological constants and $U$ is a $SU(2)$
unitary matrix. We shall use a special parametrization 
$U \equiv e^{i \zeta_j \tau_j}$,
where $\tau_j$, $j=1,2,3$, are the Pauli matrices. You find that
\be
U = e^{i \zeta T} = \one \; \cos \zeta + i \; T \; \sin \zeta
\ee
where $\zeta \equiv \sqrt{\zeta_1^2 + \zeta_2^2 + \zeta_3^2}$ 
is the unit vector of the stereographic projection and
\br
T \equiv \frac{1}{1+\u2}\; \(
\begin{array}{cc}
\u2 - 1 & -2 i u\\
2 i u^* & 1 - \u2
\end{array}\)
\er
The equations of motion are then given by 
\be
D^{\mu} B_{\mu} = \pa^{\mu} B_{\mu} + \sbr{A^{\mu}}{B_{\mu}} = 0
\lab{zcsky}
\ee
with $A_{\mu}$  given by \rf{s2pota} and
\br
B_{\mu}&\equiv & -\frac{i}{2}\, R_{\mu} T_3 +
\frac{2 \sin \zeta}{1+ \u2}\; \(
e^{i\zeta}\; S_{\mu} \; T_{+} -
e^{-i\zeta}\; S_{\mu}^* \; T_{-} \)
\lab{bsky}
\er
where
\br
R_{\mu}&\equiv& \pa_{\mu} \zeta - 8 \lambda \; \frac{\sin^2 \zeta}{\(1
  + \u2\)^2} 
\(  N_{\mu} +  N_{\mu}^*\)\nonu\\
S_{\mu}&\equiv& \pa_{\mu}u + 4 \lambda \; \( M_{\mu} - \frac{2\sin^2
\zeta}{\(1+\u2\)^2}  \; K_{\mu}\)
\lab{rsdef}
\er
and
\br
K_{\mu} &\equiv& \( \pa^{\nu} u \pa_{\nu} u^*\) \pa_{\mu} u - \(\pa_{\nu}u\)^2
\pa_{\mu} u^*\nonu\\
M_{\mu} &\equiv& \( \pa^{\nu} u \pa_{\nu} \zeta\) \pa_{\mu} \zeta
- \(\pa_{\nu}\zeta\)^2
\pa_{\mu} u\nonu\\
N_{\mu} &\equiv& \( \pa^{\nu} u \pa_{\nu} u^*\) \pa_{\mu} \zeta -
\(\pa_{\nu}\zeta \pa^{\nu} u\) \pa_{\mu} u^* \lab{kmndef} \er In order
to obtain the skyrmion sector you have to impose \cite{FSG} \be
S_{\mu} \pa^{\mu} u = 0 \; ; \qquad \qquad R_{\mu} \pa^{\mu} u = 0
\lab{const1sky} \ee or in a more restricted form \be \( \pa^{\mu} u
\)^2 = 0 \; ; \qquad \qquad \pa^{\mu} \zeta \; \pa_{\mu} u = 0
\lab{sufcond} \ee the first of which is the strong condition.  There
are families of conserved currents.  An important point is that the
skyrmions with charge $Q =1$ satisfies the above equations.  The
rational map Ansatz, widely used in numerical analysis, {\it cannot}
provide exact solutions for charges bigger than one due to the
restrictive character of the eikonal equation \cite{FSG}.  We proved
\cite{asgw-cpn} that relaxing the eikonal equation and imposing the
weaker integrability condition \rf{weaker} leads to infinitely many
conserved currents in sectors that include the skyrmion and the
rational maps of higher degree.  The geometrical interpretation in
terms of the orthogonality of the gradients of the phase and the
modulus is maintained and a general classification of possible models
is provided~\cite{ASGW1-S3}.

\subsection{Yang Mills systems}

It is clear that a formalism based to a large extent on gauge
transformation properties may be applicable to gauge dynamics.  In
fact, our first proposal \cite{Alvarez:1997ma} demonstrated that the
self-dual sectors of the Yang Mills systems could be accommodated in
the scheme although only a finite number of conservation laws were
given.  The subsequent analysis of the Skyrme Faddeev systems was
related to Yang Mills theory and it was conjectured that the system
was connected to the infrared behaviour of gluonic QCD. This is a
matter of intense debate \cite{gies,s35}.  One of the arguments was
that the Skyrme-Faddeev Lagrangian \rf{sfextendedaction} could be
obtained by a decomposition of the Yang Mills field
\cite{cho1,shabanov1}.  This nonlocal change of variables
\begin{equation}
\vec{A}_\mu =C_{\mu}\vec{n}+\partial_{\mu}\vec{n} \times \vec{n} + \vec
W_{\mu}, \lab{decomp}
\end{equation}
known as the CFNS decomposition relates the original $SU(2)$
non-Abelian gauge field with three fields: a three component unit
vector field $\vec{n}$ pointing into the color direction, an Abelian
gauge potential $C_{\mu}$ and a color vector field $W_{\mu}^a$ which
is perpendicular to $\vec{n}$.  The fields are not independent.  In
fact, if we want to keep the correct gauge transformation properties
\begin{equation}
\delta n^a=\epsilon^{abc}n^b\alpha^c, \;\;\; \delta
W^a_{\mu}=\epsilon^{abc}W^b_{\mu}\alpha^c, \;\;\; \delta
C_{\mu}=n^a\alpha^a_{\mu} \lab{gauge trans decomp}
\end{equation}
under the primary gauge transformation
\begin{equation}
\delta A^a_{\mu}=(D_{\mu} \alpha)^a = \alpha^a_{\mu} +
\epsilon^{abc} A^b_{\mu} \alpha^c \lab{gauge trans su2}
\end{equation}
then you have to impose the constraint ($n^b_{\mu}\equiv \partial_\mu
n^b$, etc.)
\begin{equation}
\partial^{\mu}W^a_{\mu}+C^{\mu}\epsilon^{abc}n^bW^c_{\mu} + n^a
W^{b\mu}n^b_{\mu}=0. \lab{decomp constrain}
\end{equation}
In a subsequent analysis we assumed a particular form for the so
called valence field $W_{\mu}^a$.  This is equivalent to a partial gauge
fixing and there is a residual local $U(1)$ gauge symmetry given by
\begin{equation} 
W^a_{\mu} =  \rho
 n^a_{\mu} + \sigma \epsilon^{abc} n^b_{\mu} n^c, \lab{W-def}
\end{equation}
where $\rho, \sigma$ are real scalars.  It is convenient to
combine these into a complex scalar $v=\rho + i \sigma$.  The Lagrange
density now takes the form ($u_\mu \equiv \partial_\mu u$, etc.)
\begin{eqnarray}
L&=&F^2_{\mu \nu}  - 2(1-|v|^2)H_{\mu\nu} + (1-|v|^2)^2 H_{\mu\nu}^2 
\nonumber\\
&+& \frac{8}{(1+|u|^2)^2} \left[ (u_{\mu} {u^\ast}^{\mu}) (D^{\nu} v
\overline{D_{\nu}v}) 
- (D_{\mu}v {u^\ast}^{\mu}) (\overline{D_{\nu}v}
u^{\nu}) \right], \lab{lagrangian}
\end{eqnarray}
where
\begin{equation}
H_{\mu \nu}= \vec{n} \cdot \left[ \vec{n}_{\mu} \times \vec{n}_{\nu}
\right]= \frac{-2i}{(1+|u|^2)^2} (u_{\mu}{u^\ast}_{\nu} - u_{\nu}
{u^\ast}_{\mu} ), \quad H_{\mu \nu}^2= \frac{8}{(1+|u|^2)^4} [
(u_{\mu} {u^\ast}^{\mu})^2 -u_{\mu}^2{u^\ast}^2_{\nu} ] \lab{H}
\end{equation}
and the covariant derivatives are $D_{\mu} v =v_{\mu} -ieC_{\mu}v$,
$\overline{D_{\mu} v}={v}^{\ast}_{\mu}+ieC_{\mu}{v^\ast}$. Here we
expressed the unit vector field by means of the stereographic
projection and $F_{\mu \nu}\equiv \partial_\mu C_\nu -\partial_\nu
C_\mu$ is the Abelian field strength tensor corresponding to the
Abelian gauge field $C_{\mu}$.  Notice that only the complex field $v$
couples to the gauge field via the covariant derivative.  A direct
application of the method developed in \cite {asgconf,we-abel} to
self-dual equations using these variables leads to an infinite family of
conserved currents \cite{sd}

\begin{equation}
j_{\mu}^{G}= i(1+|u|^2)^2 \left({\pi}{^\ast}_{\mu} \frac{\partial
G}{\partial u}- \pi_{\mu} \frac{\partial G}{\partial {u^\ast}}
\right), \lab{curr def}
\end{equation}
where $G$ is an arbitrary function of the field $u$ and $\pi_{\mu}$
the canonically conjugate momentum of $u$.  The currents \rf{curr def}
are invariant under the residual U(1) gauge transformations that
remains after the partial gauge fixing implied by the CFNS
decomposition.  These charges obey the algebra of area-preserving
diffeomorphisms
\begin{equation}
\{ Q^{G_1}, Q^{G_2} \} = Q^{G_3} \quad ,
\qquad G_3 = i (1+|u|^2)^2
(G_{1, u^\ast} G_{2,u} - G_{1,u}G_{2, u^\ast}) .
\end{equation}
on the target $S^2$ under the ordinary Poisson bracket, i.e. with the
canonical momenta of the original Yang-Mills system.  The
diffeomorphisms generated by these charges are not necessarily
symmetries of the Yang-Mills self-dual equations.  After incorporating
the nonlocal decomposition \rf{decomp} these field equations are not
derivable from a Lagrangian and therefore the Noether theorem does not
apply.
 
The Yang-Mills dilaton system has been analyzed along the same lines
\cite{YMdil} with the gauge invariant restriction $W^a_{\mu}=0$
instead of \rf{W-def}.  Infinitely many static solutions were found
that are limits of solutions of the full system known from numerical
analysis.

\section{Conclusions and outlook}
\label{sec:conclusions}
 
We have reviewed our proposal of the generalized zero curvature (GZC)
formulation of integrable field theories in any dimension.  Beginning
with the conceptual foundations, our discussion of connections in the
space of loops, has been thoroughly revised.  New concepts like
$r$-flatness were discussed and consequences of it were obtained:
$r$-flat connections are curvature local and their holonomies are
abelian.  This justifies the many implementations of the approach in
the last decade that were summarized in this review.  These approaches
are essentially based on local sufficiency conditions that a certain
covariantly closed differential form $B$ takes values in a abelian
ideal of the gauge group.  On the other hand, these results also
indicate the limitations of the approach.  In any case, our approach
has produced new {\it integrable} relativistic invariant theories in
higher dimensions with exact solutions as discussed in this review.
Well known physically relevant theories, from Skyrme to Yang-Mills systems,
have been revisited with interesting discoveries  like new integrability
conditions that lead to sectors with infinite conserved currents.
All these systems satisfy the complex eikonal equation or a weaker form
that generalize the BPS equations.  The understanding of these results in
terms of the geometry and the symmetries of the target and the base spaces has
led in turn to new applications, to classifications and additional
developments that will hopefully improve our understanding
of the approach.

The main conceptual challenges of our scheme are a deeper connection
between loop and base spaces, including the better understanding of
non-localities.  More concretely, one would like to implement our new
{\it r-flatness}, a weaker condition, to represent equations of local
field theories and to use the holonomies for new conservation laws.
Perhaps one would also understand better the implementation of
dressing methods to generate solutions, which is another remaining problem.
Besides, there are interesting open problems already at the level of
the present applications.

Among them we have a better understanding of the physical properties
of the conserved currents and charges, as well as the algebraic
properties of many of them.  Other promising problem is the emergence
of a generalization of weaker conditions than BPS to yield first order
equations in relevant theories, which cannot have the standard ones,
like Skyrme theory.  Another possibility of the approach is the
analysis of time dependence and Q balls, which has been already
initiated with interesting results.  The extension to supersymmetry
and higher rank algebras has been explored and should be developed.
Also, as the approach is rather independent of the dimensions, it
could be used for topological quantum field \cite{l} and string
theories. The special role of symmetries, specially conformal, of our approach
could provide a hint for the latter.

Another interesting development is the consideration of more
phenomenologically relevant theories and making a more detailed
analysis of the ones already discussed, like Skyrme and variations of
Yang Mills theory, as e.g. the combination with Higgs and dilaton
fields and the Einstein Yang Mills equations, which are under
investigation at present.  In this line there are other field
decompositions besides the CFNS considered here, like the color spin
\cite{fcs} separation and different abelian projections, which are
worth studying.  The stability of higher dimensional solutions is also
an important question.

A  potentially interesting phenomenological application is an exact
analysis of color fields
that could interpolate between the naive geometrical models and the
highly involved AdS correspondence used  to analyze heavy ion
collisions and color fields at high densities and high
temperatures. Analysis of elliptical flow and the breaking of the
conformal invariant regime with the dilaton has been initiated within 
our scheme.

One of the main challenges in any case is going  to the quantum level
since the approach is essentially classical. One possibility is to
consider dualities making our non-perturbative results into special
versions of quantum properties. More specifically some of the
classical solutions could dominate path integrals or world-line
methods. The analysis of the differences between 
Euclidean and Minkowski formulations is also an interesting problem to
be treated along these lines. 

Further, one can make contact, in both directions with results of
renormalization group evolution, including new terms in the classical
models describing the infrared regime, or considering the
exponentiation of additional contributions in the action of the path
integrals \cite {s35}.
 
Of course, most classical models analyzed can be seen as different
effective theories and also  one can in principle try to perform
collective coordinate quantizations.  

\section*{Acknowledgments}

The work of OA was supported in part by the National Science
Foundation under Grants PHY-0244261 and PHY-0554821.  The work of JSG
by Ministerio de Ciencia e Innovaci\'{o}n and Conseller'a de
Educaci\'{o}n under FPA2008-01177 and 2006/51. LAF is partially
supported by CNPq (Brazil).

\providecommand{\href}[2]{#2}\begingroup\raggedright\endgroup

\end{document}